\documentclass[manuscript]{aa}

\usepackage{graphicx}
\usepackage[intlimits,sumlimits]{amsmath}
\usepackage{deluxetable}
\usepackage{times,epsfig} 
\usepackage{amssymb}
\usepackage{mathrsfs}  
\usepackage{natbib}
\bibpunct{(}{)}{;}{a}{}{,} 
\usepackage{amsmath}
\usepackage[english]{babel}
\usepackage{longtable}
\usepackage{url}
\usepackage{hyperref}

\begin{document}

\title{A metallicity study of 1987A-like supernova host galaxies\thanks{Based on observations performed at the Nordic Optical Telescope (Proposal number 44-011 and 45-004, PI: F. Taddia; and 47-701), La Palma, Spain; the Very Large Telescope (Program 090.D-0092, PI: F. Taddia), Paranal, Chile.  Some data were also obtained from the ESO Science Archive Facility and from the Isaac Newton Telescope archive.}}

\author{F. Taddia\inst{1}
\and J. Sollerman\inst{1}
\and A. Razza\inst{1}
\and E. Gafton\inst{1}
\and A. Pastorello\inst{2}
\and C. Fransson\inst{1}
\and M.~D. Stritzinger\inst{3}
\and G. Leloudas\inst{4,5}
\and M. Ergon\inst{1}}

\institute{The Oskar Klein Centre, Department of Astronomy, Stockholm University, AlbaNova, 10691 Stockholm, Sweden\\ (\email{francesco.taddia@astro.su.se})
\and INAF-Osservatorio Astronomico di Padova, Vicolo dell’Osservatorio 5, 35122 Padova, Italy
\and Department of Physics and Astronomy, Aarhus University, Ny Munkegade 120, 8000 Aarhus C, Denmark
\and The Oskar Klein Centre, Department of Physics, Stockholm University, AlbaNova, 10691 Stockholm, Sweden
\and Dark Cosmology Centre, Niels Bohr Institute, University of Copenhagen, Juliane Maries Vej 30, 2100 Copenhagen, Denmark}
\date{Received / Accepted}

\abstract
{The origin of the blue supergiant (BSG) progenitor of Supernova (SN)~1987A has long been debated, along with the role that its sub-solar metallicity played. We now have a sample of SN~1987A-like events that arise from the rare core collapse (CC) of massive ($\sim$20~M$_{\odot}$) and compact ($\lesssim$~100~R$_{\odot}$) BSGs.}
{The metallicity of the explosion sites of the known BSG~SNe is investigated, as well as the association of BSG~SNe to star-forming regions.}
{Both indirect and direct metallicity measurements of 13 BSG~SN host galaxies are presented, and compared to those of other CC~SN types. 
Indirect measurements are based on the known luminosity-metallicity relation and on published metallicity gradients of spiral galaxies. In order to provide direct metallicity measurements based on strong line diagnostics, we obtained spectra of each BSG~SN host galaxy both at the exact SN explosion sites and at the positions of other \ion{H}{ii} regions.
We also observed these hosts with narrow H$\alpha$ and broad $R-$band filters in order to produce continuum-subtracted H$\alpha$ images. This allows us to measure the degree of association between BSG~SNe and star-forming regions, and to compare it to that of other SN types.} 
{BSG~SNe are found to explode either in low-luminosity galaxies or at large distances from the nuclei of luminous hosts. Therefore, their indirectly measured metallicities are typically lower than those of SNe~IIP and Ibc. This result is confirmed by the direct metallicity estimates, which show slightly sub-solar oxygen abundances (12$+$log(O/H)$\sim$8.3-8.4~dex) for the local environments of BSG~SNe, similar to that of the Large Magellanic Cloud (LMC), where SN~1987A exploded. However, we also note that two objects of our sample (SNe 1998A and 2004em) were found at near solar metallicity. 
SNe~IIb have a metallicity distribution similar to that of our BSG~SNe.
Finally, we find that the degree of association to star-forming regions is similar among BSG~SNe, SNe~IIP and IIn.}
{Our results suggest that LMC metal abundances play a role in the formation of some 1987A-like SNe. This would naturally fit in a single star scenario for the progenitors. However, the existence of two events at nearly solar metallicity suggests that also other channels, e.g. binarity, contribute to produce BSG~SNe.}


\maketitle

\section{Introduction}

Supernova (SN) 1987A, the most studied SN ever, was also the first
object whose progenitor (Sanduleak $-$69$^\circ$202) could be
directly identified \citep{gilmozzi87}. Surprisingly, the massive
(M$_{ZAMS}$~$\sim16-22$~M$_\odot$, \citealp{arnett89}) progenitor star was
not a red supergiant (RSG) with a radius of a few hundred R$_{\odot}$
as envisioned by contemporary theory, but rather a compact
(R~$\lesssim$~50~R$_{\odot}$) blue supergiant (BSG). The compactness
of its progenitor explained both its slowly rising light curve
($\approx$84~days) and its low luminosity in the first months 
\citep[e.g.][]{arnett89}. Most of the kinetic energy was spent to
adiabatically expand the star while the light curve of SN~1987A was
mainly powered by the radioactive decay of $^{56}$Ni and $^{56}$Co.

Many attempts have been made to explain the nature of the BSG
progenitor of SN~1987A (see \citealp{pod92} for a review), since
standard stellar models did not predict the explosion of massive stars
in the BSG phase \citep[e.g.][]{woosley02}. As a matter of fact, a
single non-rotating $\sim$20~M$_{\odot}$ star born in a
solar-metallicity environment is expected to be a RSG at the time of collapse
\citep{eldridge05}. Therefore, special conditions are required
for a massive star to be a BSG at the endpoint of its life. The
possible scenarios include those that involve a single progenitor
star and those that consider Sanduleak $-$69$^\circ$202 as part of a
binary system. We note that for the specific case of the progenitor of
SN~1987A, any proposed scenario not only has to explain why it was a
BSG, but must also fit other constraints such as the chemical
anomalies \citep{saio88,fransson89} and the peculiar CSM of SN~1987A
\citep{morris09}. We now briefly review the possible mechanisms that
lead to the explosion of a BSG according to theoretical models.

In the single-star framework, a first natural explanation for the BSG
nature of SN~1987A's progenitor was found in the sub-solar metallicity
(12$+$log(O/H)~$=$~8.37~dex, \citealp{russell90}) measured in
the Large Magellanic Cloud (LMC), which hosted SN~1987A
\citep{langer91}. A low metallicity can allow a blue solution in the
HR diagram for the endpoint of a massive star (see e.g., \citealp{brunish82};
\citealp{arnett87}; \citealp{hillebrandt87}; \citealp{woosley88}).
Another single star scenario that can lead to
the explosion of a BSG is that invoking a rotating star, where
rotational mixing plays an important role in pushing the stellar
evolution towards the blue part of the HR diagram (see
e.g. \citealp{weiss88}; \citealp{hirschi04}; \citealp{ekstrom12}). 

In the binary scenario, both accretion and merger models may
lead to the explosion of BSGs (see \citealp{woosley02} for
references). We note that the binary solution was considered for the
specific case of SN~1987A because it would more easily explain the
``hour-glass" shape of the CSM observed in its remnant
\citep{morris09}. As noted by \citet{woosley02}, if massive stars end
their lives as BSGs because they merge with a companion, we would
expect to find BSG SNe also at solar metallicity.

From the observational point of view, testing the role of metallicity
in the production of exploding BSGs has not been possible until
recently, because SN~1987A belonged to a class of its own for many
years, although a handful of poorly observed SNe were suggested
to be similar to it already in the months following the explosion of
SN~1987A (see e.g. \citealp{vandenbergh88}). Among these SNe,
 following \citet{pastorello12}, we consider only
 SNe~1909A and 1982F as reliable 1987A-like transients.
 Therefore, it was impossible to say
whether the explosion of  compact BSGs usually occurs at low
metallicity or not.

Recently, however, two new SNe with BSG progenitors were presented by
\citet{kleiser11}. These objects, SNe 2000cb and 2005ci, display a
long rise time and low luminosity at early epochs, closely resembling
the light curve of SN~1987A. 
The BSG origin of SN~2000cb was later corroborated by detailed
modeling \citep{utrobin11}. Before these events, SN~1998A
\citep{pastorello05} had been suggested to result from the explosion
of a BSG star. Recently we published an analysis of two additional
SN~1987A-like transients (SNe~2006V and 2006au;
\citealp{taddia12}). Through the modeling of the bolometric light
curves, we determine progenitor radii to less than
100~$R_{\odot}$. This clearly argues for compact BSG progenitors.
Furthermore, in \citet{pastorello12}, we presented another object,
SN~2009E, which also has a BSG origin. \citet{pastorello12} discussed
the sample of known SN~1987A-like objects, listing a total of 11
transients including the aforementioned
 SNe. \citet{arcavi12} added a few new examples to the family of known 
SN~1987A-like events. The complete list of BSG~SNe is given in
Table~\ref{tab:sample}, where we also report basic information
concerning their hosts, i.e. the coordinates, morphological type,
axis dimensions, position angle, absolute magnitude, Milky Way color
excess, redshift and distance modulus.

This new sample of BSG~SNe allows us to investigate the
role played by the metallicity in the origin of BSG explosions.
Studying the metallicity of SN host galaxies has become a popular line of
investigation. Indirect estimates based on galaxy luminosity
\citep{tremonti04} and known metallicity gradients
\citep[e.g][]{pilyugin04} are useful indicators, but this sort of
analysis can be substantially improved by determining the
metallicities at the exact sites of the explosion long after the SN faded away
(e.g. \citealp{leloudas11,modjaz11}) and the actual metallicity
gradient through emission line diagnostics \citep[e.g.][]{thoene09}.

The aim of this paper is to provide an observational test of the 
role of metallicity for BSG SNe, and   
we therefore perform a detailed metallicity analysis of the
galaxies that hosted a BSG~SN. In Sect.~\ref{sec:observations} details are given
 on our photometric and spectroscopic observations;
Sect.~\ref{sec:indirect} includes the results on the metallicity of
the BSG~SN explosion sites that were obtained indirectly from the galaxy
luminosity and the SN distance from its host center; in
Sect.~\ref{sec:direct} we present a metallicity study of the observed
BSG hosts using long slit spectroscopy. 
This is based on strong line diagnostics, including
metallicity estimates at the exact explosion sites and a metallicity
gradient for each host that was observed.  
Section~\ref{sec:NCR}
describes the measurements and the results concerning the association of BSG~SNe 
to star-forming regions in their hosts.  A discussion based on these results is given in
Sect.~\ref{sec:discussion} and our conclusions follow in
Sect.~\ref{sec:conclusions}.

\section{Observations and data reduction}
\label{sec:observations}

A log of all the observations that were performed and collected for this
study is presented in Table~\ref{log}.  A total of 14 BSG~SNe are
known in the literature, including SN~1987A. We gathered photometric
and spectroscopic data for 12 of them (the metallicity of SN~1987A is
known and we did not observe the host of SN~2004ek, which was very
recently identified as a BSG~SN by \citealp{arcavi12}). We rejected SN~1998bt from the sample after measuring its
redshift, which made this SN too bright to be a SN~1987A-like event
(see Sect.~\ref{spectra}). Therefore, a total of 11 BSG~SN host galaxies
were directly investigated to obtain their metallicities.

\subsection{Photometry}

Photometric observations in broad $R$ (Bessel) and narrow H$\alpha$
passbands were carried out at the Nordic Optical Telescope (NOT) using
ALFOSC, during  the nights starting on March 29 2012, April 12 $\&$ 13 2012
and May 20 $\&$ 22 2013 for the host galaxies of SNe 2004em, 2005ci,
2005dp and 2009E.

The host galaxies of SNe~1998bt and OGLE-2003-NOOS-005 were observed
on March 17 2013 at the UT1 of the Very Large Telescope (VLT), which
also provided deep images for the hosts of SNe~1998A, 2000cb, 2006V
and 2006au. These images were obtained with FORS2 in broad $R$ and
narrow H$\alpha$ filters. For each galaxy we chose the narrow
filter with the effective wavelength matching the 
redshifted wavelength of H$\alpha$ (filter names are given in Table~\ref{log}). 
For instance, the high-redshift host of OGLE-2003-NOOS-005 was observed 
with the narrow \ion{S}{ii} filter to match the observed H$\alpha$ wavelength.
 This galaxy was also
observed in the $B$ band, as was the host of SN~1998bt; the latter was not
observed in H$\alpha$ given its low luminosity. For the host of
SN~1982F  we used broad $R$ and narrow H$\alpha$ archive images
 that were taken with the Isaac Newton Telescope. The continuum-subtracted H$\alpha$ image
  of M~101 (host of SN~1909A) was obtained from \href{http://www.aoc.nrao.edu}{http://www.aoc.nrao.edu}.
  
Images obtained with ALFOSC and FORS2 are characterized by a pixel scale of 0.19~\arcsec$/$pix and 0.25~\arcsec$/$pix, respectively. This corresponds to $\approx$0.1~kpc$/$pix at the distance of 100~Mpc.

The total integration times, which are on the order of a few hundred
seconds for each galaxy, were subdivided into multiple exposures for
each filter.  In this way, after bias-subtraction and flat field
correction, the combination of multiple frames allowed for the removal
of cosmic rays.  The WCS coordinates of each final $R$-band image were
re-calibrated by matching the point sources in the frame with those in
the USNO-B catalogue \citep{monet03}. 
This was done using the software available at
\href{http://astrometry.net}{http://astrometry.net} and presented by \citet{lang10}.
This step was necessary to identify the pixel
corresponding to the SN position in the galaxy with high precision. 

Armed with the scientific frames in narrow H$\alpha$ and broad $R$
band, we proceeded to build the continuum-subtracted H$\alpha$ image
for each galaxy.  We aligned the images in the two filters (using the
task ``\textit{imalign}" in IRAF),  then we scaled the fluxes in order
to match the counts of the stars in the two frames.  Successively, the
$R$-band image was subtracted from that obtained with the narrow
filter, to remove the continuum and to leave only the
H$\alpha$(+[\ion{N}{ii}]) emission component,  characteristic of
\ion{H}{ii} regions. Bright foreground stars sometimes leave
residuals after the subtraction, and we patched these
imperfections by setting the values of these pixels equal to those of
the closest portions of the frame. 
The resulting H$\alpha$ maps were useful both to measure the association
of BSG~SNe to star-forming regions and to show the exact locations of
the \ion{H}{ii} regions that were spectroscopically observed.

We also measured the apparent magnitudes of the host of SN~1998bt from
the $B$- and $R$-band images, by using stars and galaxies of known
magnitudes (from SIMBAD) in the local field. 
We confirmed m$_{R}$~$=$~23.30$\pm$0.10~mag \citep{germany04} 
and found m$_{B}$~$=$~21.95$\pm$0.10~mag.

\subsection{Spectroscopy}\label{spectra} 

Aiming to directly measure the metallicity for the environment of each
BSG~SN through strong line diagnostics, we obtained
long exposures ($>$~1800~sec) with spectroscopy at their location and
simultaneously at the center of their host galaxies. 
The telescope field was rotated to the angle determined by the
host center and the SN position, the slit was aligned on a nearby star
 in the host galaxy field and then the proper offset
was given to intercept both the host nucleus and the explosion site.
This procedure minimizes the error on the pointing, 
and was checked with through-slit images.
For SN~1909A, 
whose position was known to be far from bright \ion{H}{ii} regions, we
placed the slit on a nearby H$\alpha$ emitting region, also to check
the consistency of our measurements with those already published. We
complemented the data in the literature also for SN~1982F, by placing
the slit close to the SN position and through bright \ion{H}{ii}
regions that were not at the center of the host.  For the hosts of
SNe~2005ci, 2005dp and 2009E we also considered the spectra included
in the Sloan Digital Sky Survey (SDSS) 
archive\footnote{through \href{http://skyserver.sdss3.org}{ http://skyserver.sdss3.org}} that were taken at 
the center of each host.

Spectra of the hosts of 
SNe 1909A, 1982F, 2004em, 2005ci, 2005db, 2006V and 2009E were taken at the 
NOT ($+$ ALFOSC with grism~$\#$4), as reported in Table~\ref{log}. This particular grism
 covers the $\sim$3500-9000~\AA\ 
wavelength range, enabling the detection of the most important
 \ion{H}{ii} region lines (see Sect.~\ref{sec:direct}). A slit of 1$\arcsec$ width was chosen to match
  the typical seeing at the NOT and
to reach a good compromise between spatial resolution and exposure times.
VLT with FORS2, grism 300V and a 1$\arcsec$ wide slit was used to get spectra of the SN~1987A-like hosts 
that were observable from the 
Southern Hemisphere (SNe~1998A, 1998bt,  2000cb, OGLE-2003-NOOS-005,  2006V, and 2006au). The
 wavelength range 
covered with this setup is 3300-8650~\AA. 
The Full-width-at-half-maximum (FWHM) spectral resolution obtained at the NOT and at the VLT is about 
16-17~\AA\ and 11-14~\AA, respectively.
Archival spectra complemented the dataset for SN~1998A. In this case, we
 reconstructed the slit position, obtaining a perfect match between the position
  of bright H$\alpha$ emitting regions in the 2D-spectra and those observed
   in the H$\alpha$ images. Since these archival spectra included the SN signal, the pointing was trivial to establish.

The spectral reduction was carried out through the following steps:
each 2D-spectrum was a) biased subtracted and b) flat-field corrected
to remove CCD inhomogeneities and minimize the fringing in the red
part of the spectrum. c) If multiple 2D-spectra were obtained, each 2D-frame was combined
 to remove cosmic rays. Then d) the trace of the
galaxy nucleus or that of a bright star or of a \ion{H}{ii} region with a bright continuum was
extracted and fit by a low-order polynomial. This trace was plotted over
the 2D-spectrum to visually check its precision. e) All the spectra of
\ion{H}{ii} regions visible in the 2D-spectrum were extracted using
the same trace, but shifted in the spatial direction to match their
positions. The extraction region was determined by looking at the
H$\alpha$ flux profile and then plotting this region over the
2D-spectrum to ensure the tracing and to include all the emission
lines. f) Each spectrum was wavelength calibrated by comparison with a
spectrum of an arc-lamp. g) The wavelength-calibrated spectrum was
flux calibrated with the sensitivity function obtained from the
spectrum of a standard star observed in the same night. Both the
\ion{H}{ii} region spectra and the standard spectrum were corrected
for the airmass and for the second order contamination following the
prescriptions of \citet{stanishev07}. 
 
 The spectra that show at least H$\alpha$ and
[\ion{N}{ii}]~$\lambda$6584 emission were included in this study. The
spatial position of the corresponding \ion{H}{ii} region was 
identified on the H$\alpha$ maps for each spectrum, to compute the
metallicity gradient (see Sect.~\ref{sec:direct}).

The spectral observation of the host of SN~1998bt indicates a
redshift of z$=$0.436, much higher than the value assumed by
\citet{germany04} and \citet{pastorello12}.  These authors 
assumed that the galaxy hosting SN~1998bt belonged
 to a cluster whose average redshift is z~$=$~0.046. Our spectrum
 shows that this galaxy is not a member of the cluster.

The new redshift was measured through the detection of strong emission lines (H$\beta$, H$\gamma$, 
[\ion{O}{iii}]~$\lambda\lambda$5007,4959 and [\ion{O}{ii}]~$\lambda$3727). 
This suggest that SN~1998bt was not a SN~1987A like event, but a much
 brighter object with $V_{\rm max}=-21.6$~mag. 
There is no spectroscopic classification of SN~1998bt, so this event could
 have been a super-luminous SN (SLSN) with a light curve similar to SN~2010gx \citep{pastorello10}, which also has a slow rising and dome-like light curve. 
Moreover, the new redshift makes the host of SN~1998bt a galaxy with a normal M$_{B}$ $= -18.58$~mag (here we applied a k-correction of $+$0.73~mag based on $m_{B}-m_{R}$, from \href{http://kcor.sai.msu.ru}{http://kcor.sai.msu.ru})
rather than a very faint galaxy with M$_{B}$ $\approx -13$~mag. We also note
 that the metallicity measured with the R23 method (see Sect.~\ref{sec:direct}) for this host galaxy is 12+log(O/H)~$=$~7.83~dex, which is
consistent with the abundances of dwarf galaxies hosting SN~2010gx-like events \citep{chen13}.
Therefore, SN~1998bt is no longer considered as part of our BSG~SN sample.

\section{Indirect metallicity measurements}
\label{sec:indirect}

It is well established that luminous galaxies have higher integrated
or central metallicity than low-luminosity galaxies.

\citet{tremonti04} present a simple relation that, given the absolute $B$-band
magnitude of the host, allows us to estimate the metallicity 
-- the oxygen abundance 12$+$log(O/H) -- at the galaxy center.  
The absolute $B$-band magnitude for each galaxy was computed from the
apparent magnitude listed in the Asiago Supernova Catalogue (ASC,
\citealp{barbon}, where the magnitudes are mainly taken from
\citealp{devauc91}), and with distance moduli and Milky Way extinction
as reported in Table~\ref{tab:sample}.  The distance was computed from
the known redshift combined with the Hubble law, where we assumed
H$_{0}$~$=$~73~km~s$^{-1}$~Mpc$^{-1}$ \citep{spergel07}. For very nearby
objects (z~$<$~0.01) we adopted redshift independent distances from
NED\footnote{\href{http://ned.ipac.caltech.edu}{http://ned.ipac.caltech.edu}}. 
The extinction due to the dust along the line of sight in the Milky Way was corrected by using the data from \citet{schlegel98}\footnote{Extinction values are obtained through the \href{http://irsa.ipac.caltech.edu/applications/DUST/}{NASA/IPAC Infrared Science Archive}}. Our sample includes a few (5) large galaxies having
absolute magnitude similar or lower than that of the Milky Way ($M_{B}^{MW}$=$-$20.2). The faintest galaxy is characterized by $M_{B}$=$-$17.5~mag.
The central metallicities from $M_{B}$ are reported in the second column of Table~\ref{metal}.

At the same time, several authors
\citep[e.g.][]{pilyugin04,pilyugin07} have shown that \ion{H}{ii} regions at
large distances from the center of their galaxy have lower metallicity
than those close to the center. Therefore, a first and indirect
approach (see e.g. \citealp{boissier09}) to infer the metallicity of 1987A-like SNe is to compute the
central metallicity of the hosts from their absolute $B$-band
magnitude and then to use the metallicity gradients in the literature
to estimate the value at the SN distance from the nucleus.

The de-projected,
normalized distance (r$_{SN}$/R$_{25}$) from the host galaxy center for
each BSG~SN was determined following \citet{hakobyan09}. Given the SN
and the host nucleus coordinates, the major (2R$_{25}$) and minor (2b)
axes along with the position angle (PA) of the galaxy are the data 
needed to derive r$_{SN}$/R$_{25}$. We also included a
correction of the inclination based on the morphological type of the
galaxy (t-type), following \citet{hakobyan12}.  Both the input data
and the resulting r$_{SN}$/R$_{25}$ are reported in
Table~\ref{tab:sample}, and further details are
given in the notes to that Table.  
The main sources for the galaxy properties were
ASC, NED, and
SIMBAD\footnote{\href{http://simbad.u-strasbg.fr/simbad}{http://simbad.u-strasbg.fr/simbad}}.
The normalized
distances that we obtained are smaller than those reported by
\citet[][their Table~4]{pastorello12}. 
This is simply because they
used galaxy radii from
HyperLeda\footnote{\href{http://leda.univ-lyon1.fr}{http://leda.univ-lyon1.fr}},
which are smaller than those reported by NED and/or by the ASC. 

By using the average gradient of the galaxies presented by 
\citet{pilyugin04} ($-$0.47~dex~R$_{25}^{-1}$), 
we then extrapolated the
metallicity at the SN normalized and de-projected distance. 
This estimate is given in the third column in
Table~\ref{metal}. We note that for NGC~4490 and M~101 their measured gradients were used, which are
published by \citet{pilyugin07} and \citet{kennicutt03} ($-$0.063~dex~R$_{25}^{-1}$ 
and $-$0.90~dex~R$_{25}^{-1}$, respectively). 
The choice of the average gradient from \citet{pilyugin04} is justified by the fact that gradients expressed in dex~R$_{25}^{-1}$ exhibit a smaller range of values than those expressed in dex~kpc$^{-1}$, as noticed by \citet{garnett97}.

Overall the indication that comes from the indirect measurements is
that BSG~SNe arise from marginally sub-solar metallicity environs
($<$ 12$+$log(O/H)$_{BSG}$ $>$~$=$~8.51$\pm$0.04~dex), as suggested by \citet{vandenbergh89} and
\citet{pastorello12}.

The obtained values for the host galaxy M$_{B}$, for the SN-host
distance and the metallicity at r$_{\rm SN}$ are more
interesting when compared to those of different SN types. To
make this comparison, we built large samples of galaxies that hosted
Type II, IIP, IIn, IIb, Ib and Ic SNe using the ASC. These samples
were built with the following criteria: a) all hosts at
redshift z~$>$~0.023 were excluded, like in \citet{kelly12}, to avoid a bias towards
very bright SNe. In the sample of 13 BSG~SNe that means the exclusion
of OGLE-2003-NOOS-005 from the comparisons. b) We also excluded
galaxies more inclined than i~$>$~75~deg to avoid large uncertainties
in the de-projected distances. c) Finally, only those galaxies with
known PA, major and minor axes, t-type and apparent $B$ magnitude were
included. 

\begin{figure}
 \centering$
\begin{array}{ccc}
\includegraphics[width=6cm,angle=0]{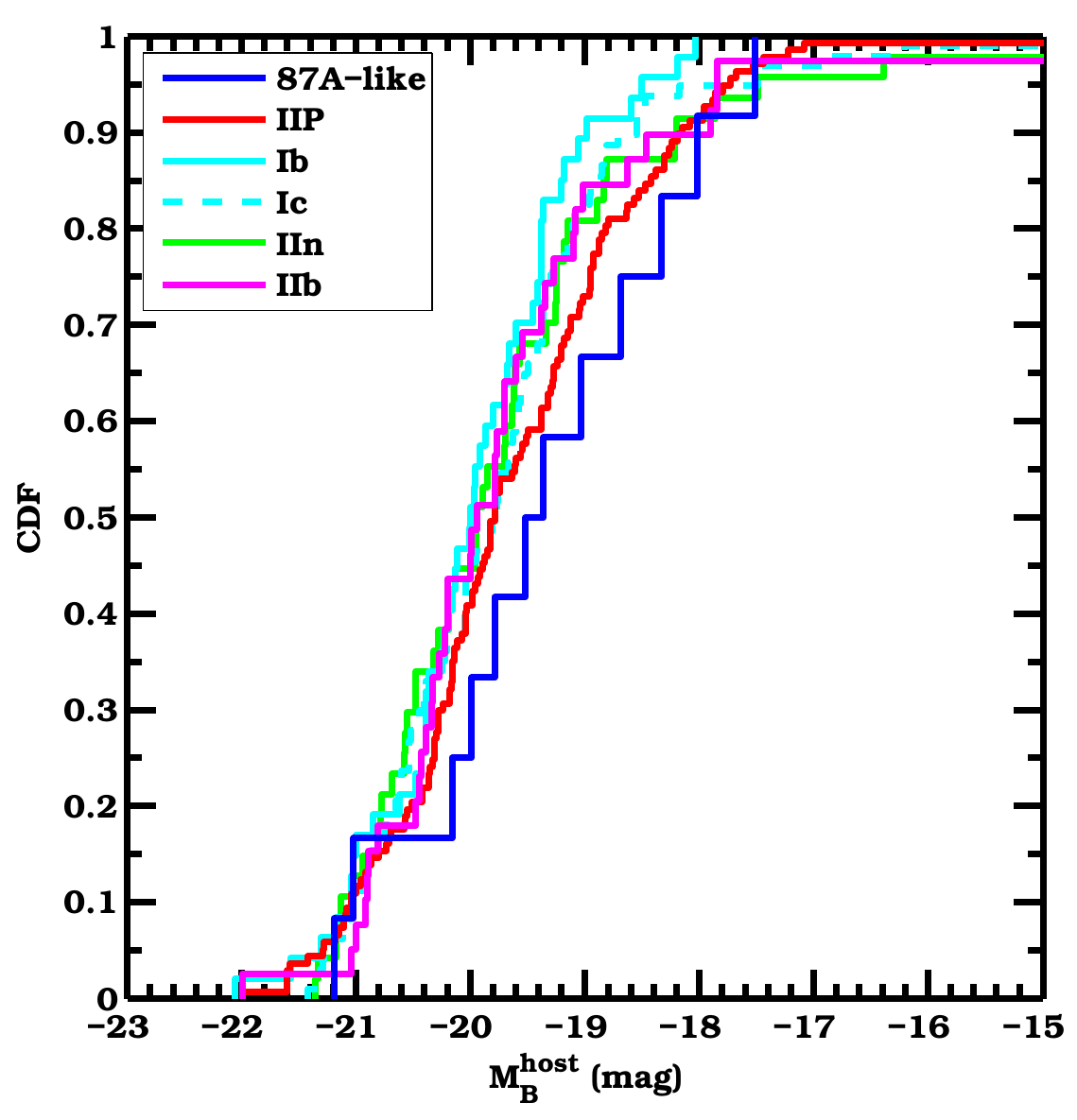}&
\includegraphics[width=6cm,angle=0]{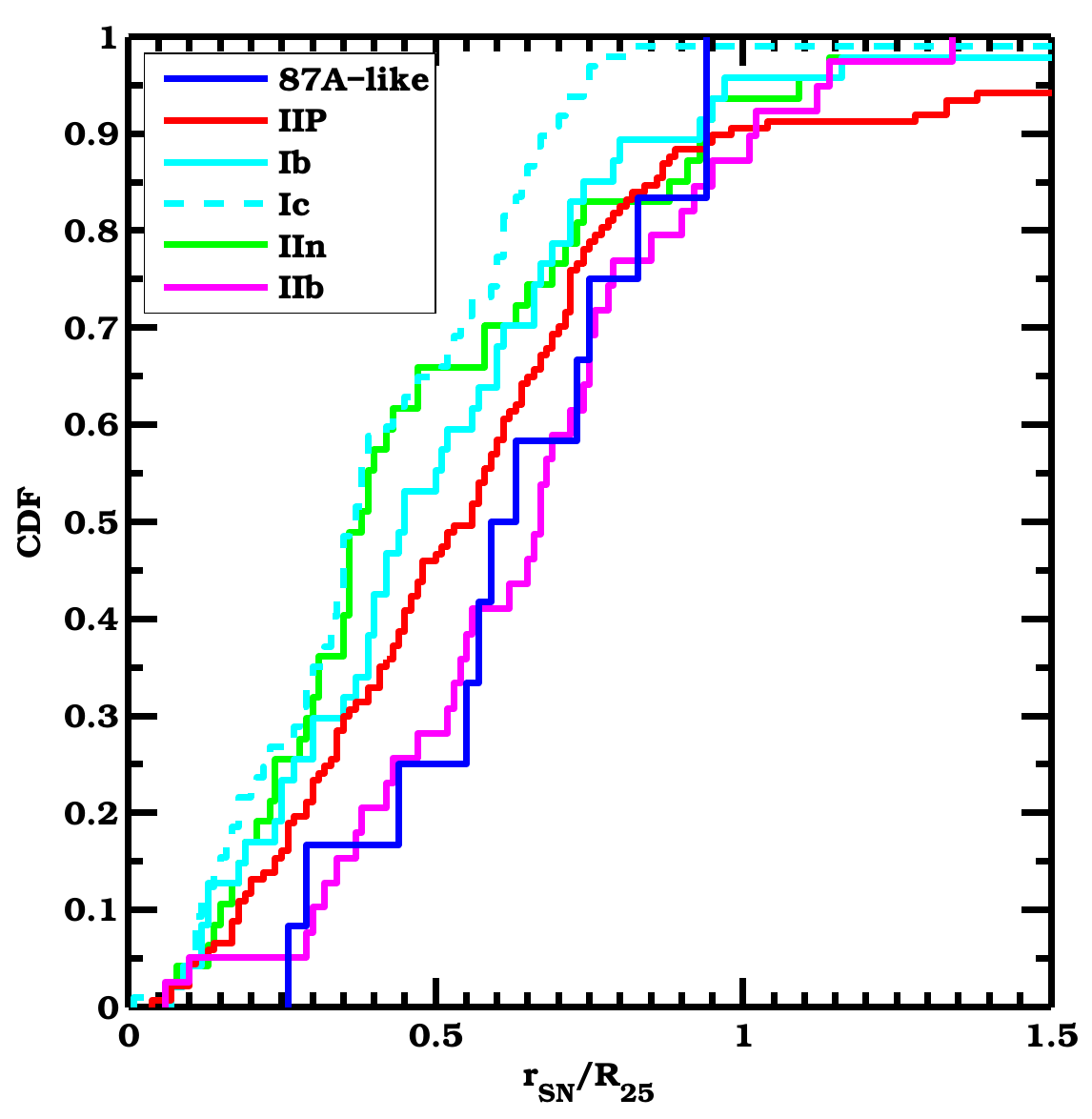}&
\includegraphics[width=6cm,angle=0]{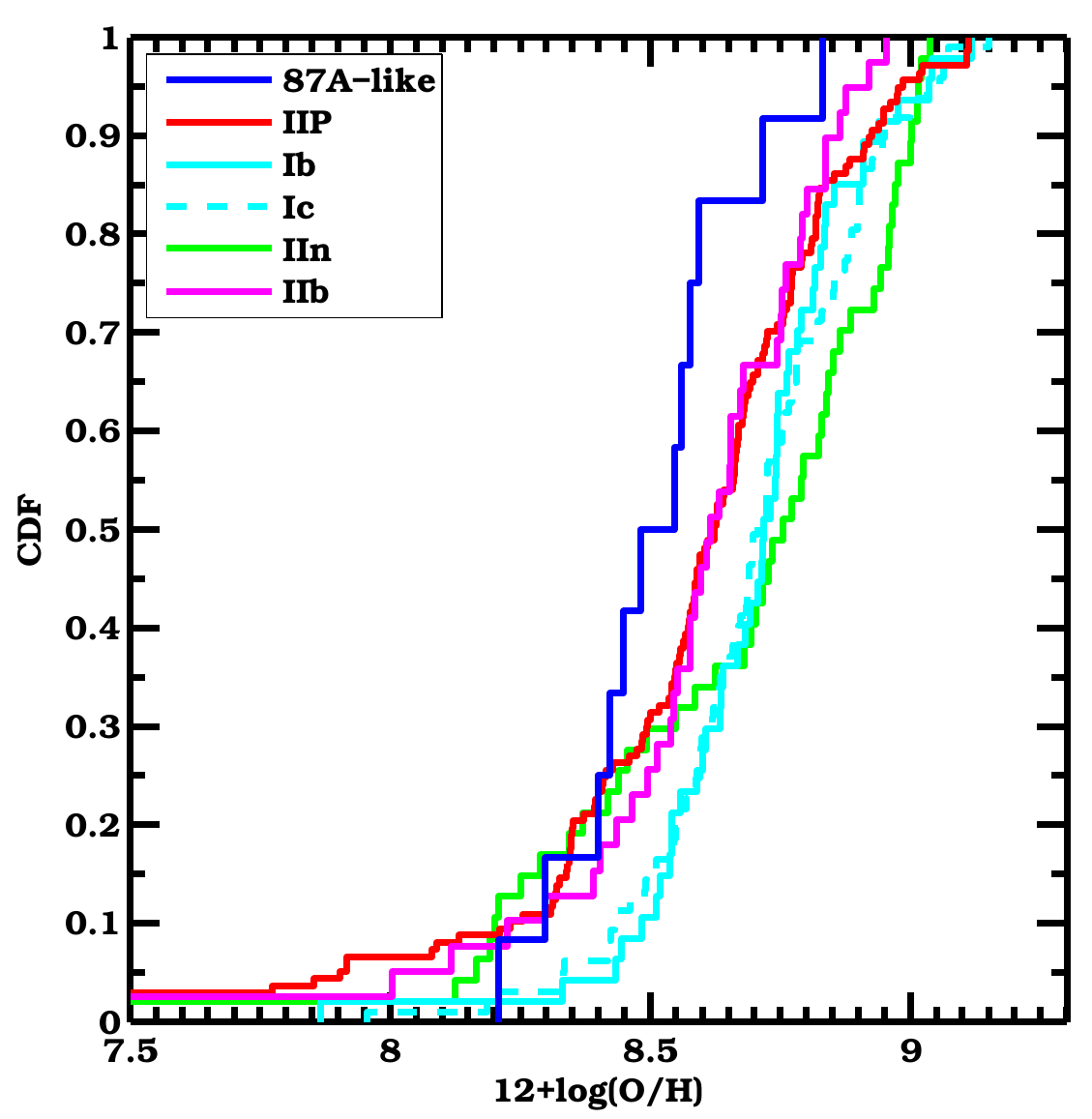}
\end{array}$
  \caption{Cumulative distributions of the host galaxy absolute $B$ magnitudes {\it(left panel)}, of the SN-host center normalized offsets \textit{(central panel)}, and of the extrapolated metallicity at the SN distance \textit{(right panel)}, for our BSG~SN sample and for other CC~SN types. The metalllicity has here been obtained by the indirect method outlined in Sect.~\ref{sec:indirect}.\label{cdfdist}}
 \end{figure}

For each host M$_{B}$, r$_{SN}$/R$_{25}$ and the
metallicity at the SN distance were computed as we did for our 1987A-like SNe,  
and for each type the cumulative distribution (CDF) of these
quantities was plotted in Fig.~\ref{cdfdist}. BSG~SNe seem to be located in
slightly fainter galaxies than those of other SN types (see left panel
of Fig.~\ref{cdfdist}). BSG~SNe are also located more
distant from the center than SNe~II and IIP, and therefore also more distant than
SNe~Ibc. A similar trend is followed by the SN~IIb population (see
central panel of Fig.~\ref{cdfdist}). Finally, BSG~SNe present a
lower (indirect) metallicity at their positions (see right panel of
Fig.~\ref{cdfdist}).  We note that our results concerning the other SN
types (in particular the SN-host center distances) are consistent with
those found by \citet{kelly12}. The difference in metallicity
between 1987A-like SNe and SNe~IIP is not statistically significant 
(a Kolmogorov-Smirnov test, hereafter K-S, gives a p$-$value of 0.46). 
The difference in distance have higher
statistical significance (p$-$value$=$0.32). Obviously these
differences are larger when we compare BSG~SNe and
SNe~Ibc, which are known to be located even more centrally than SNe~IIP
\citep[e.g.][]{hakobyan08}.

\begin{figure}
 \centering
\includegraphics[width=8cm,angle=0]{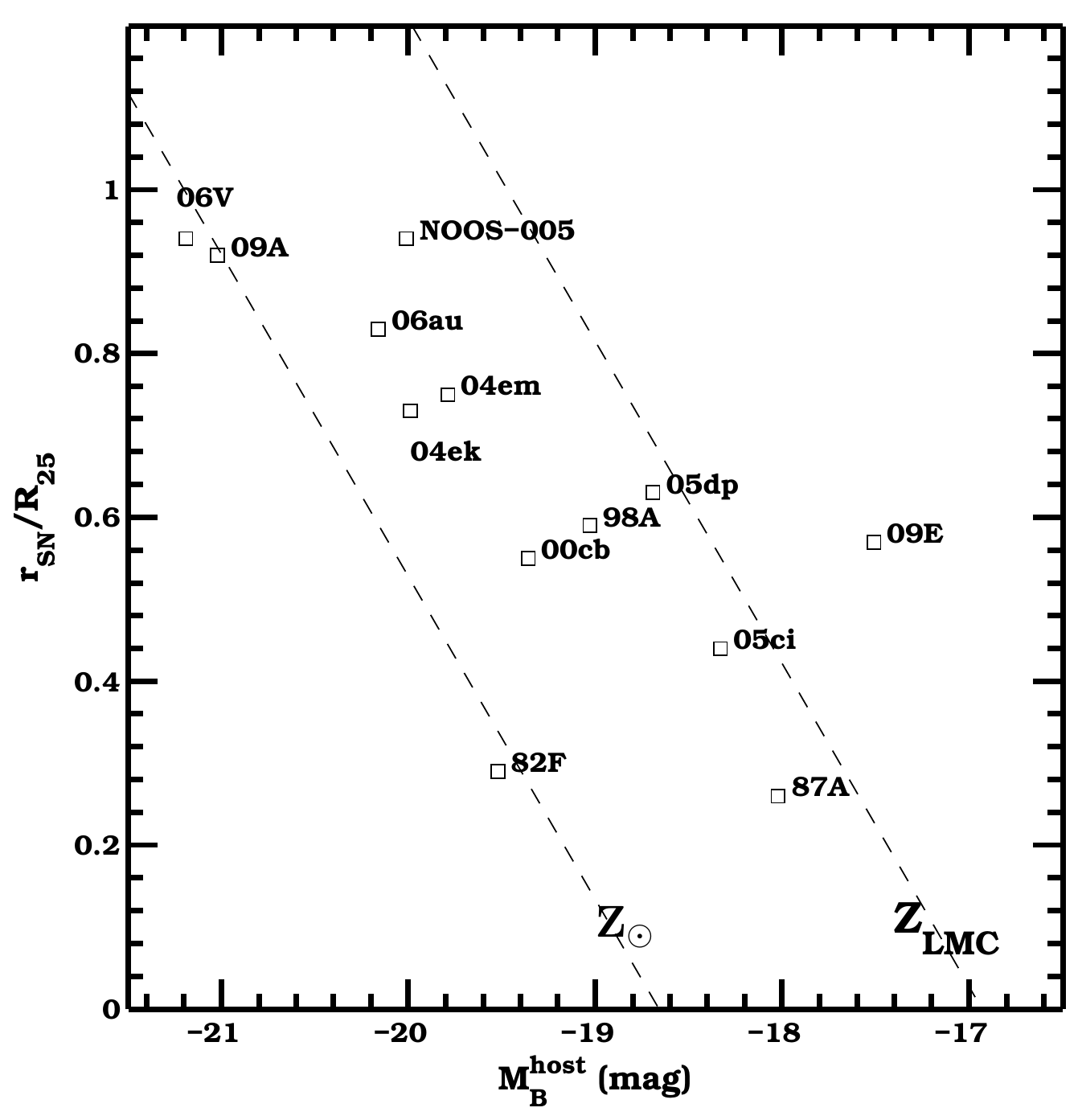}
  \caption{Normalized SN-host center offset vs host galaxy absolute $B$ magnitude for each BSG~SN in our sample. These SNe either exploded in low-luminous galaxies or in the outer parts of brighter hosts. Also shown are the iso-metallicity curves for Z$_{\odot}$ and Z$_{\rm LMC}$, where we assume the luminosity-metallicity relation and the mean gradient described in Sect.~\ref{sec:indirect}.\label{distvsMb}}
 \end{figure}

Figure~\ref{distvsMb} illustrates that BSG~SNe are at lower metallicity
because either they are located in faint galaxies, which are
intrinsically metal poor, or they are placed in the metal-poor
outskirts of bright spiral galaxies. SN~1982F is located quite close
to the center of its bright host. However, in Sect.~\ref{sec:82F} we
will see that this galaxy is in fact intrinsically metal-poor.

\section{Direct metallicity measurements}
\label{sec:direct}

We make use of the emission-line diagnostic named N2 \citep{pettini04}
to obtain direct metallicity measurements of 11 BSG~SN host
galaxies. This diagnostic has been used in similar studies, for
instance in \citet{thoene09}, \citet{leloudas11}, \citet{anderson10},
 \citet{sanders12} and \citet{stoll12}.
 
N2 is the logarithm of the flux ratio between
[\ion{N}{ii}]~$\lambda6584$ and H$\alpha$.  Since these two lines are
close in wavelength, the method is neither affected by extinction effects
 nor by differential slit losses
 due to lack of an atmospheric dispersion corrector (ADC). We note 
 that FORS2 is equipped with an ADC \citep{adc}, whereas the ADC was not used when we observed with ALFOSC.
The expression to obtain the oxygen abundance from N2 is
12$+$log(O/H)~$=$~9.37$+$2.03$\times$N2$+$1.2$\times$N2$^2+$0.32$\times$N2$^3$
\citep{pettini04}.  The systematic N2 uncertainty on 12$+$log(O/H) is
0.18~dex and dominates the overall error that affects the flux
measurements. Therefore, we assume a total uncertainty of 0.2 dex on
the single metallicity measurement. The same choice was done by
\citealp{thoene09}, who used the same observational setup.

\begin{figure}
 \centering
\includegraphics[width=12cm,angle=0]{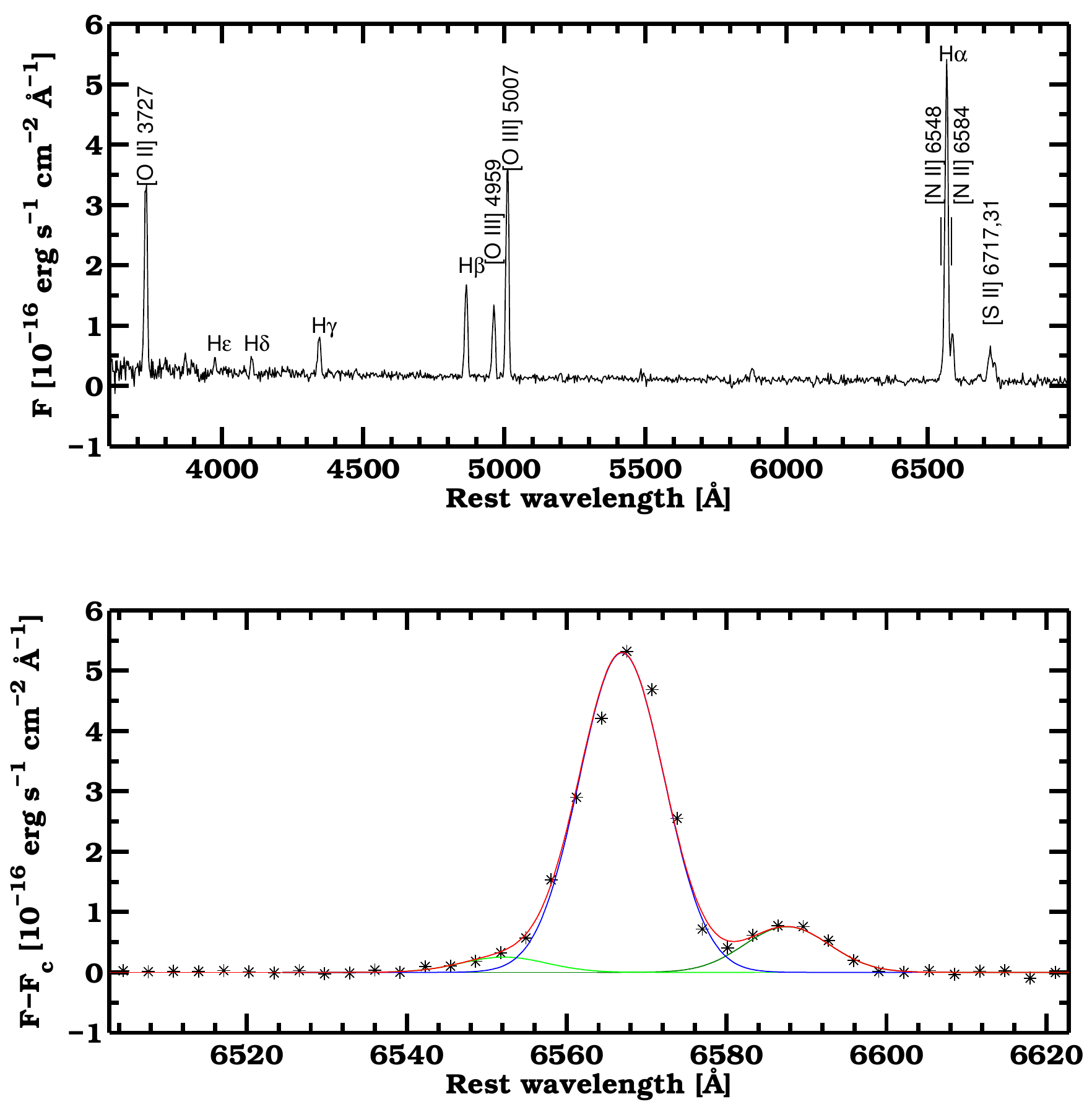}
  \caption{\textit{(Top panel)} An example of a bright \ion{H}{ii} region spectrum (in UGC~6510), where we identify the strongest detected lines. \textit{(Bottom panel)} The continuum subtracted H$\alpha$ and [\ion{N}{ii}] lines are shown along with the
  multiple Gaussian fit that allows us to deblend the three different features and to measure their fluxes. 
   In red we indicate the total fit. In bright green, dark green and blue we show the single components fitting [\ion{N}{ii}]~$\lambda$6548, [\ion{N}{ii}]~$\lambda$6584, and H$\alpha$, respectively.
  The difference in wavelength between the centroids of each line are fixed, a single FWHM is used and the [\ion{N}{ii}]~$\lambda$6584 to [\ion{N}{ii}]~$\lambda$6548 flux ratio is fixed to 3.\label{blend}}
 \end{figure}

The [\ion{N}{ii}]~$\lambda6584$ and H$\alpha$ fluxes were determined in
the following way: a) we carefully fit the continuum around the lines. We notice that, despite the careful fit of the 
continuum, the underlying H$\alpha$ stellar absorption might affect the result, pushing the metallicity to higher 
values; however, the N2 measurements of three host galaxy nuclei obtained through the SDSS line fluxes (see Sect.~\ref{sec:05ci}, 
\ref{sec:05dp} and \ref{sec:09E}), which were measured after 
subtracting the stellar absorption \citep{bolton12}, do not differ significantly from our measurements at the same positions.
b) Then, three Gaussians with a single FWHM (determined by the
spectral resolution) and fixed wavelength centroids (known from the
atomic physics) were simultaneously fit to [\ion{N}{ii}]~$\lambda6584$,
[\ion{N}{ii}]~$\lambda6548$ and H$\alpha$.  Since
[\ion{N}{ii}]~$\lambda6548$ is difficult to fit given its faintness,
its flux was fixed to be 1/3 of [\ion{N}{ii}]~$\lambda6584$ \citep[see][]{osterbrock89}. The fit
of this faint and blended line is done to optimize the measurement of
the H$\alpha$ flux and remove possible contaminations, especially for
the NOT spectra, which are of lower resolution. This procedure allowed
us to measure the two important lines in a robust way.
In Fig.~\ref{blend} we show the spectrum of a bright and
metal-poor \ion{H}{ii} region in UGC~6510 (top panel), and the
triple-Gaussian fit that was performed on its H$\alpha$ and
[\ion{N}{ii}] lines (bottom panel).  
In the case of SN~2005dp, the host galaxy was observed with
 ALFOSC using both grism~$\#$4 and grism~$\#$8 (which has a 
 higher resolution), to compare the line ratios obtained at low
  (with the line deblending) and high resolution. We basically
  obtained the same line ratios, with an average
   difference smaller than 1\%.

\begin{figure}
 \centering
\includegraphics[width=12.5cm,angle=90]{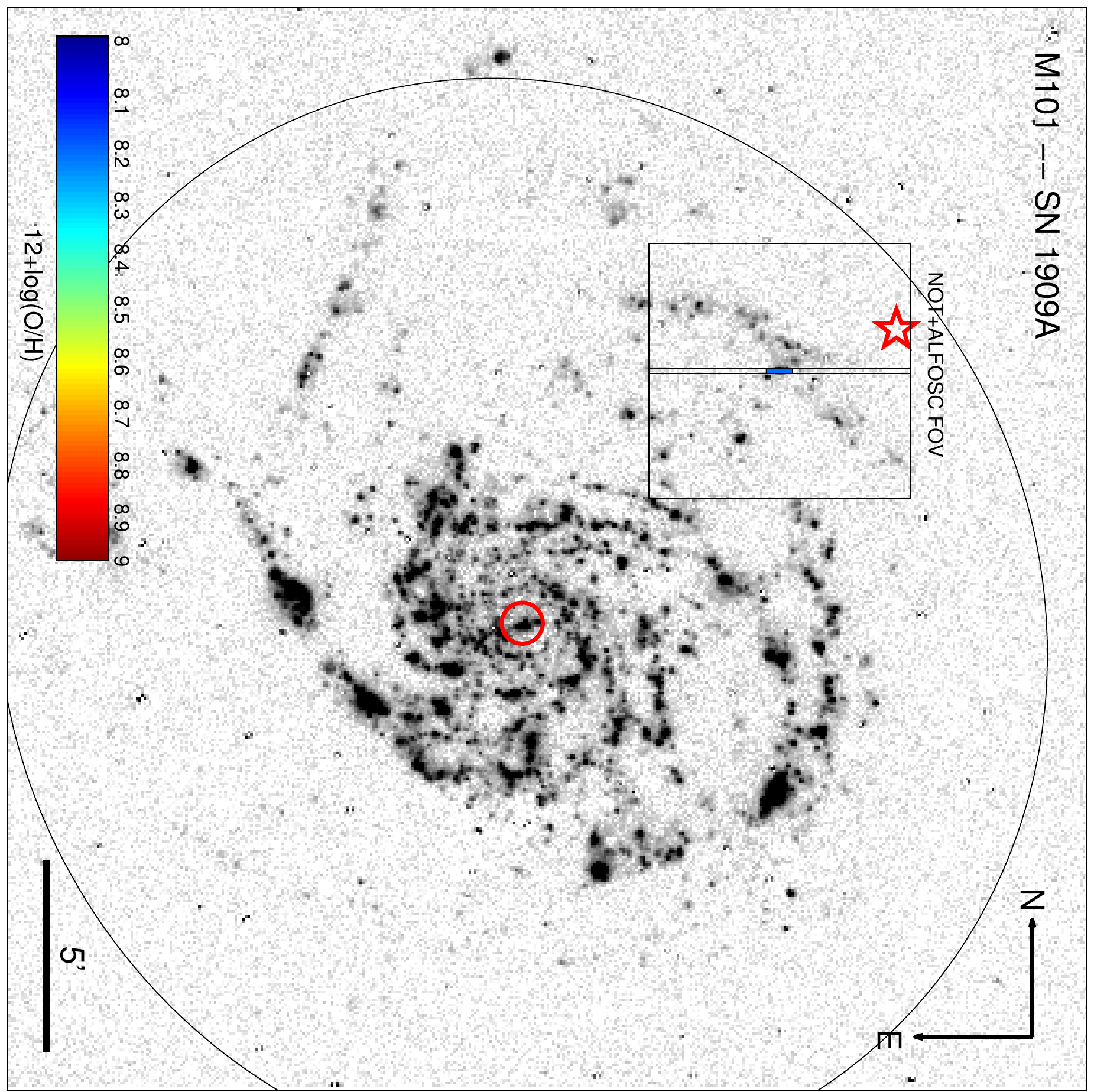} 
\includegraphics[width=12.5cm,angle=0]{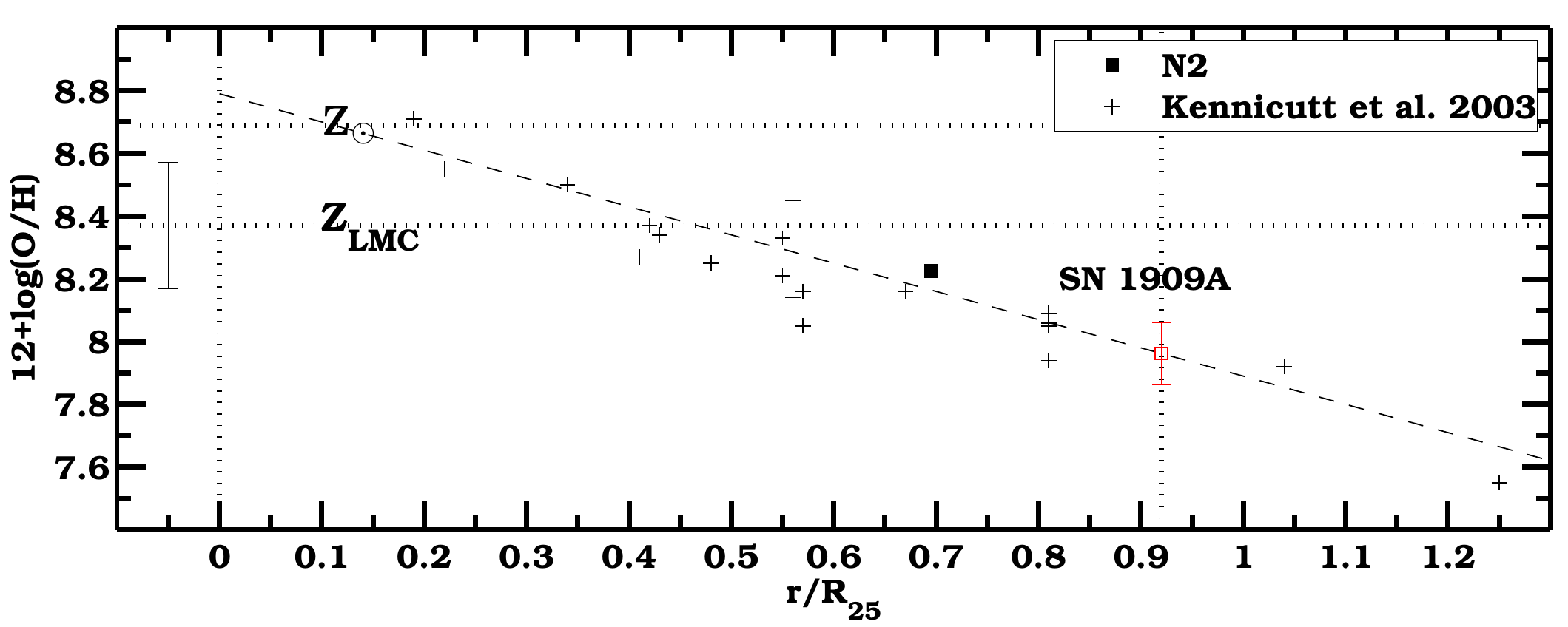}
 \caption{\textit{(Top panel)} Continuum-subtracted H$\alpha$ image of M~101, obtained from \href{http://www.aoc.nrao.edu}{http://www.aoc.nrao.edu}. The 25th $B$-band magnitude elliptic contour is shown by a black solid line, along with the position of SN~1909A (marked by a red star) and the center of the galaxy (marked by a red circle). We also show the field of view (FOV) of ALFOSC and the slit position. The width of the slit (1$\arcsec$) has been enhanced in the figure to show, through a color code, the measured metallicity at the position of a bright \ion{H}{ii} region that we inspected. \textit{(Bottom panel)} Metallicity gradient of M~101 from \citet{kennicutt03}. In the plot we include our measurement, which matches with those that are already published. The linear fit from \citet{kennicutt03} is shown by a dashed line. The interpolated metallicity at the SN distance is marked by a red square and its uncertainty corresponds to the fit error. The error bar ($\pm$0.2~dex) for our N2 measurement is shown aside. The positions of SN and nucleus are marked by vertical dotted lines. The solar metallicity \citep{asplund09} and the LMC metallicity \citep{russell90} are indicated by two horizontal dotted lines.\label{sn09A}}
 \end{figure}

\begin{figure}
 \centering$
\begin{array}{cc}
\includegraphics[width=7.2cm,angle=0]{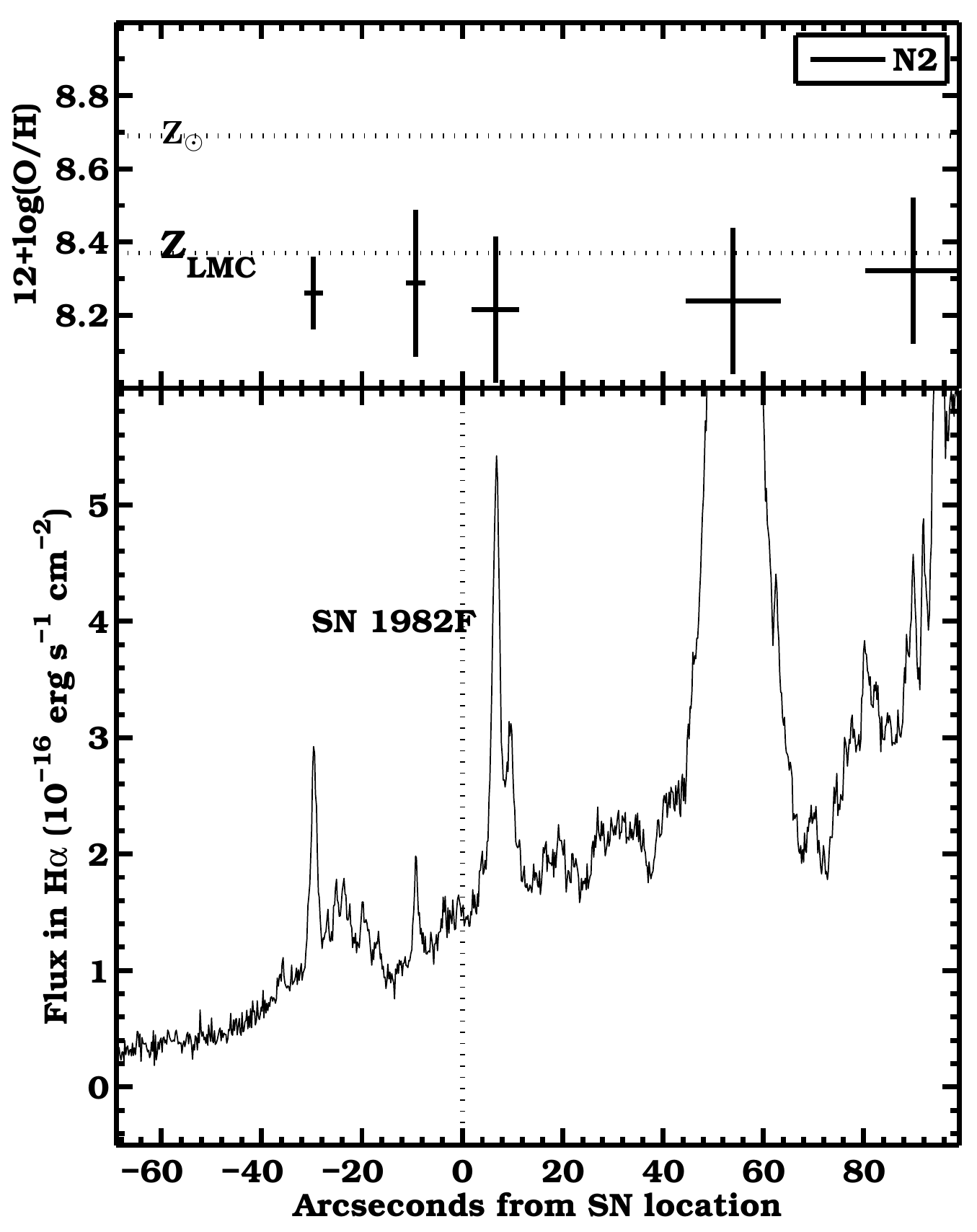}&
\includegraphics[width=9cm,angle=90]{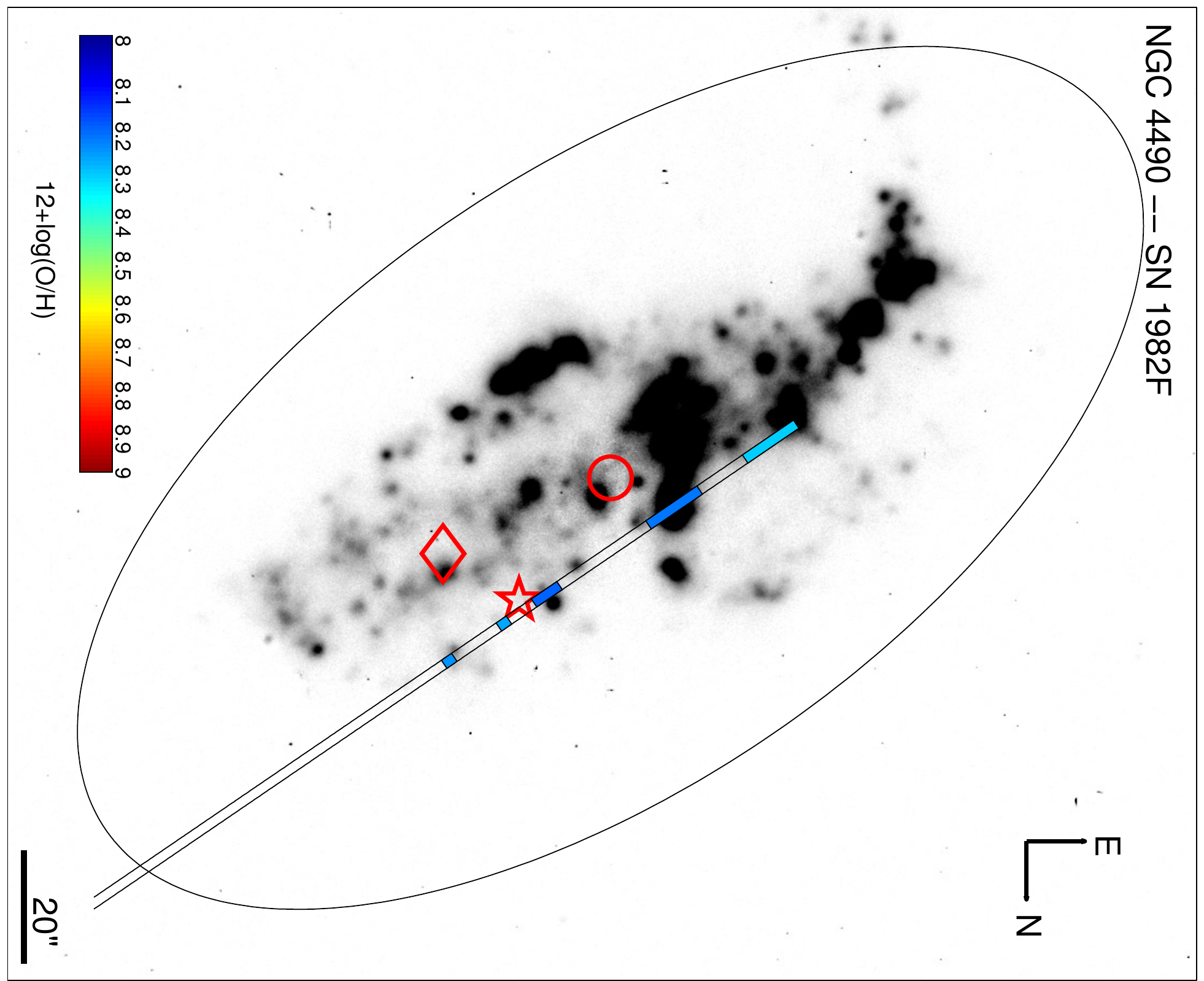} \\
\end{array}$
\includegraphics[width=15cm,angle=0]{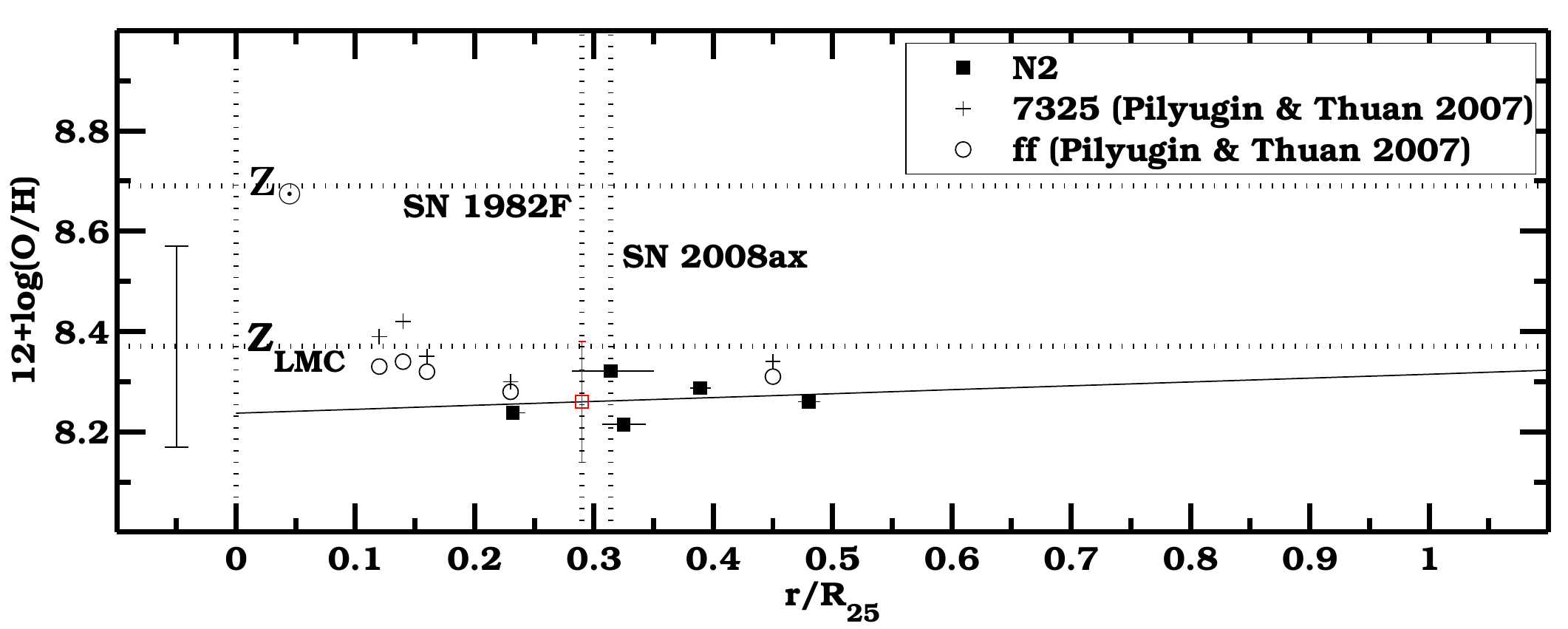}
  \caption{\textit{(Top-right panel)} Continuum-subtracted H$\alpha$ image of NGC~4490. The 25th $B$-band magnitude elliptic contour is shown by a black solid line, along with the positions of SN~1982F (marked by a red star), SN~2008ax (marked by a red diamond) and the center of the galaxy (marked by a red circle). The slit position is shown, and a color code is used to present the N2 metallicity measurements at the position of each \ion{H}{ii} region that we inspected. The slit width (1\arcsec) has been enhanced to better show the colored patches. \textit{(Top-left panel)} Flux at the H$\alpha$ wavelength along the slit, shown as a function of the distance from the location of SN~1982F (marked by a dashed line). The N2 measurements are shown at the corresponding positions in the upper sub-panel. \textit{(Bottom panel)} Metallicity gradient of NGC~4490. In the plot we include the measurements from \citet{pilyugin07}, which match our N2 metallicity estimates. The linear fit on our measurements is shown by a solid line. The interpolated metallicity at the distance of SN~1982F is marked by a red square and its uncertainty corresponds to the fit error. The error bar ($\pm$0.2~dex) for our N2 measurements is shown aside. The positions of SN~1982F, SN~2008ax and nucleus are marked by vertical dotted lines. The solar metallicity \citep{asplund09} and the LMC metallicity \citep{russell90} are indicated by two horizontal dotted lines.\label{sn82F}}
 \end{figure}

For each \ion{H}{ii} region falling into the slit, the
metallicity was measured with the N2 method. These values are reported on
the continuum-subtracted H$\alpha$ images at the position of each of
these regions using a color map (see the top or top-right panels of
Figs.~\ref{sn09A}-\ref{sn09E}). The width of each colored patch
corresponds to the actual width of the spectral extraction.  In the
continuum-subtracted H$\alpha$ images we also show the slit aperture
and the 25th $B$-band magnitude contour, from the values reported in
Table~\ref{tab:sample}, as well as the image orientation and its scale
in arcseconds. The center of the galaxy is indicated with a
circle and the SN position with a star.

The flux at the H$\alpha$ rest wavelength versus the distance in
arcseconds from the SN position (or nucleus position) is shown in the
top-left or central panels of Figs.~\ref{sn82F}-\ref{sn09E}. Here, we
also report the metallicity values at their positions. The spatial
uncertainty of each metallicity value corresponds again to the width
of the spectral extraction.

Using the same formulae that were employed to de-project the distance of
each SN from its host center, we compute the deprojected distance of
each \ion{H}{ii} region with measured metallicity from the galaxy nucleus. The
metallicity values versus these de-projected distances (normalized by the
galaxy radius) are shown in the bottom panels of Figs.~\ref{sn09A}-\ref{sn09E}.
 Fitting a line to these points (we performed a chi-square fit), the metallicity was found to 
 decrease as the distance from the center increases (with the exception of NGG~4490, which is an interacting galaxy).
These line fits were used to interpolate the metallicity at the exact 
SN position (the SN distance is marked by a dotted, vertical line) when
the SN does not fall onto a bright \ion{H}{ii} region. We found
that the interpolated or extrapolated metallicity always matches
within the error with the directly measured metallicity when the SN
coincides in position with a bright \ion{H}{ii} region (see the 5th
and 6th columns of Table~\ref{metal}). Therefore, for each SN it was 
assumed that the value from the linear fit is the best estimate of the
N2 metallicity. These values are
marked by a red, void square in the bottom panels of
Figs.~\ref{sn09A}-\ref{sn09E} and summarized in the 5th column of
Table~\ref{metal}. The uncertainty on the extrapolated value
corresponds to the standard deviation of 1000 extrapolations from
linear fits that were Monte Carlo simulated based on the uncertainty
on the slope and on the central metallicity.  These uncertainties
 do not consider the systematic error included in the N2 method. They
  only take into account the scatter of each 
 metallicity estimate from the best linear fit, since for each estimate we observed a small scatter around the linear fit 
 (see the bottom panels of Figs.~\ref{sn09A}-\ref{sn09E}) compared 
    to the systematic 0.18~dex reported by \citet{pettini04}. We also note that our measurements
    involve a range of oxygen abundances (12+log(O/H)$\approx$8.2$-$8.7~dex) that is
     limited compared to that investigated by 
    \citet{pettini04}. In this interval, the scatter of the N2 relation
     is indeed smaller than 0.18~dex, as visible in their Figure~1.

The strong-line diagnostic method that has been employed is not the only
possibility. Other methods that are often used in the literature are
the so called O3N2 and R23. O3N2 is defined as
O3N2$=$log[([\ion{O}{iii}]~$\lambda$5007$/$H$\beta$)/([\ion{N}{ii}]~$\lambda6584$/H$\alpha$)]
and it is linked to the oxygen abundances by the following relation:
12$+$log(O/H)$=$8.73$-$0.32~O3N2 \citep{pettini04}. Like N2, it is
based on ratios of lines that are close in wavelength. However, it is not always possible to detect all the four
needed lines, and especially [\ion{O}{iii}]~$\lambda$5007 is faint at low
metallicity.  R23 is a method based on oxygen lines and H$\beta$
(R23 $=$ ([\ion{O}{ii}]~$\lambda$3727$+$[\ion{O}{iii}]~$\lambda$4959$+$[\ion{O}{iii}]~$\lambda$5007) $/$ H$\beta$). It is sensitive to the extinction, that it is usually
estimated from the Balmer decrement \citep{osterbrock89}, and to differential slit losses. The
parameter R23 is related to the metallicity by a polynomial relation
given by e.g. \citet{nagao06}.  Also in this case, it is not
always possible to detect all the lines that are needed to measure the
R23 parameter. In the case of SN~2005ci (see
Fig.~\ref{sn05ci}, center-left panel), we show the metallicities
obtained with N2, O3N2 and R23 for each \ion{H}{ii} region. 
We also present the values of $E(B-V)$ as estimated from the Balmer
decrement, which are needed to compute the R23 parameter.
The R23 estimates differ from those obtained with the other two 
methods by $\lesssim$~0.1~dex. This is not surprising, since there are offsets 
between the methods which are discussed in the literature \citep[e.g.][]{kewley08}.

In summary, we chose the N2 method for the following reasons: I) it is easier to 
observe the lines needed for this method than those needed for the other methods.
II) N2 is less affected by extinction and differential slit losses.
III) N2 has often been used in the literature to measure the metallicity of 
other SN types. Therefore, choosing N2
for our sample allows us to directly compare our results to those of other SN classes 
(see Sect.~\ref{sec:metcomp}).
IV) N2 is calibrated against metallicity measured with the electron temperature
 method (see \citealp{pettini04}).
This method is found to match the metallicities measured by the modeling of 
stellar spectra better than other methods \citep{bresolin09}. Therefore, since the
 solar metallicity (Z$_{\odot}$) estimate is also based on spectral modeling 
 \citep{asplund09}, N2 is appropriate to compare our results to Z$_{\odot}$.

In the following sections, each host is individually discussed
together with its metallicity measurements.

\subsection{M~101 -- SN~1909A}

M~101 is a nearby spiral galaxy that covers several square arc-minutes
on the northern sky.  SN~1909A exploded in the outskirts of this
galaxy, far from any bright \ion{H}{ii} region. Since the metallicity
gradient of this galaxy is well studied and presented by \citet{kennicutt03}, we
simply tested our N2 measurement on a bright \ion{H}{ii} region
relatively close (de-projected distance of 7.66~kpc) 
to the position of SN~1909A (top panel of
Fig.~\ref{sn09A}) against the published metallicity gradient (bottom
panel of Fig.~\ref{sn09A}). We found a good match and therefore we
proceeded with the extrapolation of the 12$+$log(O/H) to the SN
distance, obtaining the lowest value of our sample
(12$+$log(O/H)$=$7.96$\pm$0.10~dex).  
SN~1909A is thus a typical example of a BSG~SN
that exploded far from the metal-rich nucleus of a bright galaxy. The
explosion site of the SN has a low oxygen abundance due to the metallicity gradient in the
host.

\subsection{NGC~4490 -- SN~1982F\label{sec:82F}}

NGC~4490 (see Fig.~\ref{sn82F}) is a nearby, interacting galaxy that hosted
SN~1982F and also the Type IIb SN~2008ax \citep{taubenberger11}. \citet{pilyugin07} published some metallicity measurements that we have complemented with our
N2 estimates. In this case we placed the slit through a small
\ion{H}{ii} region close to the location of SN~1982F and through a bright
\ion{H}{ii} region closer to the center. All the obtained 12$+$log(O/H) values 
fall below solar metallicity (consistently with
the results by \citealp{pilyugin07}), implying an extrapolated SN
oxygen abundance of 8.2-8.3~dex (similar to that found in the LMC).
The metallicity gradient is found to be unusually flat, likely because this galaxy is interacting with another one (NGC~4485). \citet{perez11} show that the central metallicity of interacting galaxies is diluted by tidally induced low-metallicity gas inflows, resulting in a flatter oxygen abundance gradient.
We notice that using O3N2, \citet{kelly12} found 12$+$log(O/H)$=$8.21~dex for
SN~2008ax, confirming that NGC~4490 produces core-collapse (CC) SNe at
relatively low metallicity.  SN~1982F is an example of a BSG~SN
that is located not too far (r$_{SN}$/R$_{25}=$0.30) from the center of
a galaxy that has an intrinsically low metallicity.

\subsection{IC~2627 -- SN~1998A}

IC~2627 is a southern object, characterized by strong star formation,
as shown by the bright H$\alpha$ knots that are mostly distributed
along two arms.  SN~1998A exploded in the outer region of one of these
arms and was located on a bright \ion{H}{ii} region (see
Fig.~\ref{sn98A}, top panel).  This galaxy also hosted the normal 
Type~II SN~1994R,  whose position we marked with a diamond. The galaxy was observed
 along the two directions defined by the host center and the
two SNe (see the slits marked with VLT 98A and VLT 94R in
Fig.~\ref{sn98A}, top panel). Furthermore, we used two spectra from
the ESO archive that included the flux 
of SN~1998A\footnote{from ESO program 60.D-0281, PI: J. Spyromilio.}.  
By reconstructing the slit positions of these two spectra, they are found to be placed
 almost through the galaxy center (see the slit marked 
NTT~1 in Fig.~\ref{sn98A}, top panel), and on the bright arm that
hosted SN~1998A (slit marked NTT~2 in Fig.~\ref{sn98A}, top
panel). The three different metallicity measurements at the exact
position of SN~1998A converge to 8.62-8.68~dex,  which is the highest
value in our sample, consistent with solar abundance
(8.69$\pm$0.05~dex; \citealp{asplund09}).  The other measurements across the galaxy
are all solar or super-solar, and overall the galaxy exhibits a very
shallow metallicity gradient (Fig.~\ref{sn98A}, bottom panel). The
region between SN~1994R and the host center is that with the highest
N2 metallicity, close to 9.0~dex (Fig.~\ref{sn98A}, center panels).
SN~1998A represents the rare counterexample of SN~1909A, since it is
located in a solar-metallicity environment despite its relatively
large distance (r$_{SN}$/R$_{25}=$0.59) from the nucleus.

\subsection{IC~1158 -- SN~2000cb}

IC~1158 is another spiral galaxy with diffuse star formation,
particularly concentrated in its southern and central part (see
Fig.~\ref{sn00cb}, top-right panel). The slit, which was placed on the
SN explosion site and through the galaxy nucleus, reveals that there
is no significant H$\alpha$ emission at the SN position. However, two
bright knots are located a few arcseconds (the nearest at a de-projected distance of 1.24~kpc) 
from it and a total of 8 \ion{H}{ii} regions fell into the slit (see
Fig.~\ref{sn00cb}, top-left panel).  This allowed us to map the
metallicity gradient, as shown in the bottom panel of
Fig.~\ref{sn00cb} and to estimate the oxygen abundance at the SN
position to be about 8.5~dex. This is consistent with the result of
\citet{anderson10}.

\subsection{2MASX J05553978-6855381 -- OGLE-2003-NOOS-005}

2MASX J05553978-6855381 is a relatively distant galaxy (z$=$0.03),
which hosted OGLE-2003-NOOS-005. This SN exploded very far from the
host center (r$_{SN}$/R$_{25}=$0.94) and not on a \ion{H}{ii} region,
although close (2.44~kpc) to a bright one (top panels of
Fig.~\ref{ogle}). For this galaxy, we measured the extent of the axes directly
 on the $B$-band image that
we obtained at the VLT. The N2 measurements reveal a super-solar central
metallicity and a gradient that seems 
to flatten in the outer part of the galaxy (bottom panel of Fig.~\ref{ogle}). We
note that the H$\alpha$ line measurements at the exact host center is
made difficult by significant stellar absorption.  At the SN
distance, the oxygen abundance is moderately sub-solar ($\sim$8.4~dex).

\subsection{IC~1303 -- SN~2004em}

The spiral galaxy IC~1303 is placed in a region of the sky crowded with foreground stars.
SN~2004em exploded in its outer part, close (1.16~kpc) to a relatively bright 
\ion{H}{ii} region (see top panels of Fig.~\ref{sn04em}). We measured the metallicity 
at the host center and at four other positions, revealing a super-solar central 
metallicity and a shallow gradient (see bottom panel of Fig.~\ref{sn04em}). 
At the SN distance, the oxygen abundance is 12$+$log(O/H)$=$8.56~dex, 
which is consistent with solar metallicity within the errors.

\subsection{NGC~5682 -- SN~2005ci}
\label{sec:05ci}
NGC~5682 is a strongly tilted (i$=$70~deg) galaxy, with two arms characterized by
significant  star formation.  SN~2005ci is located quite close to the
galaxy center (r$_{SN}$/R$_{25}$=0.44), on a bright \ion{H}{ii}
region (top panel of Fig.~\ref{sn05ci}). We placed one slit along the
SN-host center direction, and one along the major axis (see top panel
of Fig.~\ref{sn05ci}) The eight metallicity measurements that we obtained
with two slits (see central panels of Fig.~\ref{sn05ci}) show a steep
gradient and a LMC-like metallicity at the SN position (bottom panel
of Fig.~\ref{sn05ci}). We have also included the N2 measurement at the
center of the host from SDSS, finding a perfect match to our results.
As explained in Sect.~\ref{sec:direct}, for the \ion{H}{ii} regions
that have fallen into the slit at the SN position we also measured the
O3N2 and R23 metallicities, which are presented in the bottom and
center-left panels.

\subsection{ NGC~5630 -- SN~2005dp}
\label{sec:05dp}
NGC~5630  is a strongly tilted (i$=$72~deg) galaxy at low redshift (z$=$0.008). It
shows star formation mainly in the central region and in the eastern
part. The slit was first placed (Fig.~\ref{sn05dp}, central-left panel)
at the SN position and on a reference star in the north of the
galaxy. The second spectrum (Fig.~\ref{sn05dp}, central-right panel)
was taken along the major axis of the galaxy, as shown in
Fig.~\ref{sn05dp} (top panel). A total of 16 metallicity
measurements were obtained, revealing slightly sub-solar metallicity at the
center and a lower, LMC metallicity in the outer parts. The SN, which
is located at r$_{SN}$/R$_{25}=$0.63,  exploded in an environment with
12$+$log(O/H)$=$8.27$\pm$0.07~dex (Fig.~\ref{sn05dp}, bottom panel).  
Our central measurements are perfectly consistent with the N2 measurement from the
 SDSS spectrum, which is reported in the bottom panel of Fig.~\ref{sn05dp}.  



\subsection{UGC~6510 -- SN~2006V}

UGC~6510 is a bright, face-on galaxy that is formed by multiple arms
and characterized by a significant H$\alpha$ luminosity.  SN~2006V
occurred in the outermost part of the galaxy and not on a detectable
\ion{H}{ii} region. We observed this host with both the VLT and the
NOT. The NOT--ALFOSC slit was placed in order to catch the flux of the
brightest \ion{H}{ii} region close to the SN position (see the top and
center-left panels of Fig.~\ref{sn06V}).  The VLT--FORS2 slit was
positioned at the exact SN position and through the host center (see
the top and center-right panels of Fig.~\ref{sn06V}).  Metallicities
obtained with the VLT and with the NOT at similar deprojected
distances do match very well, confirming that the deblending of
H$\alpha$ and [\ion{N}{ii}]  lines (necessary for the NOT spectra)
gives correct values for their fluxes.  A total of 14 \ion{H}{ii}
regions were investigated, allowing for a precise determination of the
metallicity gradient (see the bottom panel of Fig.~\ref{sn06V}). The
extrapolated metallicity at the distance of
SN~2006V is 12$+$log(O/H)~$=$~8.35~dex.

\subsection{UGC~11057 -- SN~2006au}

UGC~11057 is a spiral galaxy that appears strongly tilted (i$=$66~deg) on the
sky (see the top-right panel of Fig.~\ref{sn06au}).  The brightest
H$\alpha$ luminosity comes from the outer part of the arms that form
this host. A bright, saturated foreground star affected the quality
of the continuum-subtracted H$\alpha$ image, leaving some residuals
that we manually removed. SN~2006au exploded in spatial coincidence with a
bright and extended \ion{H}{ii} region (see the top panels of
Fig.~\ref{sn06au}), onto which we placed the slit, at
$r_{SN}/R_{25}=$0.83. A total of eight metallicity
measurements were obtained and from the computed gradient we obtained
12$+$log(O/H)$=$8.49~dex at the SN position, consistent with
 the single-measurement value at the exact SN position (see
the bottom panel of Fig.~\ref{sn06au}). Similar to the host of OGLE-2003-NOOS-005,
UGC~11057 shows a gradient that tends to flatten in the outer regions.

\subsection{NGC~4141 -- SN~2009E}
\label{sec:09E}
NGC~4141 is the barred spiral galaxy that hosted SN~2009E. It was observed 
with the NOT, by placing the slit through the SN position and
through a very bright \ion{H}{ii} region closer to the host center
(see the top panels of Fig.~\ref{sn09E}). A total of five metallicity
measurements were obtained, including that at the exact SN
position. The galaxy has a moderately sub-solar central metallicity
(as also found by SDSS) that further decreases with the distance from
the center (see the bottom panel of Fig.~\ref{sn09E}). At the SN
position the value is lower than the LMC metallicity.

\subsection{Metallicity results for the SN~1987A-like sample and 
comparisons to other SN types\label{sec:metcomp}}

The average metallicity that we obtain from the direct measurements for our sample is
$<$~12$+$log(O/H)$_{\rm BSG}$~$>$~$=$~8.36$\pm$0.05~dex (where the uncertainty is
 given by $\sigma/\sqrt{N}$ and N$=$12). This value is moderately
sub-solar and compatible with that of the LMC, and therefore similar to that
 of SN~1987A. However, we also
find SN~1987A-like transients at solar or almost solar metallicity,
like SNe~1998A and 2004am.

It is of interest to compare the metallicity of our sample to those of
other subtypes. To make a proper comparison, we use the data from
\citet{anderson10} and \citet{sanders12} that were obtained with the N2 method. 
From \citet{anderson10}, in order to prevent gradient effects, we selected only
those SNe~IIP, Ib and Ic that are closer than 3 kpc from the inspected \ion{H}{ii} region. 
SN~2000cb was also excluded
 from the sample of SNe~IIP. From \citet{sanders12} we selected only the SNe~IIb which
 have a \ion{H}{ii} region spectrum at the exact SN position.
  We computed the cumulative distributions of the metallicity for our
sample (12 objects) and for the selected SNe~IIP (23 objects), for the
SNe~Ib (10 objects), for the SNe~Ic (13 objects) and for the SNe~IIb (5 objects).
These are presented in Fig.~\ref{cdfmetal}.
It is evident from the comparison that the population of 1987A-like
SNe is associated with lower metallicity environs than those of normal
SNe~IIP ($<$~12$+$log(O/H)$_{\rm IIP}$~$>$~$=$~8.54$\pm$0.04~dex). A K-S test
 between 1987A-like and IIP events
gives a p-value of 0.006, revealing that this difference is
statistically significant. SNe~Ib and Ic are at even higher
metallicities than SNe~IIP, as already shown by \citet{anderson10} and
\citet{kelly12}.  Interestingly, SNe~IIb were found to have a metallicity distribution that is very similar ($<$~12$+$log(O/H)$_{\rm IIb}$~$>$~$=$~8.42$\pm$0.05~dex) to those
 of our BSG~SNe, as suggested by the fact that they are also preferentially located in the 
 outer regions of their hosts (see Fig.~\ref{cdfdist}, central panel). However, the sample of SNe~IIb is
 rather small.
We note that the systematic uncertainties on the single measurements were not taken into account
when we determined the statistically significant difference between BSG~SNe and
SNe~IIP. To estimate the uncertainty on the p$-$value, N$=$10000 Monte Carlo simulations
 of the metallicity measurements were made for our BSG~SNe
  and for the SNe~IIP,
based on the fit errors given in Table~\ref{metal} (column 5) for our BSG~SNe and assuming 0.1~dex uncertainty
 for the SNe~IIP, which is the average uncertainty for our sample. For each couple of
  simulations we computed the K-S test 
 obtaining N p$-$values.
Then, the cumulative distribution of these N p$-$values was built and 68\% 
of them were found to be lower than 0.033. This can be taken to represent the error
 on the p$-$value and would confirm 
that the difference between BSG~SNe and
SNe~IIP is statistically significant ($>$ 2$\sigma$ level). On the other hand, if we assume the typical 0.2~dex error
 for each N2 metallicity estimate, 68\% of the p$-$values sits below 0.189, and
  only 35\% of them are found below 0.05. 
  However, as outlined in Sect.~\ref{sec:direct},  we think that 0.2~dex could 
  be an overestimate of the actual uncertainty on the SN metallicity. This is supported by the fact that we are investigating a narrow metallicity range where the dispersion in the \citet{pettini04} relation is smaller than 0.18~dex and also
by the observed dispersion around the metallicity linear fit (bottom panels of Figs.~\ref{sn09A}-\ref{sn09E}), which is clearly smaller than 0.18~dex for most of the galaxies.

The metallicity gradient of our galaxy sample (see the last column of
Table~\ref{metal}) has an average of $-$0.399$\pm$0.072~dex~R$_{25}^{-1}$.
The host galaxies of SNe~1982F, 1998A, and 2004em show a much flatter behavior, whereas
M~101 and the host of OGLE-2003-NOOS-005 exhibit a steeper slope. The
average gradient that we found is very similar to that found by
\citet{pilyugin04} for their larger sample of galaxies.
 
The measured central metallicity (see the 4th column of
Table~\ref{metal}) is higher for the bright galaxies than for the
faint ones (with the exception of the interacting galaxy NGC~4490), as expected by previous
studies (e.g \citealp{tremonti04}). We plot M$_{B}$ versus central
metallicity from N2 in Fig.~\ref{mbvscentrmet}. In this figure we also
plot the N2 metallicity at the SN position, showing the effect of the
metallicity gradient that is particularly important for those SNe that
were hosted by bright, metal rich galaxies.

A comparison between indirect and direct measurements at the host center reveals
a general agreement within $\sim$0.2~dex, with the exception of NGC~4490 (host of 
SN~1982F) that is bright but intrinsically metal poor. We note that the metallicity-luminosity relation
that we use \citep{tremonti04} is known to be affected by a large scatter.
When we compare indirect and direct measurements at the SN position, we found a good correspondence
within $\sim$0.2~dex for the majority of the objects. The gradient that we used to extrapolate
 the indirectly-measured metallicity at the SN distance is a good approximation for most of the hosts (it is close to our directly-measured average gradient). However, some hosts have a shallower or steeper gradient, and this contributes to the discrepancies found for some events. 
 Overall, we confirm that the indirect approach to measure the metallicity is useful to suggest 
possible differences among SN samples. However, it is necessary to directly measure 
the metallicity through strong-line diagnostics to obtain robust results for individual objects.
 
\section{Association with star-forming regions}
\label{sec:NCR}

The spectral observations of the BSG SN host galaxies revealed that
some of our objects are not spatially coincident with bright
\ion{H}{ii} regions. Therefore, armed with the continuum-subtracted
H$\alpha$ images, we proceeded to quantify the spatial
association of our SNe (a total of 11 hosts were observed in $R$ and
H$\alpha$) to star-forming regions, which are H$\alpha$ emitters.  We
made use of the ``Normalised Cumulative Rank pixel function" (NCR),
first presented by \citet{james06} and then used by \citet{anderson09} and \citet{anderson12}
to compare the association of different CC SN types to
\ion{H}{ii} regions. This method consists of the following steps: a)
each pixel in the continuum-subtracted H$\alpha$ image is ranked
according to its flux value; b) then, the cumulative distribution of the pixel values is computed; c)
the value of this cumulative distribution at the rank value of the SN
pixel corresponds to the NCR value for the SN. The NCR is 0 if the SN
is not associated to \ion{H}{ii} regions, and 1 if the SN falls onto the
brightest pixel.  Following \citet{james06} and \citet{anderson12}, the NCR analysis was performed
on the $3\times3$ rebinned continuum-subtracted
H$\alpha$ image, in order to avoid possible errors in the
determination of the SN position. 
The pixels that we
used for the statistic were only those located within the 25th $B$
magnitude iso-contour. We note that internal extinction is not accounted for in the NCR analysis.

The NCR values that we obtained are reported in the last column of Table~\ref{metal} and
compared to those obtained by \citet{anderson12} for other types of CC
SNe in Fig.~\ref{NCRfig}. We notice that the NCR of both SN~2005dp and SN~1982F were already
computed by \citet{anderson09} and \citet{anderson12}. For these two
SNe, we obtained values which are very close to theirs (within 0.08
and 0.02 respectively).
 
Interestingly, the cumulative distribution of our NCR values follows
that of SNe~IIP and IIn, indicating a moderate association to
star-forming regions for our 1987A-like SNe.

\section{Discussion}
\label{sec:discussion}

This study is based mainly on local metallicity measurements.  
We have shown that it is important to take the metallicity gradient into account, but also
remark that, even though our measurements are made at the exact SN position,  the
size of the \ion{H}{ii} regions that we inspected is always on the
order of $\sim$0.1-1~kpc. 
It is possible that any particular SN sits in a small pocket of lower (or higher) metallicity 
within the larger \ion{H}{ii} region. Some apparent associations between SN position 
and \ion{H}{ii} region may also be due to projection effects.

By observing more targets, we reduce the impact of potential local inhomogeneities in the metallicity estimates and the results obtained are valid in a statistical sense.
However, it is worth noting that we do not find any
strong metallicity difference among \ion{H}{ii} regions that sit
close to each other, and that there are no large deviations from the
metallicity gradients that we measure. This suggests that the
metallicity varies smoothly across the galaxy and that the presence of
significant inhomogeneities is likely to be rare.

An important result from our study is that, {\it on average},
the BSG~SN population is located at lower metallicity than that of
normal SNe~IIP arising from RSGs. There may be biases in
the samples of BSG SNe and SNe~IIP that we use for the comparison and
in the following we discuss them. 

Most of the BSG~SNe were
discovered by amateur astronomers or by targeted surveys
\citep{pastorello12}, which preferentially observe bright, and
therefore metal-rich galaxies. However, this is also the case for the
SNe~IIP that were taken from \citet{anderson12}. This
effect might bias both samples towards higher
metallicities. However, we still see a difference in
metallicity between the two populations.  

   The vast majority of our hosts and of those observed by
\citet{anderson12} are nearby galaxies (recessional velocity $v$~$<$~6000
km~s$^{-1}$), and this distance cut reduces the bias of the two samples towards intrinsically very bright objects. However, the light curves of SNe~IIP exhibit a long plateau phase (3-4 months), whereas the peak of SN~1987A-like events lasts a few weeks. This could bias the SN~1987A-like population that might contain a higher fraction of faint, undetected objects. In fact, many of the discovered BSG~SNe are brighter \citep{pastorello12} than their prototype, SN~1987A. The average peak brightness of BSG~SNe is $<$~$M_{R}^{max}$~$>$~$\approx$~$-$17 (data from \citealp{pastorello12} and \citealp{arcavi12}), which does not differ considerably from that of SNe~IIP (\citealp{kelly12} indicate $<$~$M_{R}^{max}$~$>$~$\approx$~$-$16 for the Type II SN class). 
We looked for possible
correlations between the metallicity of BSG~SNe and their light-curve properties. It was found
 that the peak brightness of our transients \citep[from][]{pastorello12} does not correlate with the measured metallicity. This is also true for the light-curve shape. For example, SN~2000cb is
at almost solar metallicity and SN~2005ci is clearly at LMC metallicity, but their light-curve shapes are very similar \citep{kleiser11,arcavi12}.


In summary, BSG~SNe are -- on average -- at lower metallicity
compared to those of SNe~IIP and Ibc and the moderately low
metallicity is a characteristic of the majority of BSG~SNe.
 Although it is clear from stellar evolution theory that also other effects
can influence the star - such as rotation, mixing and binarity - our simple observational test indicates that metallicity likely plays a role in the formation
of exploding BSGs, at least for a large fraction of them.
This would find a natural explanation if the progenitor was a single star. According to single-star evolution models, LMC metallicity can induce a $\sim$20~$M_{\odot}$ star to explode in the blue part of the HR diagram \citep{brunish82}.

However, our results also indicate that two SN~1987A-like SNe exploded at
near solar metallicity (SN~2004em and SN~1998A), and it is also clear that even if the
mean metallicity is low, it is not {\it very} low. 
This supports the idea that, at least for some BSG~SNe, a moderately low metallicity is not a {\it necessary} ingredient. We note that theoretical models show that a star
in a binary system \citep{pod92} or a single, fast rotating star at
solar metallicity \citep{hirschi04,ekstrom12} can end its life in the
BSG stage. 

Furthermore, the rarity of BSG~SNe (\citealp{pastorello12} estimate that they represent 1$-$3\% of all CC SNe) suggests that the slightly low metal abundance is probably one of the ingredients or channels but not the only one.
The main-sequence progenitor mass range that potentially gives rise to 
an exploding BSG is approximately 16-22~$M_{\odot}$ (\citealp{arnett89} report this range for the progenitor of SN~1987A). Assuming an initial mass function (IMF) characterized by $\alpha$~$=$~$-$2.35 \citep{salpeter55}, this corresponds to $\sim$15\% of all CC~SNe (here we adopted 8.5~$M_{\odot}$ as the lower limit to produce a CC~SN, see \citealp{smartt09}).
In order to match the rate presented by \citet{pastorello12}, this $\sim$15\% must be reduced by a factor of $\sim$10, which is likely too high to be explained only in terms of moderately sub-solar metallicity.

Finally, we discuss the relatively weak association of BSG~SNe to H$\alpha$
emitting regions. BSG~SNe were found to have a degree of associations
to \ion{H}{ii} regions that is similar to those of SNe~IIP and IIn,
with many events sitting relatively far from bright star-forming regions. The
association of a SN type to star-forming regions has been interpreted
in terms of progenitor mass. The stronger the association, the
larger is the typical initial progenitor mass \citep{anderson12} of that SN
type. However, since we know from modeling \citep[e.g.][]{utrobin11,taddia12}
that most of our BSG~SNe arise from stars that are more massive
(M$_{ZAMS}\sim$20~M$_{\odot}$) than the typical RSG progenitors of
SNe~IIP ($\sim$10~M$_{\odot}$), it seems that this mass difference
is too small to be detected by the NCR method (see also discussion in
\citealp{crowther13}). The weak association with \ion{H}{ii} regions for
the massive progenitors of 1987A-like SNe could also be interpreted
 in terms of binary scenarios giving rise to longer lifetimes.

\section{Conclusions}
\label{sec:conclusions}

The main conclusions of this work are:
\renewcommand{\labelitemi}{$\bullet$}
\begin{itemize}
\item{SN~1987A-like events are either located in the outskirts of bright, metal rich galaxies or in faint, intrinsically metal poor galaxies.}

\item{Most of the 1987A-like SNe show moderately sub-solar oxygen abundances 
($<$~12$+$log(O/H)~$>$~$=$~8.36$\pm$0.05~dex), comparable to that of the LMC. 
Compared to normal SNe~IIP that arise from RSGs, they have a lower metallicity. The metallicity distribution of our BSG~SNe
 is rather similar to that of SNe~IIb.}
 
\item{Two of our objects do indicate a solar or almost solar metallicity (SN~1998A and SN~2004em).}

\item{BSG~SNe show a moderate association to star-forming regions, similar to that of SNe~IIP and IIn.}

\item{The moderately low metallicity that characterizes the majority of the SNe in the sample likely plays a role in the formation of at least some 1987A-like SNe. This might be explained in the context of a single-star progenitor scenario. However, the nearly-solar abundances of two BSG~SNe indicate that also other ingredients and/or channels (possibly binarity) contribute to produce some of these transients.}

\end{itemize}

\begin{acknowledgements}
M.~D. Stritzinger gratefully acknowledges generous support provided by the Danish
 Agency for Science and Technology and Innovation  
realized through a Sapere Aude Level 2 grant.
G. Leloudas is supported by the Swedish Research Council through grant No. 623-2011-7117.
The Dark Cosmology Centre is funded by the Danish National Research Foundation.
The Oskar Klein Centre is funded by the Swedish Research Council.
The Nordic Optical Telescop is operated by the Nordic Optical Telescope Scientific Association at the Observatorio del Roque de los Muchachos, La Palma, Spain, of the Instituto de Astrofisica de Canarias.
This research has made use of the NASA/IPAC Extragalactic Database (NED) which is operated by the Jet Propulsion Laboratory, California Institute of Technology, under contract with the National Aeronautics and Space Administration. This research has made use of the SIMBAD database,
operated at CDS, Strasbourg, France. We acknowledge the usage of the HyperLeda database.
\end{acknowledgements}

\bibliographystyle{aa} 

\onecolumn

\clearpage

\begin{deluxetable}{lll|lll|cc|ccc|cc|c|cc}
\rotate
\tabletypesize{\scriptsize}
\tablewidth{0pt}
\tablecaption{The sample of 14$^*$ BSG~SNe known in the literature and their host galaxies.\label{tab:sample}}
\tablehead{
\colhead{SN} &
\colhead{RA} &
\colhead{DEC} &
\colhead{Galaxy} &
\colhead{RA} &
\colhead{DEC} &
\colhead{t-type} &
\colhead{Galaxy type} &
\colhead{major axis} &
\colhead{minor axis} &
\colhead{PA} &
\colhead{M$^{gal}_{B}$} &
\colhead{$E(B-V)_{MW}$} &
\colhead{r$_{\rm SN}$/R$_{25}$} &
\colhead{z} &
\colhead{$\mu$}\\
\colhead{} &
\colhead{(hh:mm:ss)} &
\colhead{(dd:mm:ss)} &
\colhead{} &
\colhead{(hh:mm:ss)} &
\colhead{(dd:mm:ss)} &
\colhead{} &
\colhead{} &
\colhead{(arcsec)} &
\colhead{(arcsec)} &
\colhead{(deg)} &
\colhead{(mag)} &
\colhead{(mag)} &
\colhead{} &
\colhead{} &
\colhead{(mag)}}
\startdata
 1909A  & 14:02:02.78 & $+$54:27:56.9 &M~101  & 14:03:12.54 & $+$54:20:56.22 & 5.9 & SABc    &  1728    &  1614    & 28.5   & -21.02   &0.0086    &   0.92 &  0.000804  & 29.18\\
 1982F  & 12:30:38.70 & $+$41:38:00.0 &NGC~4490  & 12:30:36.24  & $+$41:38:38.03  & 6.9 &   SBcd  &  378     &  186     & 125    &-19.52    &0.0219    &      0.29&  0.001885&29.52  \\ 
 1987A  & 05:35:28.01 & $-$69:16:11.6 &LMC       & 05:23:34.52   & $-$69:45:22.1   & 9.0 &   SBm  &  38739   &  32928   & 170    &-18.02    &0.0319   &    0.26&  0.000927& 18.50 \\ 
 1998A  & 11:09:50.33 & $-$23:43:43.1 &IC~2627   & 11:09:53.39   & $-$23:43:33.4   & 4.4 &   SABc  &  161.50  &  140.50  & 20     &-19.03    &0.1215    &    0.59&  0.006971 &31.13 \\
 1998bt\tablenotemark{*}  &   13:25:41.9 & $-$26:46:56      &A132541-2646     & 13:25:41.9 & $-$26:46:56     &\ldots & \ldots    &    \ldots  &  \ldots  & \ldots &-19.56    &0.0607    & \ldots &0.436    & 41.26 \\
 2000cb & 16:01:32.15 & $+$01:42:23.0 &IC~1158   & 16:01:34.08   & $+$01:42:28.2   & 5.0 & SABc    &  147.30  &  100.16  & 137    &-19.36    &0.1103    &   0.55 &   0.006971 & 32.69\\
 NOOS...\tablenotemark{**}  & 05:55:38.07 & $-$68:55:47.1 &2MASX...\tablenotemark{**}       & 05:55:39.78   & $-$68:55:38.4   & 2.2 &S     &  30.86   &  14.20    & 60     & -20.01    &0.1600    & 0.94   &  0.030608  & 35.46\\
 2004ek & 01:09:58.51 & $+$32:22:47.7 &UGC~724   & 01:09:59.40   & $+$32:22:06.4   & 3.0 &  Sab   &  125.40  &  104.08  & 15     &-19.99    &0.0644    &  0.73  & 0.017275   & 34.25\\
 2004em & 19:31:31.11 & $+$35:52:15.7 &IC~1303   & 19:31:30.06   & $+$35:52:35.8   & 5.0 &   Sc  &  77.30   &  48.70   & 115    &-19.79    &0.1099    &    0.75&   0.014894 & 33.93\\
 2005ci & 14:34:44.88 & $+$48:40:19.8 &NGC~5682  & 14:34:44.98   & $+$48:40:12.9   & 3.0 &   Sb  &  99.60   &  33.86   & 127    &-18.33    &0.0309    &   0.44 &   0.007581 & 32.80\\
 2005dp & 14:27:36.62 & $+$41:15:15.0 &NGC~5630  & 14:27:36.61   & $+$41:15:27.9   & 8.0 &  Sd   &  134.30  &  40.29   & 98     &-18.69    & 0.0115    &   0.63 &  0.008856  &32.57 \\
 2006V  & 11:31:30.01 & $-$02:17:52.2 &UGC~6510  & 11:31:32.08   & $-$02:18:33.1   & 5.5 &  SABc   &  114.30  &  106.30  & 10     &-21.19    &0.0284    &    0.94&   0.015828 & 34.06\\
 2006au & 17:57:13.56 & $+$12:11:03.2 &UGC~11057 & 17:57:14.93   & $+$12:10:46.1   & 6.0 &   Sc  &  119.70  &  49.08   & 90     &-20.16    &  0.1709   &   0.83 &  0.009580  &32.80 \\
 2009E  & 12:09:49.56 & $+$58:50:50.3 &NGC~4141  & 12:09:47.31  & $+$58:50:57.06  & 6.0 &   SBc  &  77.30   &  54.88   & 75     & -17.51   &  0.0188   &   0.57 &   0.006328 & 32.07\\
 \enddata
\tablenotetext{*}{SN~1998bt is removed from the sample given the new redshift (see discussion in the text).}
\tablenotetext{**}{Full name OGLE-2003-NOOS-005 and 2MASX J05553978-6855381 for its host galaxy.}
\tablecomments{SN coordinates (J2000) 
are from the ASC, except for SN~1909A \citep[from][]{israel75}, SN~1982F (from NED), OGLE-2003-NOOS-005 (from \citealp{pastorello12}).\\
Galaxy coordinates (J2000) are as listed in NED, with the exception of 2MASX J05553978-6855381 \citep{pastorello12}.\\
For all the galaxies the morphological t-type is taken from the ASC, except for UGC724 (t-type from \citealp{hakobyan12}) and 2MASX J05553978-6855381 (t-type from HyperLeda). The galaxy type is as reported in \citet{pastorello12}.\\
Major and minor axes are taken from NED, except for M~101 (from SIMBAD) and 2MASX J05553978-6855381 (we remeasured the major and minor axes on our $B$-band image).\\
PA for most of the galaxies is from NED. For IC2627, the PA is taken from SIMBAD. For M~101, the PA is from \citet{jarrett03}.\\
The $E(B-V)_{MW}$ values are taken from \citet{schlegel98} and the redshifts  (z) from NED. The distance moduli ($\mu$)
and M$_{B}$ are computed as explained in Sect.~\ref{sec:indirect}. Absolute magnitudes, distance moduli and $E(B-V)$ for LMC and 2MASX J05553978-6855381 are from \citet{pastorello12}.} 
 \end{deluxetable}


\begin{deluxetable}{ll|ccc|ccc}
\rotate
\tabletypesize{\scriptsize}
\tablewidth{0pt}
\tablecaption{Log of the photometric and spectroscopic observations.\label{log}}
\tablehead{
\colhead{SN} &
\colhead{Galaxy} &
\colhead{}&
\colhead{Photometry} &
\colhead{}&
\colhead{}&
\colhead{Spectroscopy}&
\colhead{}\\
\colhead{}&
\colhead{}&
\colhead{Telescope}&
\colhead{Instrument+Filter}&
\colhead{Integration (sec)}&
\colhead{Telescope}&
\colhead{Instrument+Grism+Slit}&
\colhead{Integration (sec)}}
\startdata
 1909A  &                    M 101                  & NOT       & ALFOSC$+$R\tablenotemark{*}                &    3$\times$200         & NOT     & ALFOSC$+\#$4$+$1.0$\arcsec$  & 1$\times$1800  \\ 
           &                                              & NOT       & ALFOSC$+$H$\alpha(\#21)$\tablenotemark{*}  &   3$\times$40          &         &                     &                \\  
 \hline                                                                                                      
 1982F  &                    NGC 4490                  & INT       & WFC$+$R                   & 1$\times$200  & NOT     & ALFOSC$+\#$4$+$1.0$\arcsec$  & 1$\times$1800  \\ 
           &                                              & INT       & WFC$+$H$\alpha(\#197)$           & 1$\times$750 &         &                     &                \\
 \hline
  1998A   &            IC 2627                   & VLT          &FORS2$+$R &5$\times$10 & VLT       &FORS2$+$300V$+$1.0$\arcsec$    &   3$\times$600      \\
(1994R)           &                                              & VLT &FORS2$+$H$\alpha/2500+60$ &5$\times$80 & VLT        &FORS2$+$300V$+$1.0$\arcsec$    &   3$\times$600      \\
           &                                              & & & & NTT     & EMMI $+\#$3$+$1.5$\arcsec$     &   2$\times$300      \\  
           &                                              & & & & NTT        &EMMI$+\#$3$+$1.5$\arcsec$    &   2$\times$300      \\
  \hline                                                                                                                                                       
 1998bt                   &                         A132541-2646                      & VLT          &FORS2$+$R &3$\times$150 & VLT       &FORS2$+$300V$+$1.0$\arcsec$    &   3$\times$1800      \\
                            &                                                           & VLT          &FORS2$+$B &3$\times$300 &           &                              &               \\                     
  \hline                                                                                          
2000cb  &                    IC 1158                   & VLT          &FORS2$+$R &10$\times$10          & VLT           &FORS2$+$300V$+$1.0$\arcsec$    &   3$\times$1200      \\
           &                                              & VLT          &FORS2$+$H$\alpha/2500+60$ &10$\times$80  &         &    &      \\
  \hline                                                                                          
 NOOS...\tablenotemark{*}  &2MASX...\tablenotemark{*} &  VLT         & FORS2$+$R           & 2$\times$10+3$\times$20            & VLT        & FORS2$+$300V+1.0$\arcsec$  &  4$\times$900 \\
                               &                          &   VLT        &FORS2$+$SII 2000($+$63)         &5$\times$80              &         &                     &                \\
                               &                          &   VLT        &FORS2$+$B         &2$\times$20+3$\times$10              &         &                     &                \\                               
  \hline                                                                                                    
 2004em &                    IC 1303                   & NOT       & ALFOSC$+$R                &3$\times$40  &  NOT    & ALFOSC$+\#$4$+$1.0$\arcsec$  & 2$\times$1800  \\                                                                    
           &                                              & NOT       & ALFOSC$+$H$\alpha(\#50)$  &3$\times$200 &         &                     &                \\
  \hline                                                                                          
 2005ci &                    NGC 5682                  & NOT       & ALFOSC$+$R                &3$\times$40  &  NOT    & ALFOSC$+\#$4$+$1.0$\arcsec$  & 3$\times$1800   \\                                                                    
           &                                              & NOT       & ALFOSC$+$H$\alpha(\#77)$  &3$\times$200 &     NOT    & ALFOSC$+\#$4$+$1.0$\arcsec$  & 3$\times$1800   \\                                                                    
  \hline                                                                                          
 2005dp &                    NGC 5630                  & NOT        & ALFOSC$+$R                &3$\times$300&        NOT    & ALFOSC$+\#$4$+$1.0$\arcsec$  & 3$\times$1800   \\                                                                    
           &                                              & NOT        & ALFOSC$+$H$\alpha(\#49)$    &2$\times$600&             NOT    & ALFOSC$+\#$4$+$1.0$\arcsec$  & 3$\times$1800  \\
                       &                                              &         &   & &             NOT    & ALFOSC$+\#$8$+$1.0$\arcsec$  & 1$\times$1800  \\
  \hline                                                                                          
 2006V   &                  UGC 6510                   & VLT          &FORS2$+$R &5$\times$10 & VLT     &FORS2$+$300V$+$1.0$\arcsec$    &   3$\times$1200      \\
            &                                             & VLT &FORS2$+$H$\alpha/4500+61$ &5$\times$80  &  NOT    & ALFOSC$+\#$4$+$1.0$\arcsec$  &  3$\times$1800      \\         
  \hline                                                                                          
 2006au &                    UGC 11057                  & VLT          &FORS2$+$R &10$\times$5 & VLT           &FORS2$+$300V$+$1.0$\arcsec$    &   3$\times$1200      \\
           &                                              & VLT &FORS2$+$H$\alpha/2500+60$ &10$\times$40 &         &    &      \\            
  \hline                                                                                          
 2009E  &                    NGC 4141                  & NOT       & ALFOSC$+$R                &3$\times$40  &  NOT    & ALFOSC$+\#$4$+$1.0$\arcsec$  &  3$\times$1800  \\
           &                                              & NOT       & ALFOSC$+$H$\alpha(\#77)$  &3$\times$200 &         &                     &                \\
 \enddata
\tablenotetext{*}{With the NOT we imaged M~101 only within the FOV including the positions of SN~1909A and of the \ion{H}{ii} region that we
 spectroscopically observed (see the top panel of Fig.~\ref{sn09A}). The
continuum-subtraced H$\alpha$ image of the entire galaxy was obtained from \href{http://www.aoc.nrao.edu}{http://www.aoc.nrao.edu}.} 
\tablenotetext{**}{Full name OGLE-2003-NOOS-005 and 2MASX J05553978-6855381 for its host galaxy.}
 \end{deluxetable}

\clearpage

\begin{deluxetable}{c|c|c|c|c|cc|c|c}
\tabletypesize{\scriptsize}
\tablewidth{0pt}
\tablecaption{Metallicity estimates and Normalised Cumulative Rank (NCR) pixel index for the sample of BSG SNe.\label{metal}}
\tablehead{
\colhead{SN} &
\colhead{12$+$log(O/H)}&
\colhead{12$+$log(O/H)}&
\colhead{12$+$log(O/H)}&
\colhead{\textbf{12$+$log(O/H)}}&
\colhead{12$+$log(O/H)}&
\colhead{\ion{H}{ii} region--SN}&
\colhead{12$+$log(O/H) gradient}&
\colhead{NCR}\\
\colhead{}&
\colhead{(indirect}&
\colhead{(indirect}&
\colhead{(N2 at the}&
\colhead{\textbf{(N2 at r$\mathbf{_{SN}}$, dex})}&
\colhead{(N2 of the \ion{H}{ii}}&
\colhead{distance}&
\colhead{(dex~R$_{25}^{-1}$)}&
\colhead{}\\
\colhead{}&
\colhead{at the host center,}&
\colhead{at r$_{\rm SN}$, dex)}&
\colhead{host center, dex)}&
\colhead{}&
\colhead{region closest}&
\colhead{(kpc)}&
\colhead{}&
\colhead{}\\
\colhead{}&
\colhead{ dex)}&
\colhead{}&
\colhead{}&
\colhead{}&
\colhead{to the SN, dex)}&
\colhead{}&
\colhead{}&
\colhead{}}
\startdata
1909A & 9.13&  8.30 & 8.79$\pm$0.06 & \textbf{7.96$\pm$0.10}&8.23$\pm$0.20 &7.66 &-0.900$\pm$0.085&0\\
1982F & 8.85&8.83 & 8.24$\pm$0.09 &\textbf{8.26$\pm$0.12}&8.21$\pm$0.20 &0.36 &0.078$\pm$0.256&0.113\\
1987A &   8.57  & 8.45   &  8.46$\pm$0.06\tablenotemark{**}    &     \textbf{8.37\tablenotemark{***}}     & -       & -&-0.323$\pm$0.215\tablenotemark{**}&-\\
1998A & 8.76&8.48 & 8.78$\pm$0.03 &\textbf{8.68$\pm$0.06} &8.63$\pm$0.20 &0 &-0.157$\pm$0.076&0.608\\
2000cb& 8.82&8.56 & 8.68$\pm$0.03 &\textbf{8.45$\pm$0.06}&8.39$\pm$0.20  &  1.24 &-0.411$\pm$0.096&0\\
NOOS...\tablenotemark{*}   & 8.94& 8.50 & 9.02$\pm$0.13 & \textbf{8.39$\pm$0.28} & 8.57$\pm$0.20&2.44 & -0.677$\pm$0.250&0.076\\
2004ek &  8.94&8.59    & - &  - &  - & - &-&-\\
2004em & 8.90&8.55& 8.72$\pm$0.05 &\textbf{8.56$\pm$0.11 } &8.59$\pm$0.20 &1.16 &-0.213$\pm$0.122&0.024\\
2005ci & 8.63&8.42& 8.49$\pm$0.02 &\textbf{8.31$\pm$0.04}   &8.31$\pm$0.20 &0 &-0.412$\pm$0.089&0.179\\
2005dp & 8.70&8.40& 8.49$\pm$0.03 &\textbf{8.27$\pm$0.07 }& 8.37$\pm$0.20   &3.38 &-0.336$\pm$0.094&0.510\\
2006V  & 9.16&8.72& 8.81$\pm$0.05 &\textbf{8.35$\pm$0.11}   & 8.36$\pm$0.20 &6.03 &-0.487$\pm$0.101&0\\
2006au & 8.97&8.58&  8.95$\pm$0.06 &\textbf{8.49$\pm$0.12}  & 8.54$\pm$0.20 & 0 &-0.562$\pm$0.113&0.073\\
2009E & 8.48&8.21& 8.44$\pm$0.05 &\textbf{8.22$\pm$0.08}   & 8.22$\pm$0.20 &0 &-0.391$\pm$0.112&0.857\\
$<$BSG~SNe$>$&8.83$\pm$0.06  & 8.51$\pm$0.04& 8.66$\pm$0.07 &\textbf{8.36$\pm$0.05}   & 8.40$\pm$0.05 &2.02$\pm$0.80 &-0.399$\pm$0.072&0.222$\pm$0.089\\
 \enddata
\tablenotetext{*}{Full name OGLE-2003-NOOS-005.}
\tablenotetext{**}{From \citet{pagel78}.}
\tablenotetext{***}{From \citet{russell90}; it is the average metallicity of the \ion{H}{ii} regions in the LMC.}
\tablecomments{In the 2nd and 3rd columns we report the {\it indirect} metallicity estimates at the host center and at the SN distance.
The {\it direct} metallicity measurements at the host center and at the SN distance are reported in the 4th and 5th columns. The quoted error is the fit uncertainty. The systematic error for the N2 method is 0.18~dex. Columns 6 and 7
include the oxygen abundance of the \ion{H}{ii} region closest to the SN position, and the distance in kpc between this \ion{H}{ii} region and the SN position. The last two columns show the measured host metallicity gradient and the NCR index for each SN. 
 The mean value and the uncertainty of each quantity for the entire sample is reported in the last row of the table. } 
 \end{deluxetable}

\clearpage

\clearpage

\clearpage
\onlfig{6}{
\begin{figure}
 \centering
\includegraphics[height=8cm,angle=90]{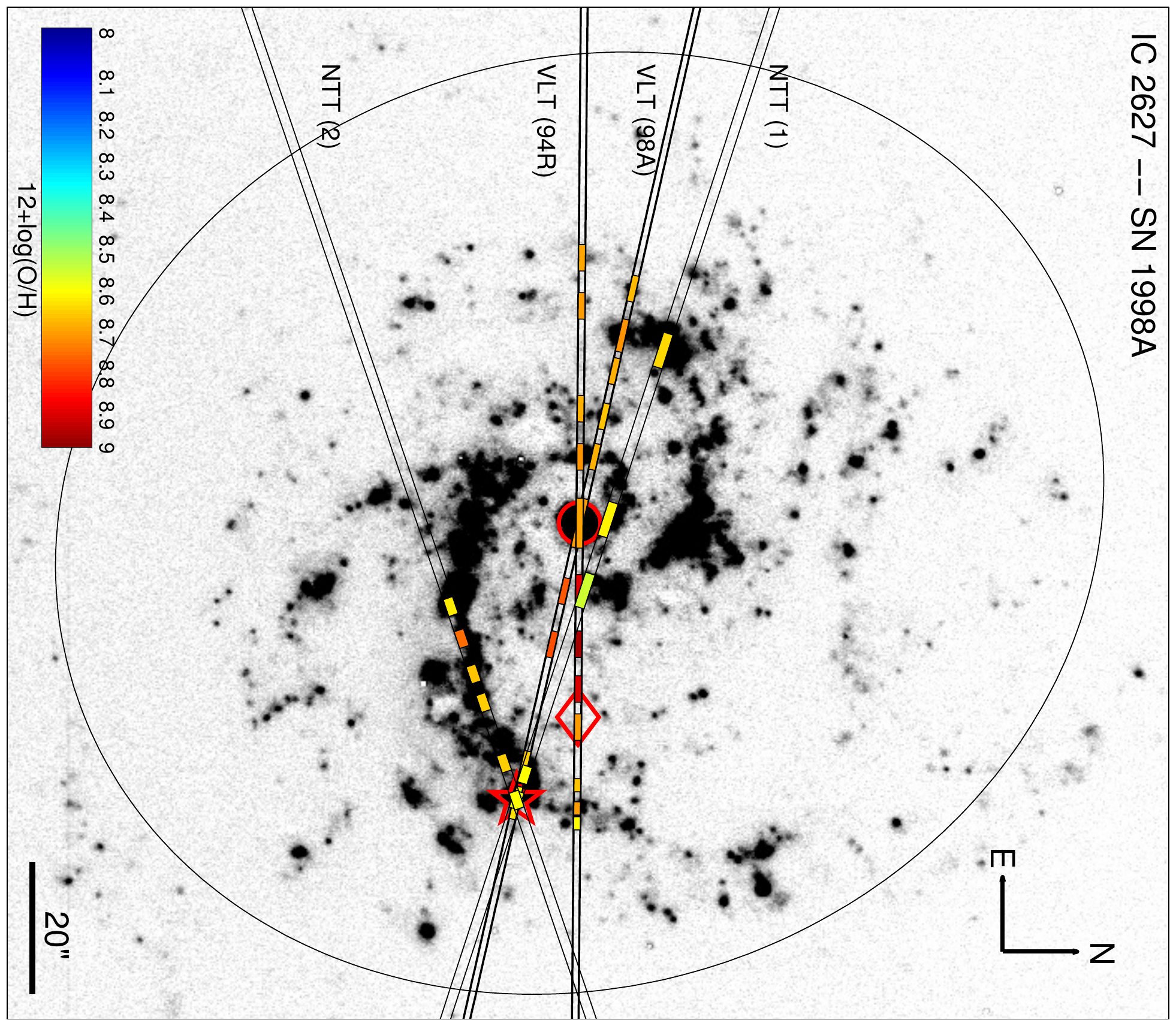}\\
  $\begin{array}{cc}
 \includegraphics[width=4.0cm,angle=0]{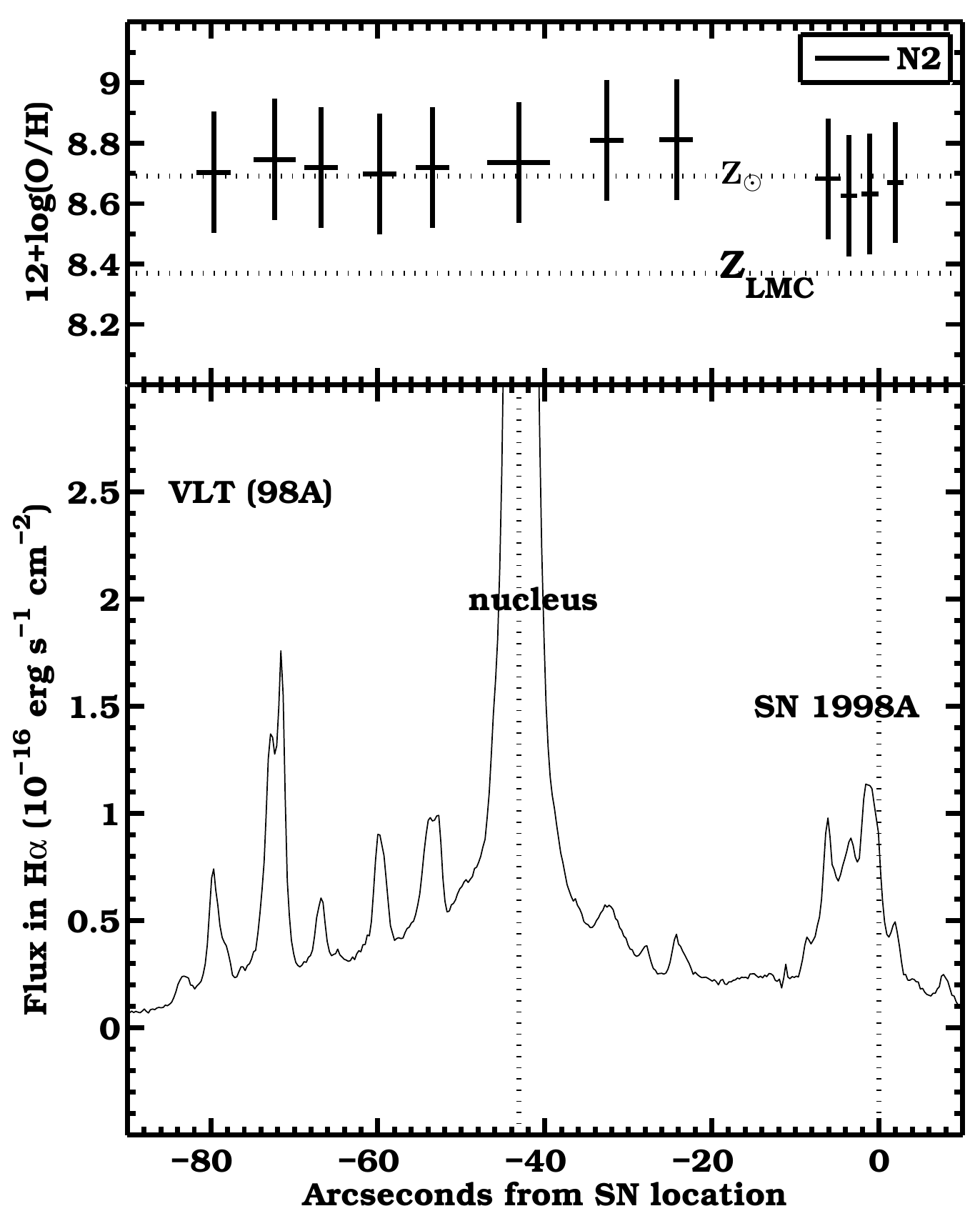}&
 \includegraphics[width=4.0cm,angle=0]{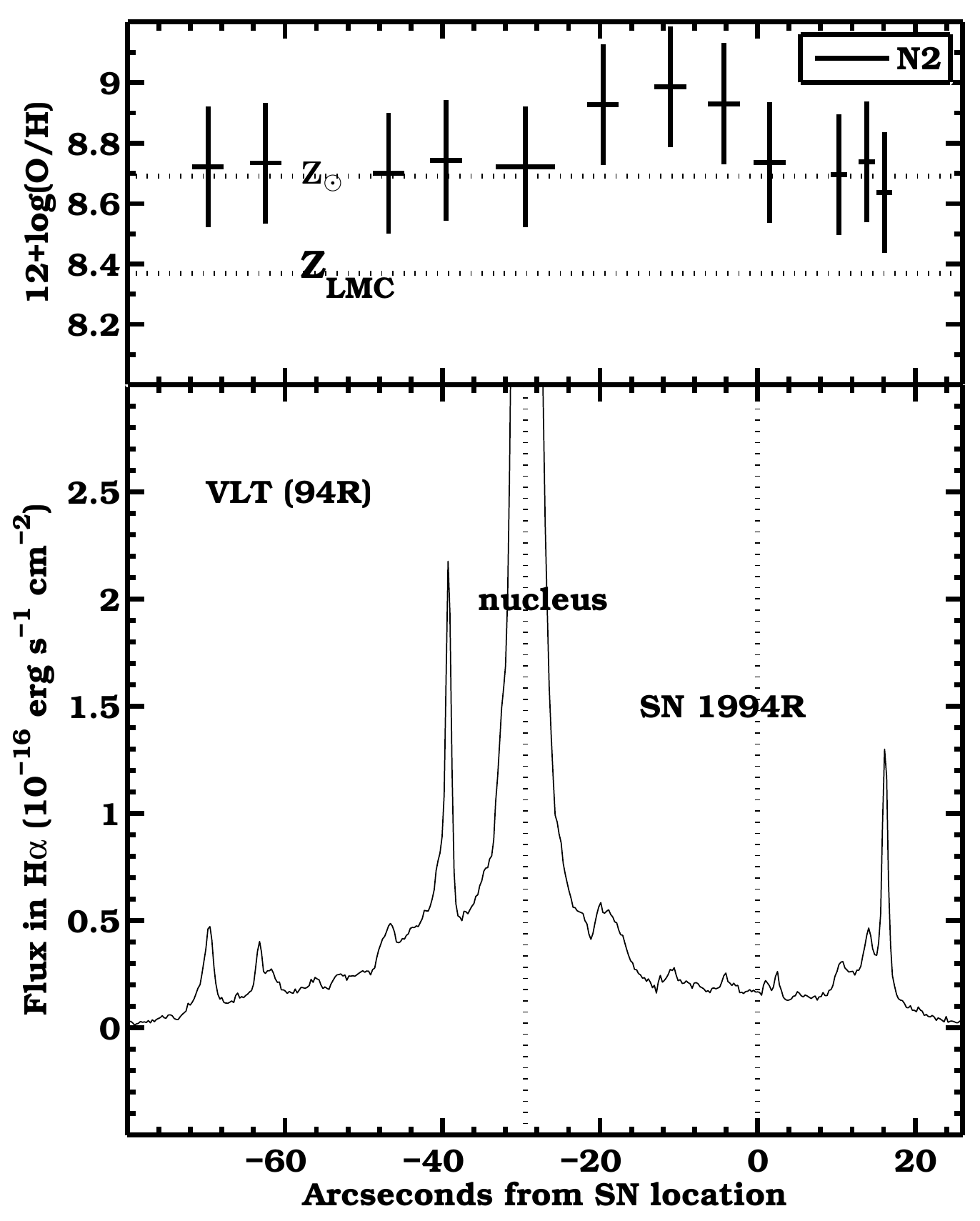}\\
 \includegraphics[width=4.0cm,angle=0]{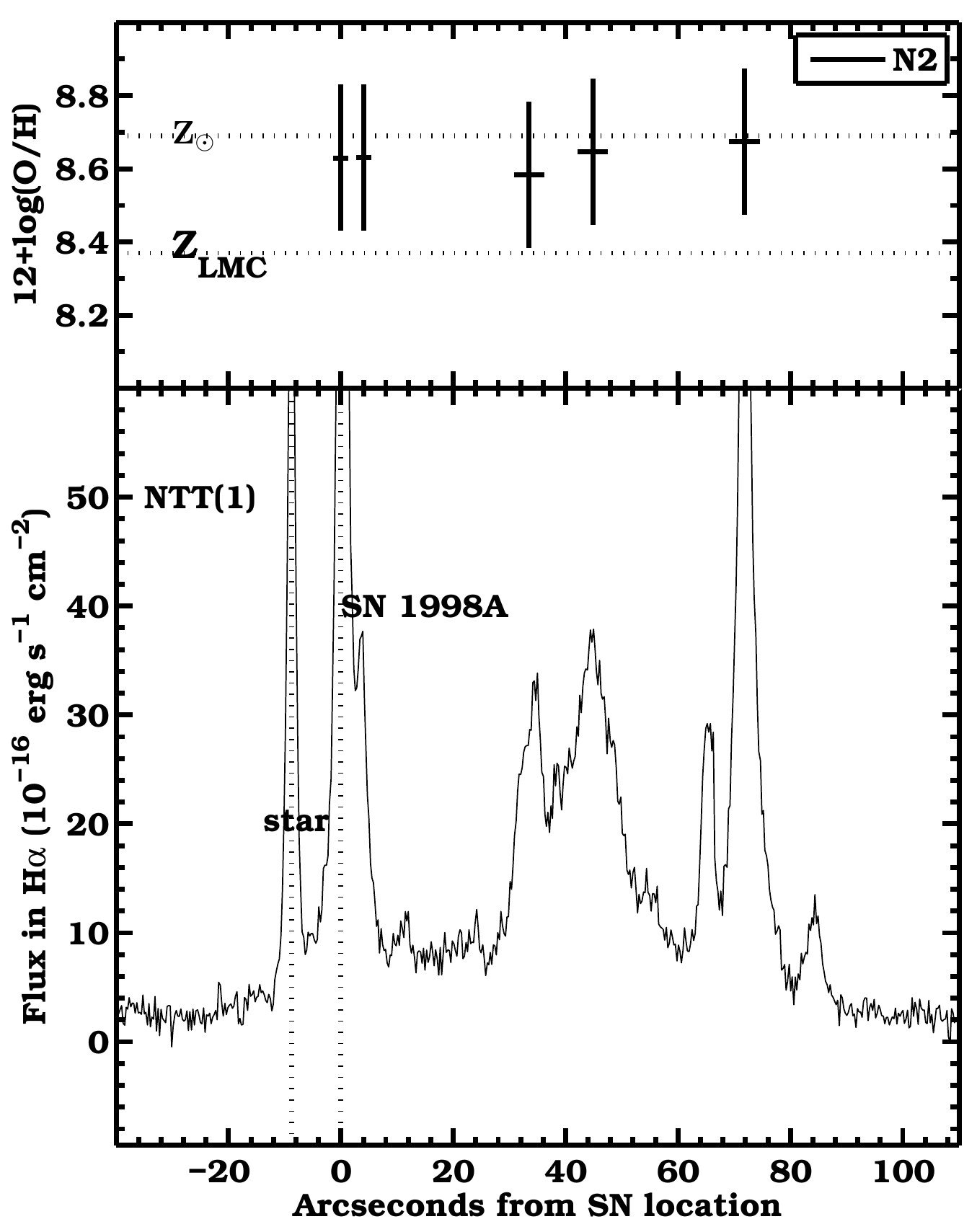}&
  \includegraphics[width=4.0cm,angle=0]{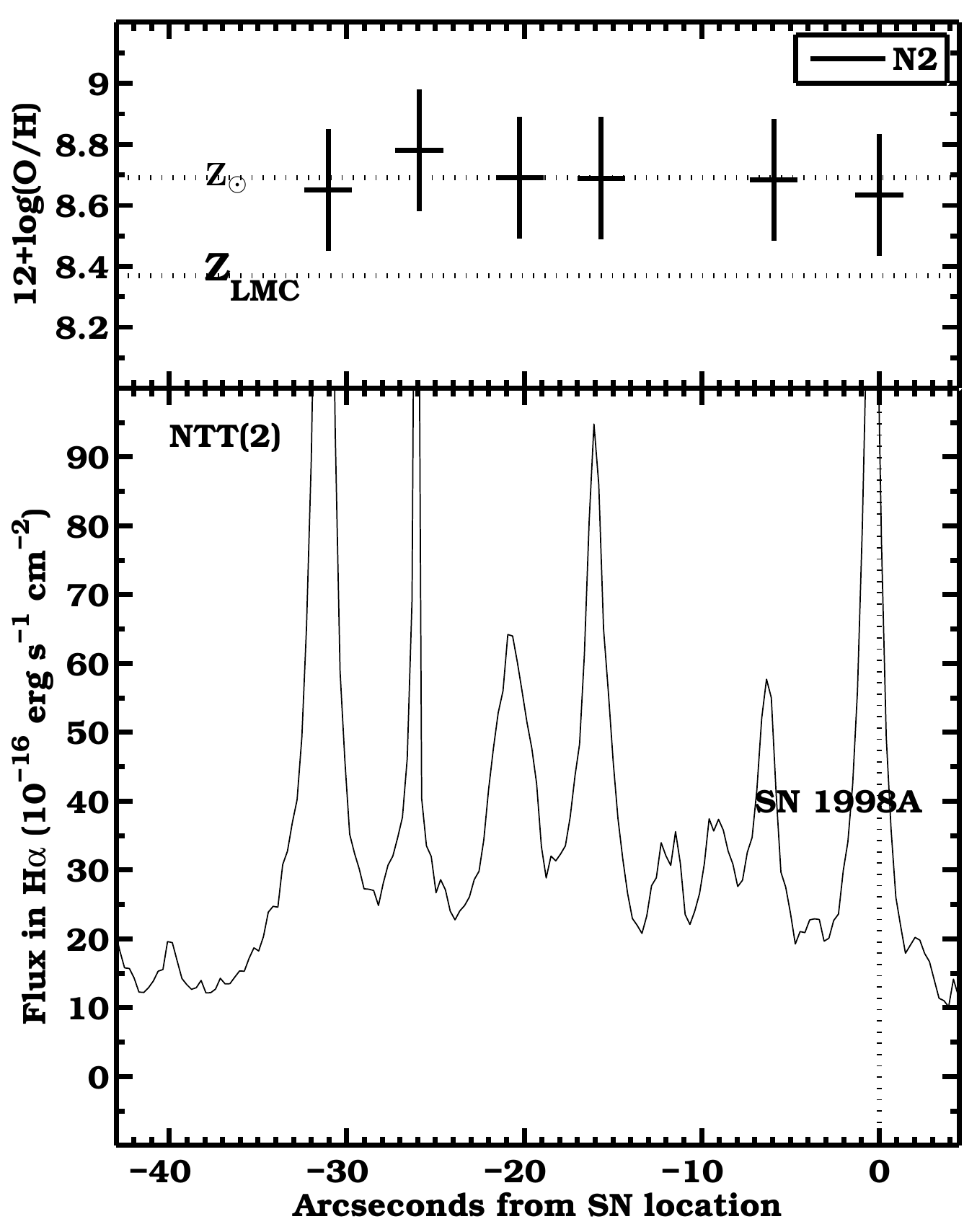} 
 \end{array}$\\
\includegraphics[width=8cm,angle=0]{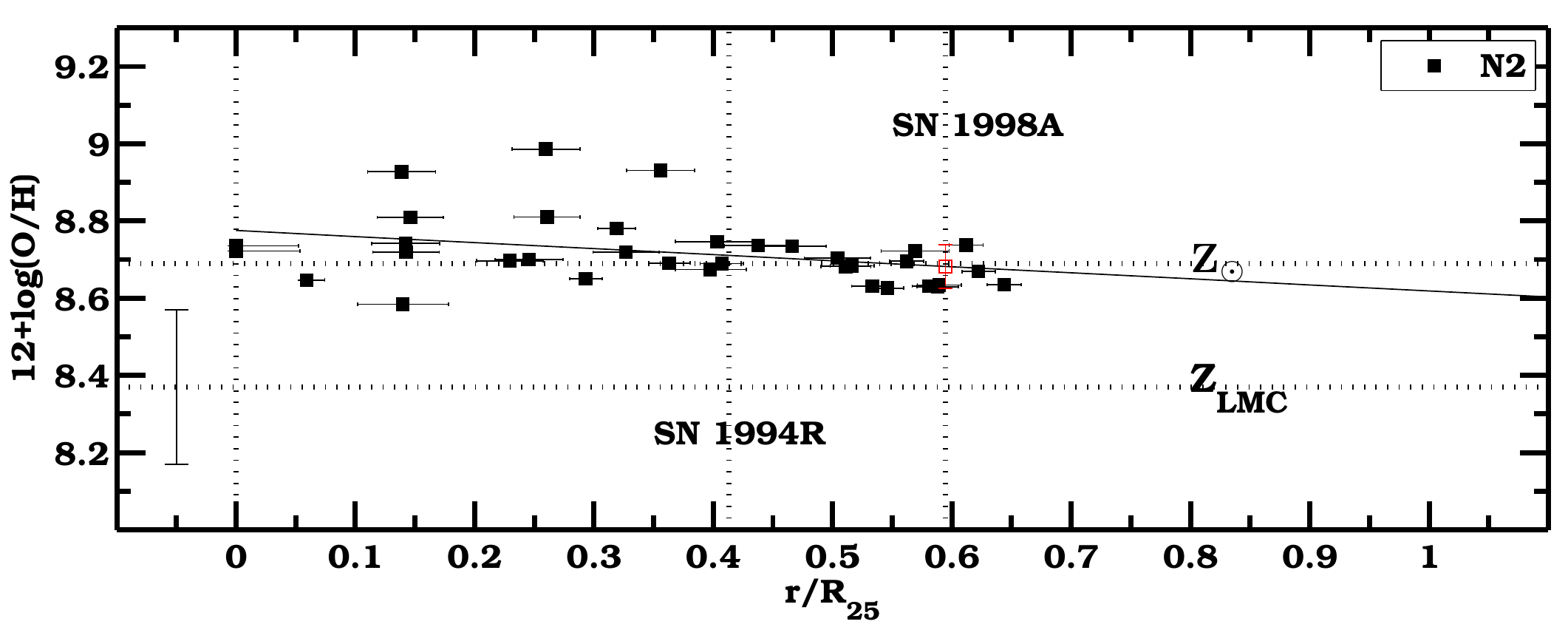}
  \caption{\textit{(Top panel)} Continuum-subtracted H$\alpha$ image of IC~2627. The 25th $B$-band magnitude elliptic contour is shown by a black solid line, along with the position of SN~1998A (marked by a red star), the center of the galaxy (marked by a red circle) and the position of SN~1994R (marked by a red diamond). The positions of the four slits are shown, and a color code is used to present the N2 metallicity measurements at the position of each \ion{H}{ii} region that we inspected. \textit{(Center panels)} Flux at the H$\alpha$ wavelength along the four slits, shown as a function of the distance from the SN location. These positions are marked by dotted lines. The N2 measurements for each \ion{H}{ii} region are shown at the corresponding positions in the upper sub-panels. \textit{(Bottom panel)} Metallicity gradient of IC~2627. The linear fit on our measurements is shown by a solid line. The interpolated metallicity at the SN distance is marked by a red square and its uncertainty corresponds to the fit error. The error bar ($\pm$0.2~dex) for our N2 measurements is shown aside. The positions of SN and nucleus are marked by vertical dotted lines. The solar metallicity \citep{asplund09} and the LMC metallicity \citep{russell90} are indicated by two horizontal dotted lines.\label{sn98A}}
 \end{figure}}

\clearpage
\onlfig{7}{
\begin{figure}
 \centering
  $\begin{array}{cc}
\includegraphics[width=7cm,angle=0]{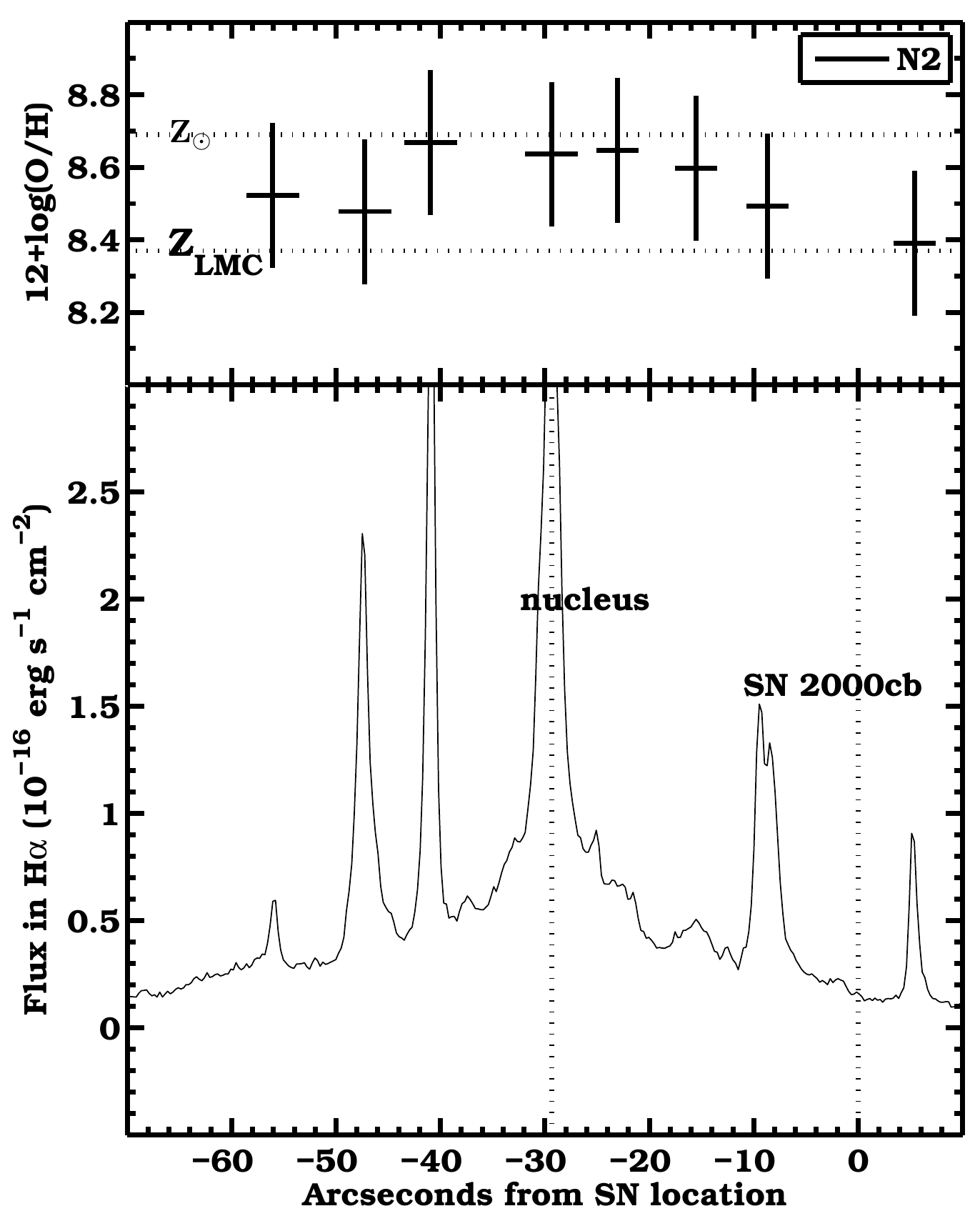}&
\includegraphics[width=8.6cm,angle=90]{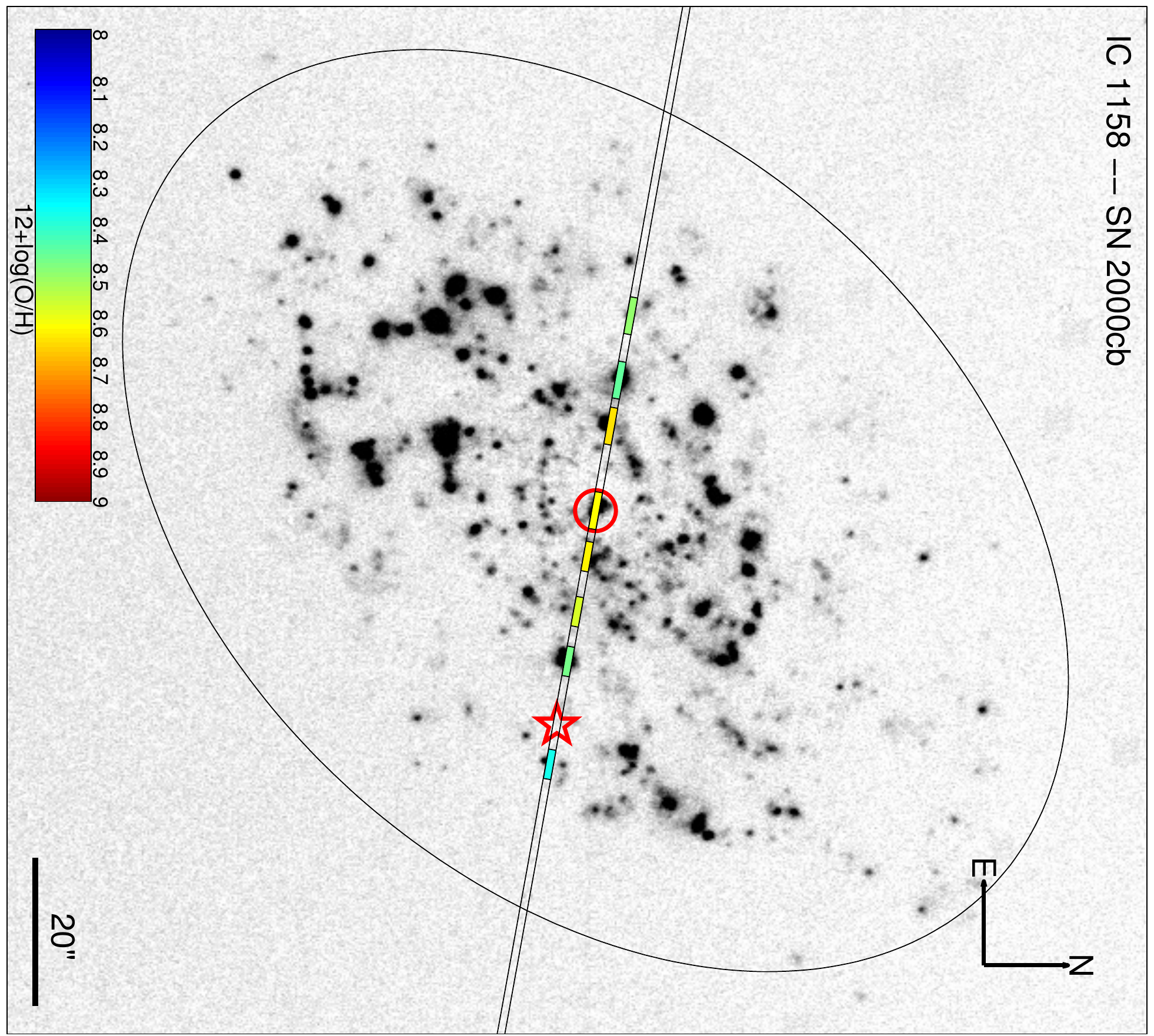}
\end{array}$
\includegraphics[width=15.1cm,angle=0]{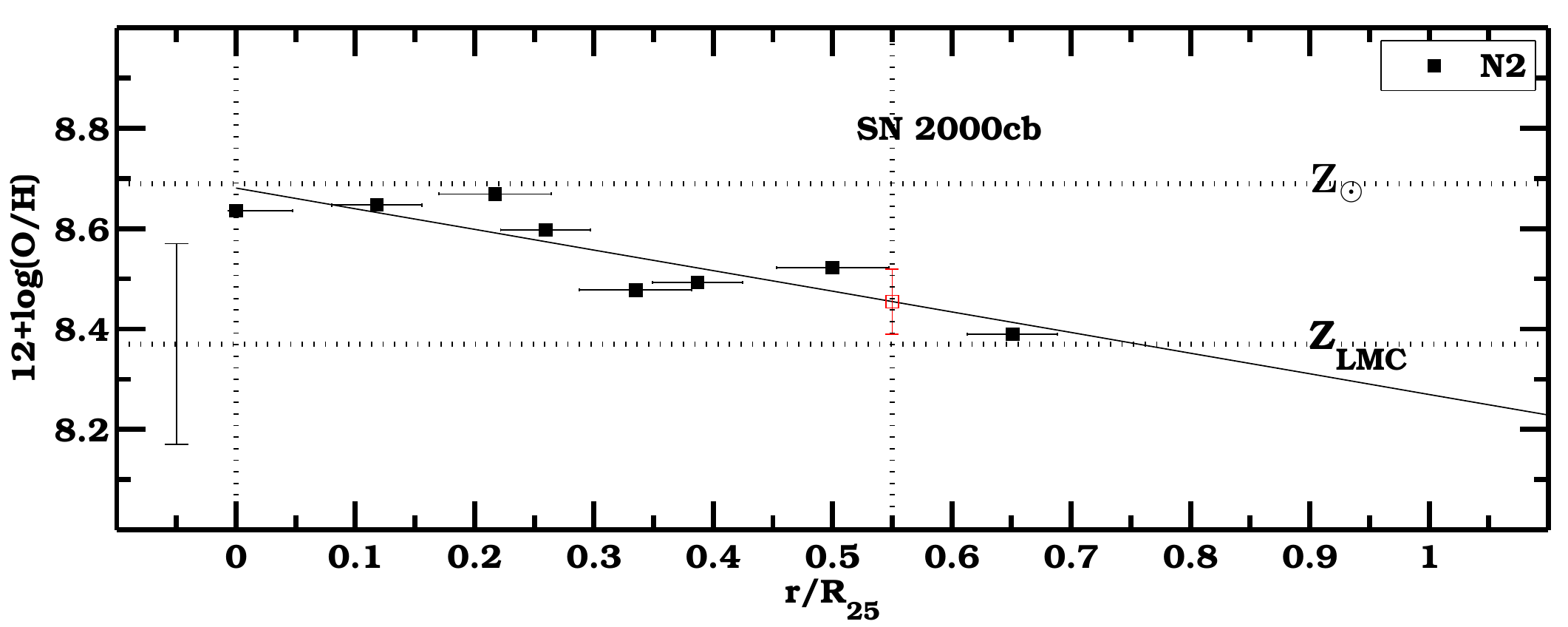}
  \caption{\textit{(Top-right panel)} Continuum-subtracted H$\alpha$ image of IC~1158. The 25th $B$-band magnitude elliptic contour is shown by a black solid line, along with the position of SN~2000cb (marked by a red star) and the center of the galaxy (marked by a red circle). The slit position is shown, and a color code is used to present the N2 metallicity measurements at the position of each \ion{H}{ii} region that we inspected. \textit{(Top-left panel)} Flux at the H$\alpha$ wavelength along the slit, shown as a function of the distance from the SN location (marked by a dotted line, like the nucleus position). The N2 measurements are shown at the corresponding positions in the top sub-panel. \textit{(Bottom panel)} Metallicity gradient IC~1158, see the bottom-panel caption of Fig.~\ref{sn98A} for details.\label{sn00cb}}
 \end{figure}}

\clearpage
\onlfig{8}{
\begin{figure}
 \centering
 $\begin{array}{cc}
\includegraphics[height=6cm,angle=0]{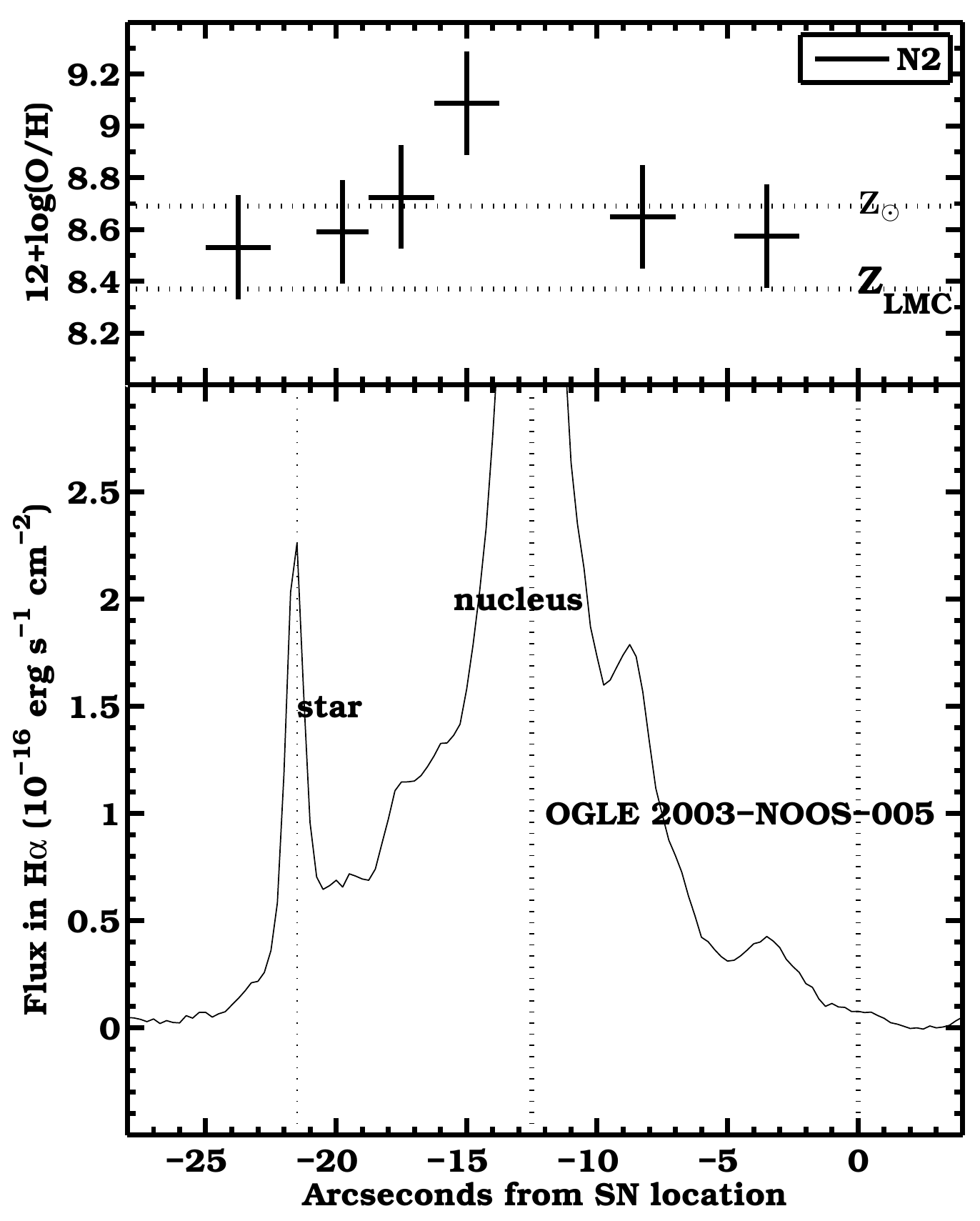}&
\includegraphics[height=9.8cm,angle=90]{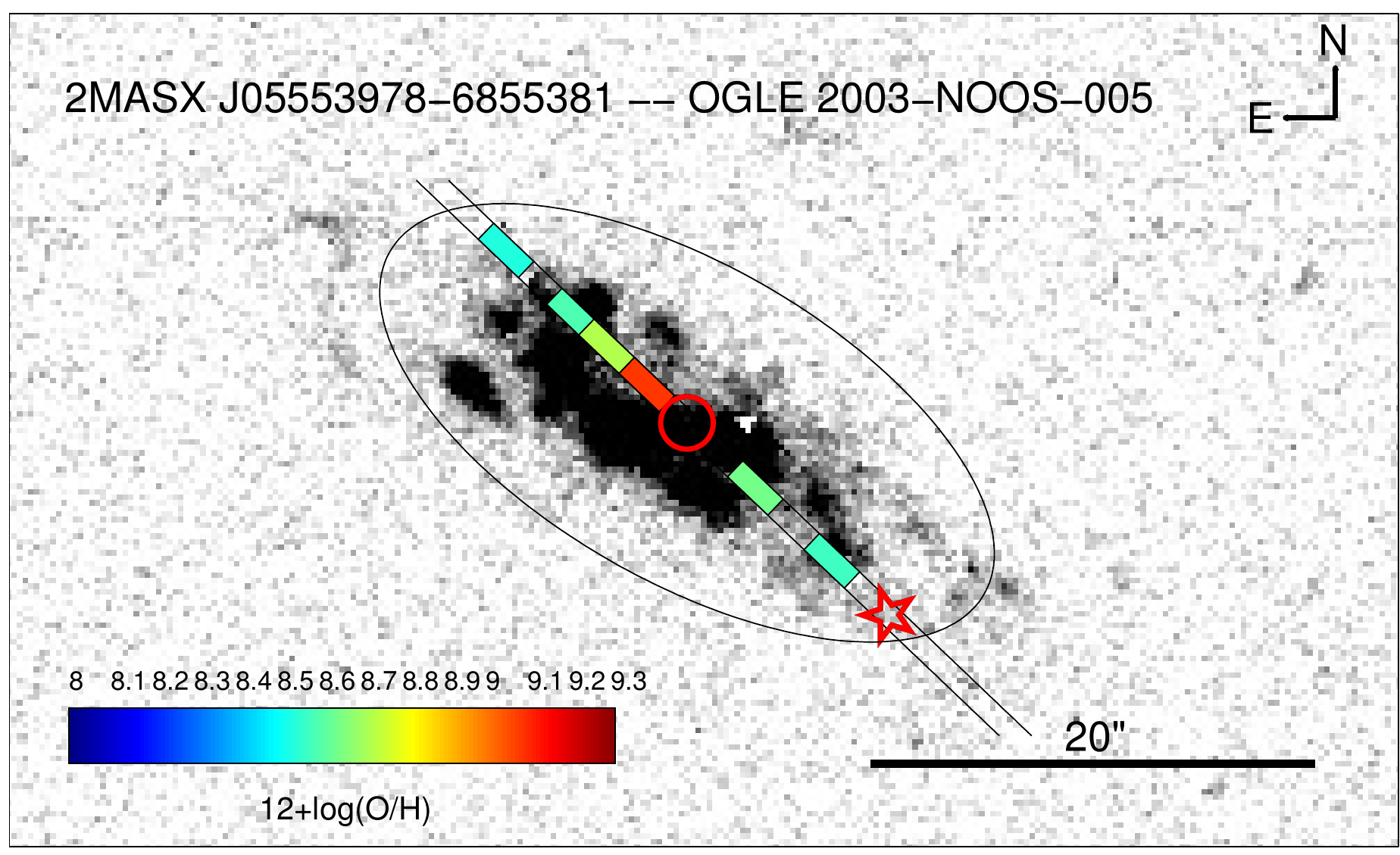}
\end{array}$
\includegraphics[width=15.1cm,angle=0]{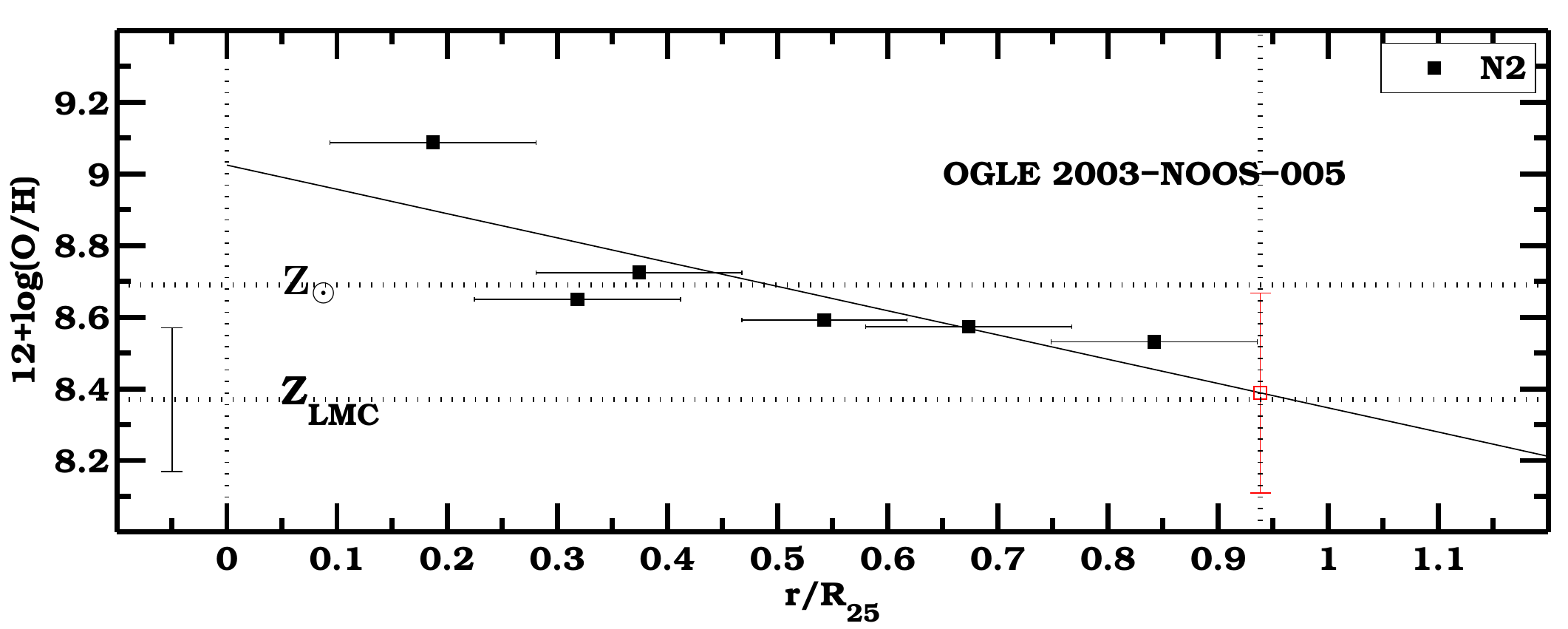}
 \caption{\textit{(Top-right panel)} Continuum-subtracted H$\alpha$ image of 2MASX J05553978-6855381, see the top-right panel caption of Fig.~\ref{sn00cb} for details. \textit{(Top-left panel)} Flux at the H$\alpha$ wavelength along the slit, see the top-left panel caption of Fig.~\ref{sn00cb} for details. \textit{(Bottom panel)} Metallicity gradient of 2MASX J05553978-6855381, see the bottom-panel caption of Fig.~\ref{sn98A} for details.\label{ogle}}
 \end{figure}}

\clearpage
\onlfig{9}{
\begin{figure}
 \centering$
\begin{array}{cc}
\includegraphics[width=7cm,angle=0]{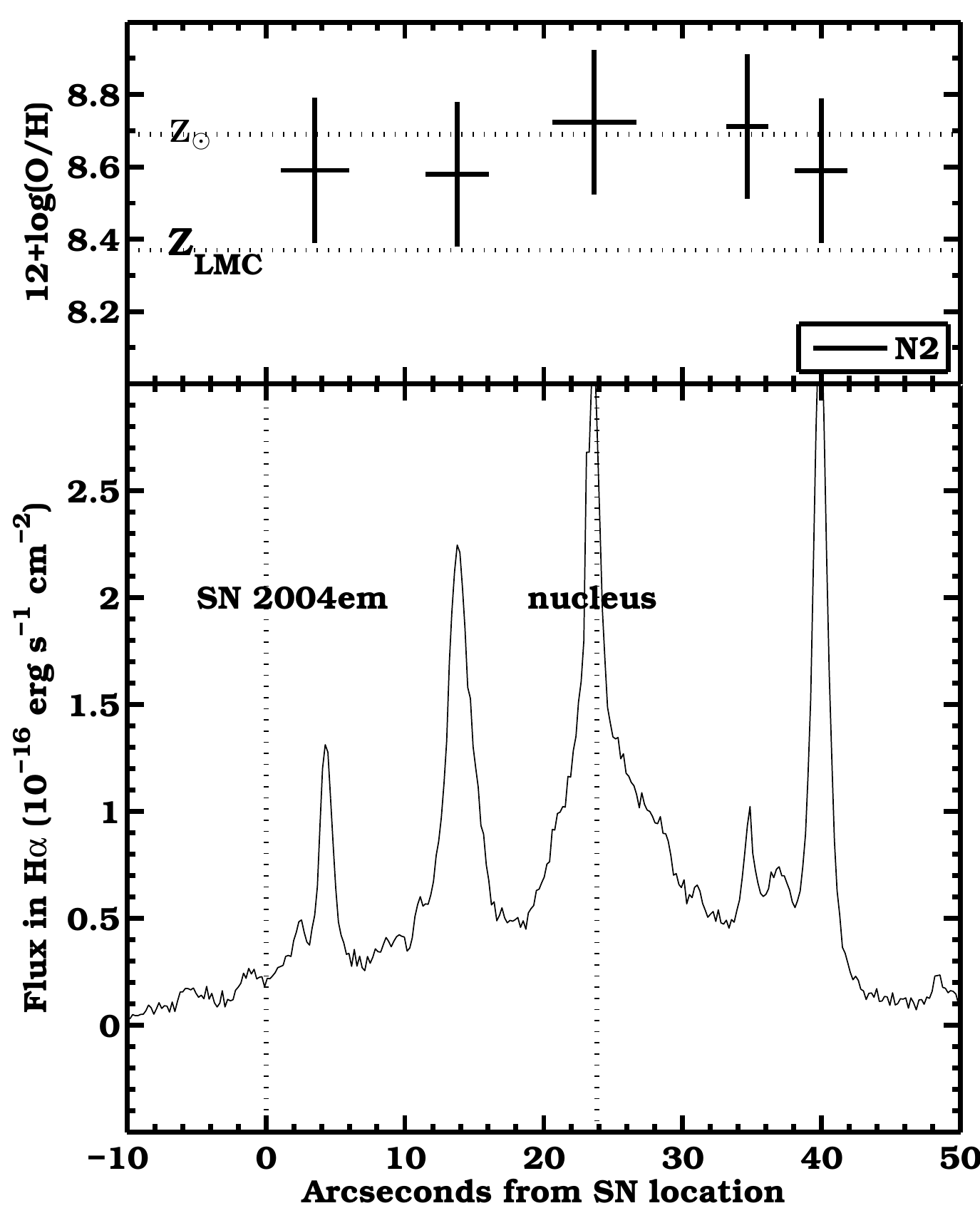}&
\includegraphics[width=8.6cm,angle=90]{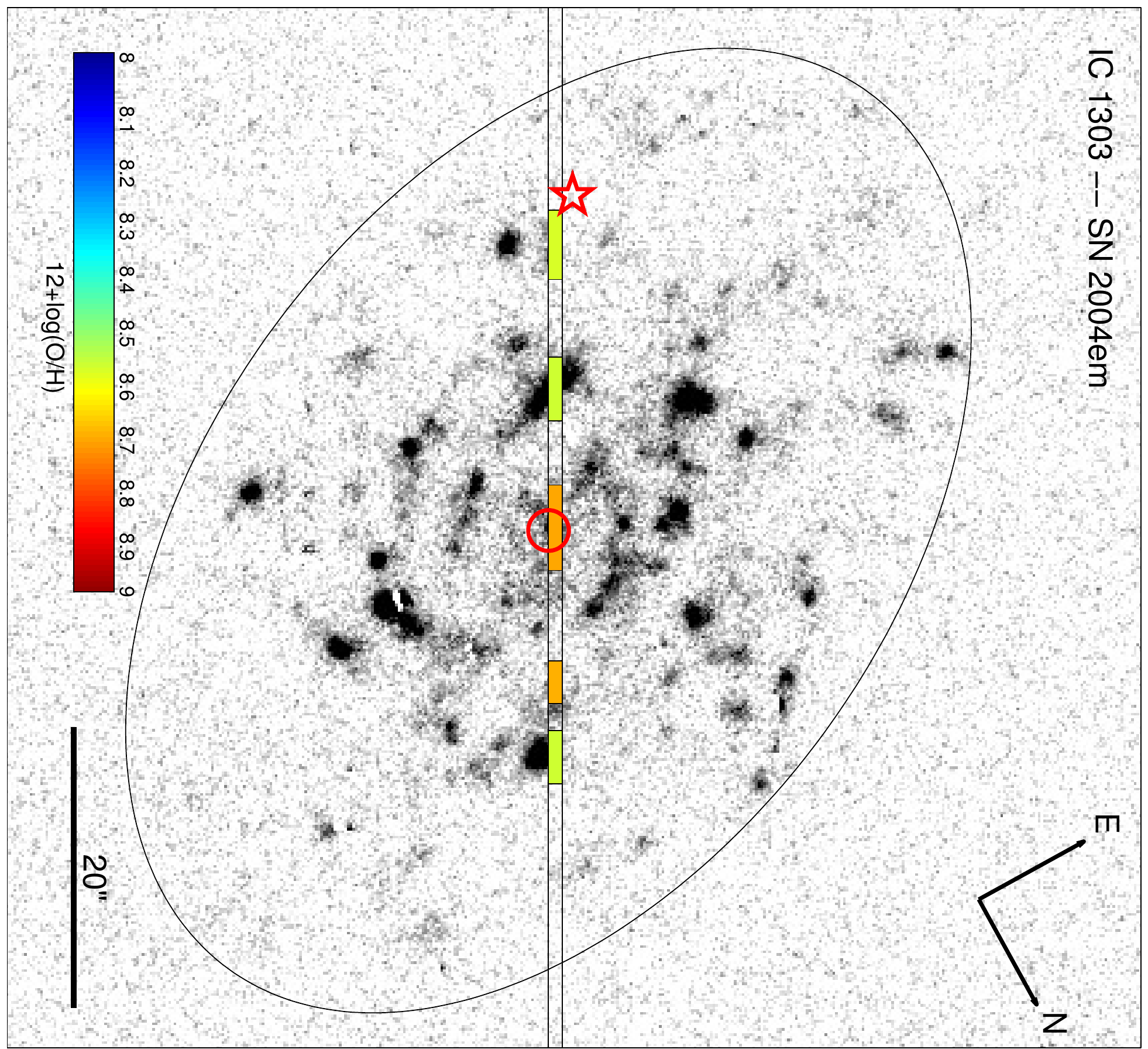} \\
\end{array}$
\includegraphics[width=15.1cm,angle=0]{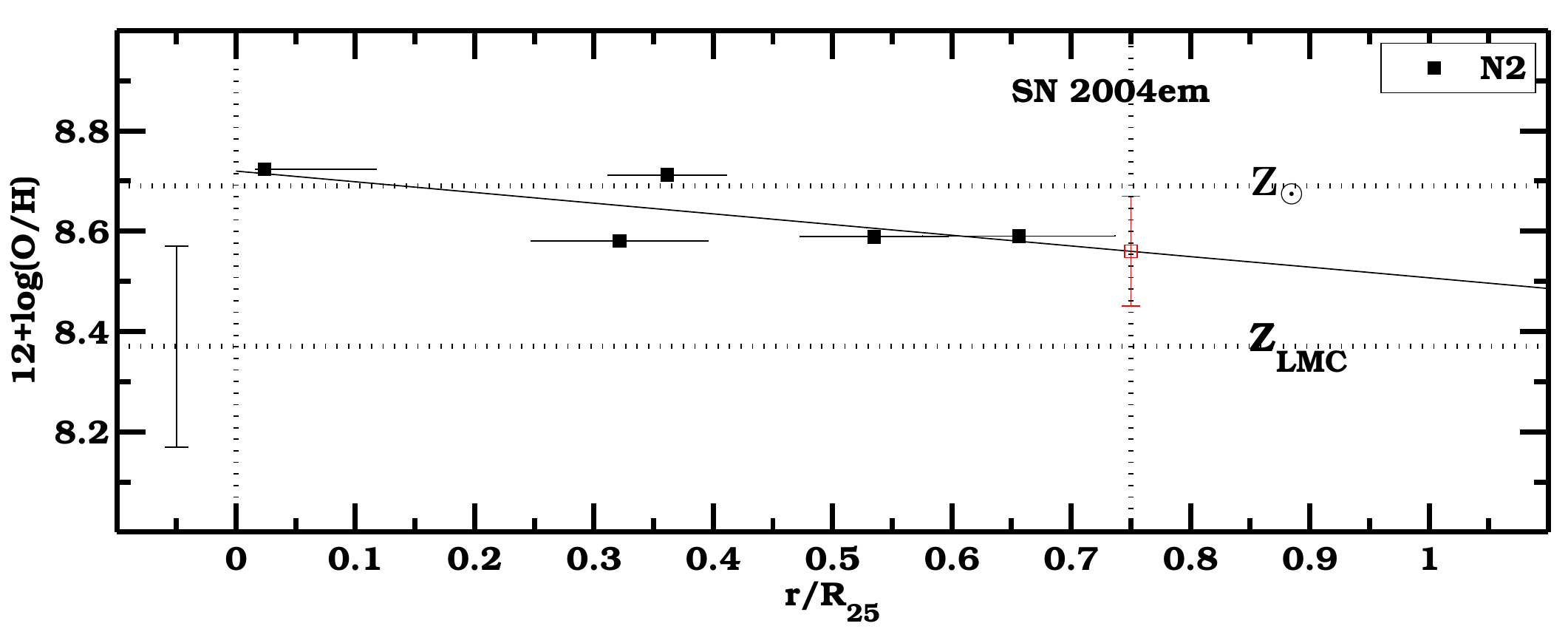}
 \caption{\textit{(Top-right panel)} Continuum-subtracted H$\alpha$ image of IC~1303, see the top-right panel caption of Fig.~\ref{sn00cb} for details. \textit{(Top-left panel)} Flux at the H$\alpha$ wavelength along the slit, see the top-left panel caption of Fig.~\ref{sn00cb} for details. \textit{(Bottom panel)} Metallicity gradient of IC~1303, see the bottom-panel caption of Fig.~\ref{sn98A} for details.\label{sn04em}}
 \end{figure}}

\onlfig{10}{
\begin{figure}
 \centering
 \includegraphics[width=11cm,angle=90]{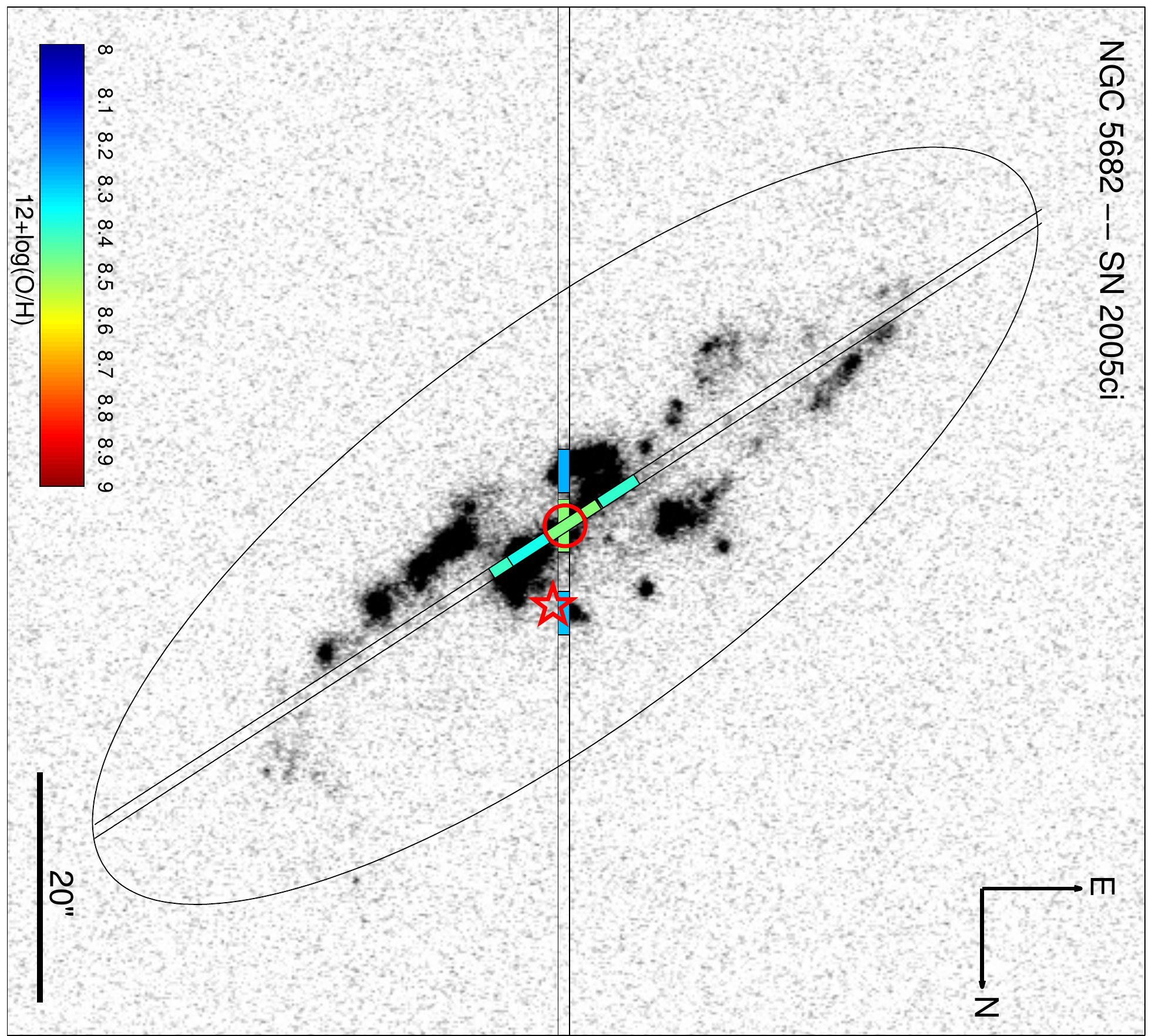} 
 $\begin{array}{cc}
\includegraphics[width=5cm,angle=0]{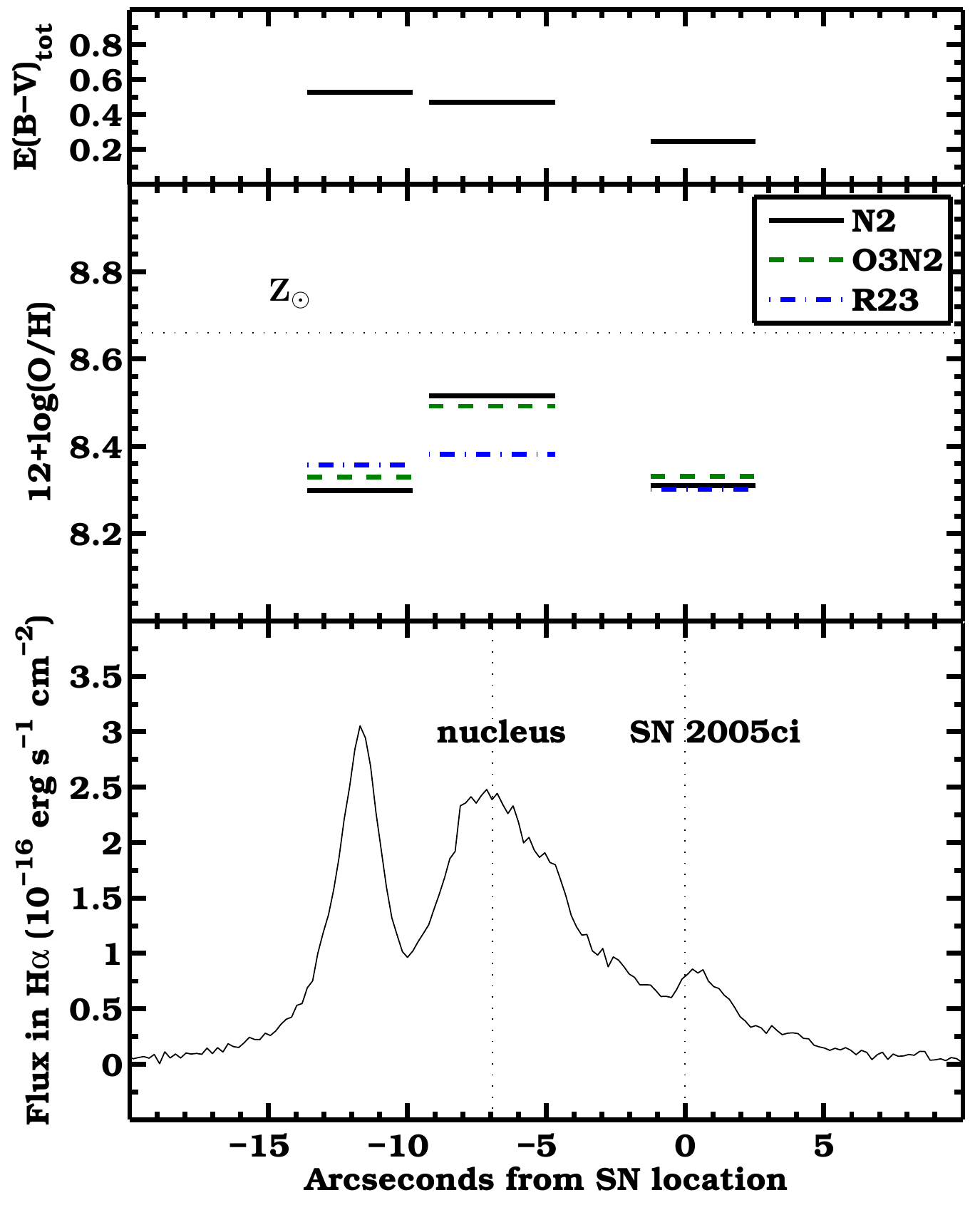}&
\includegraphics[width=4.95cm,angle=0]{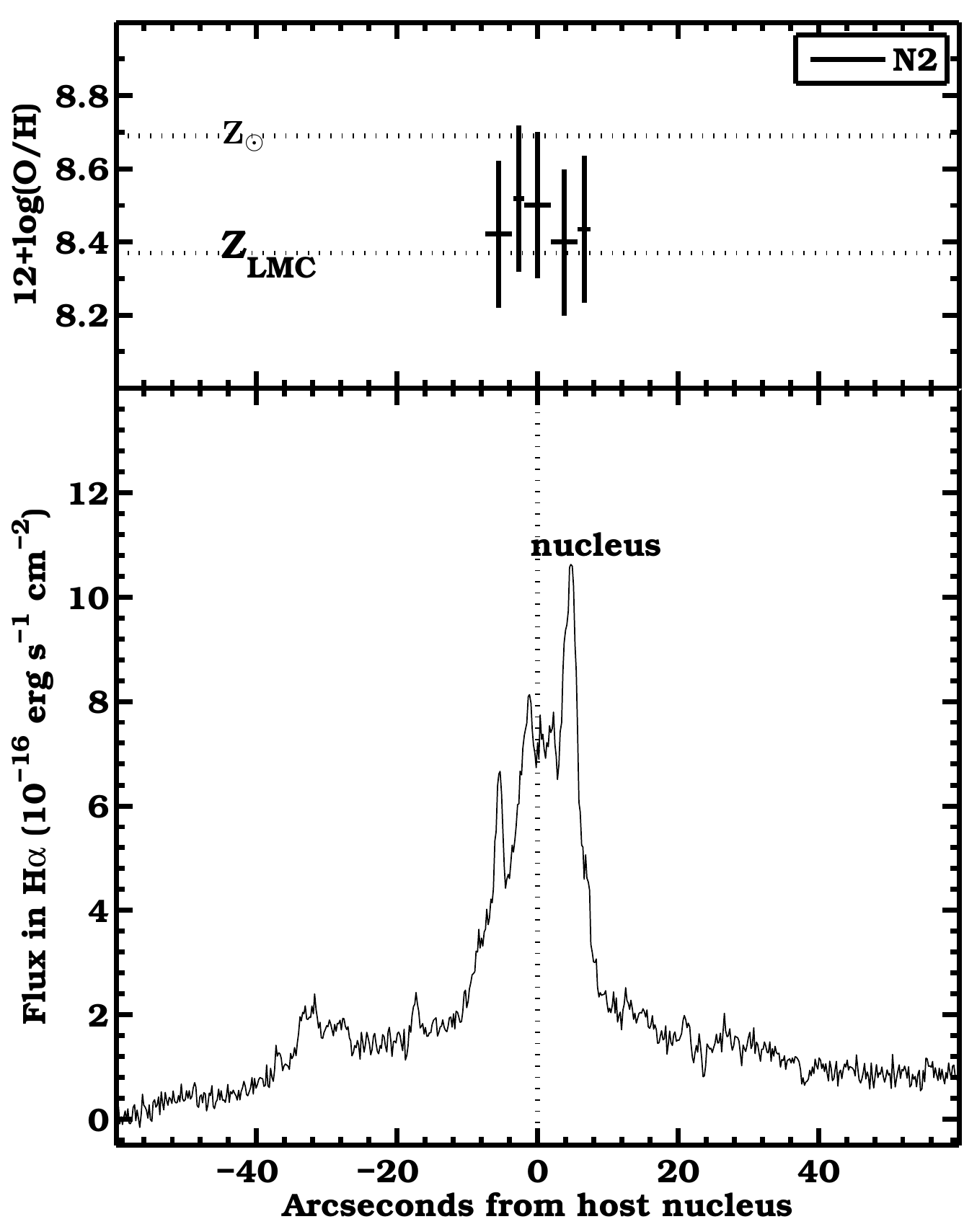}\\
\end{array}$
\includegraphics[width=10cm,angle=0]{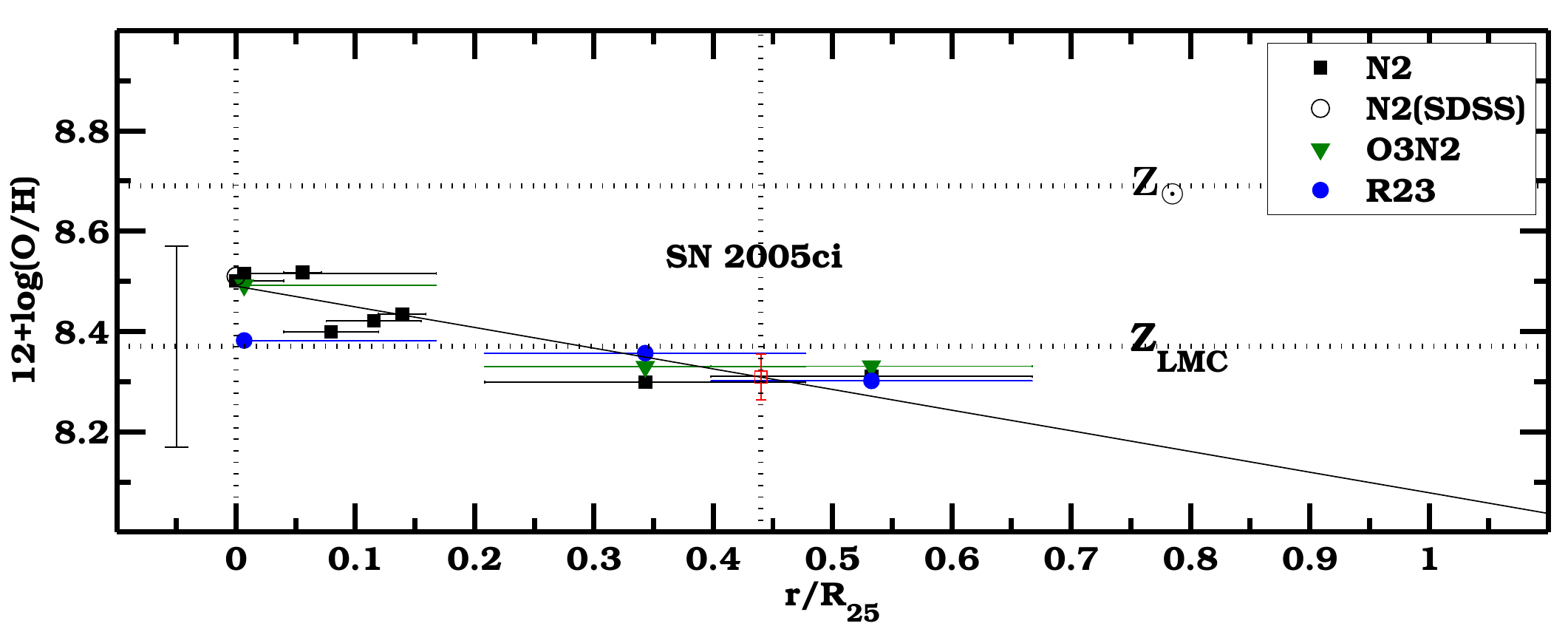}
 \caption{\textit{(Top panel)} Continuum-subtracted H$\alpha$ image of NGC~5682, see the top-right panel caption of Fig.~\ref{sn00cb} for details. Here we show the positions of two slits. \textit{(Center panels)} Flux at the H$\alpha$ wavelength along the two slits, shown as a function of the distance from the SN location and from the nucleus. These positions are marked by vertical dotted lines. The N2 measurements for each \ion{H}{ii} region are shown at the corresponding positions in the top sub-panels. For the slit at the SN position we also show the O3N2 and R23 metallicity estimates, and the $E(B-V)$ values that we computed through the measured Balmer decrements. \textit{(Bottom panel)} Metallicity gradient of NGC~5682, see the bottom-panel caption of Fig.~\ref{sn98A} for details.
  We also reported the N2 metallicity from the spectrum taken by SDSS at the center of the galaxy. \label{sn05ci}}
 \end{figure}}

\onlfig{11}{
\begin{figure}
 \centering
 \includegraphics[width=7.9cm,angle=90]{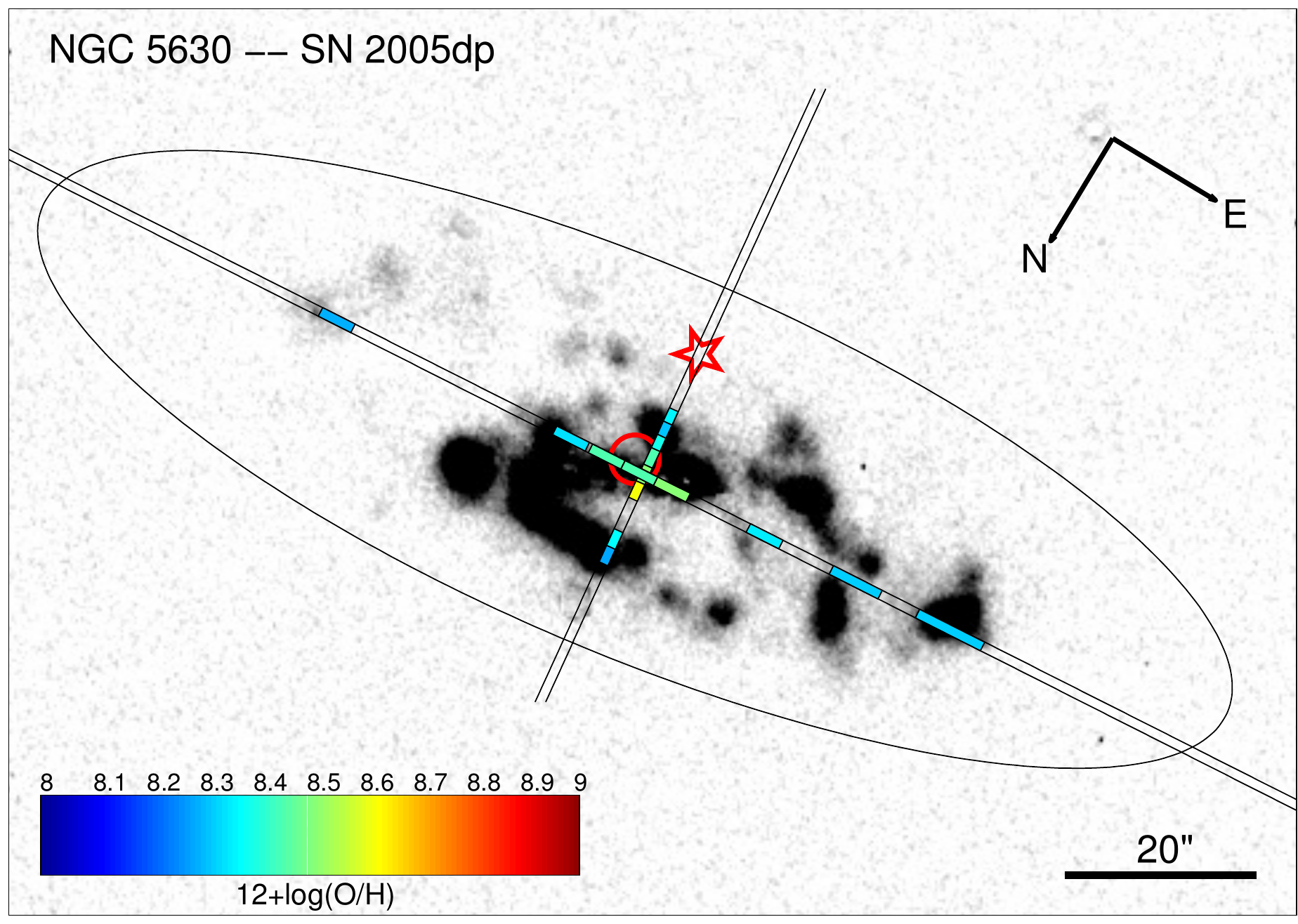}
$\begin{array}{cc}
 \includegraphics[width=5.55cm,angle=0]{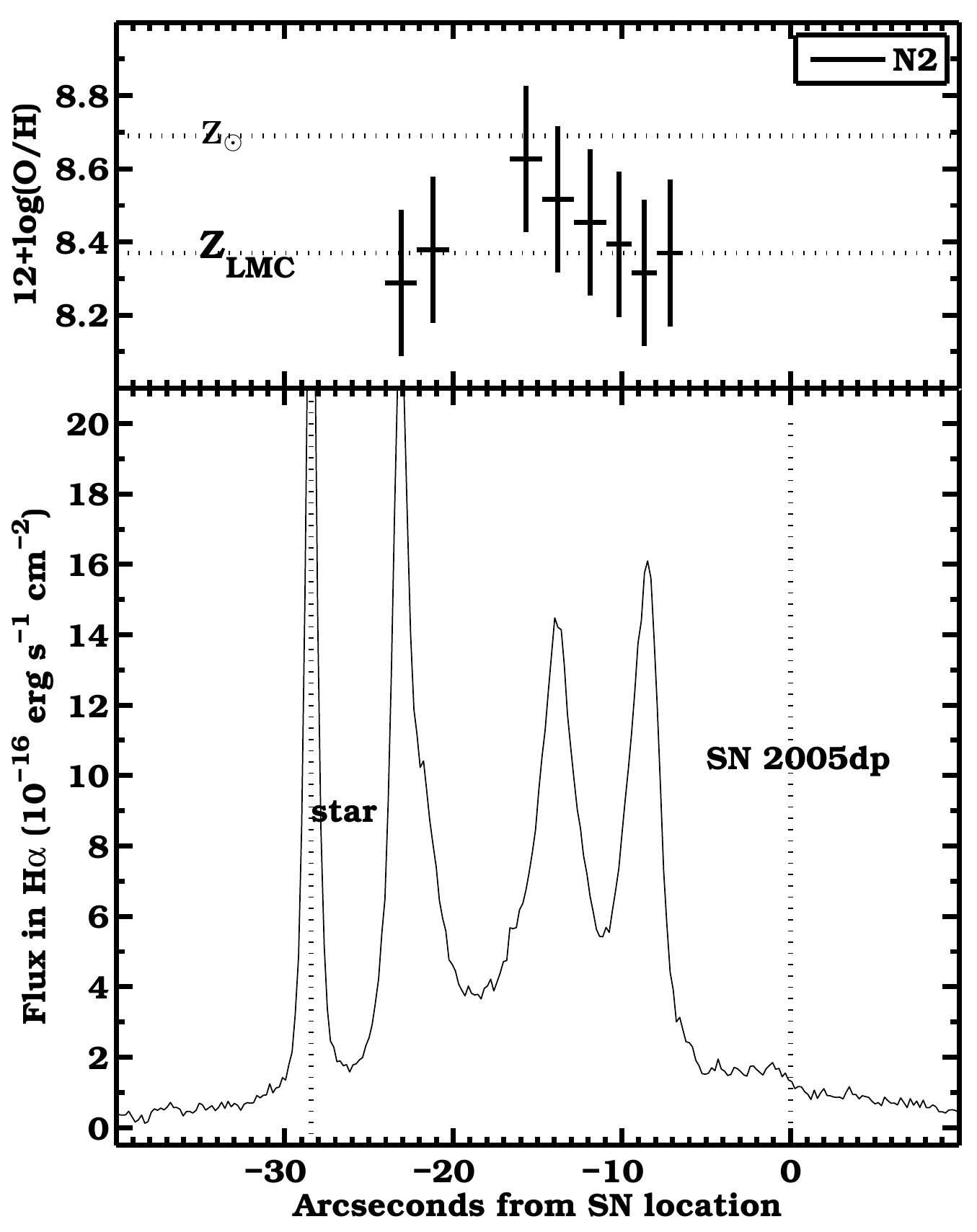}&
\includegraphics[width=5.55cm,angle=0]{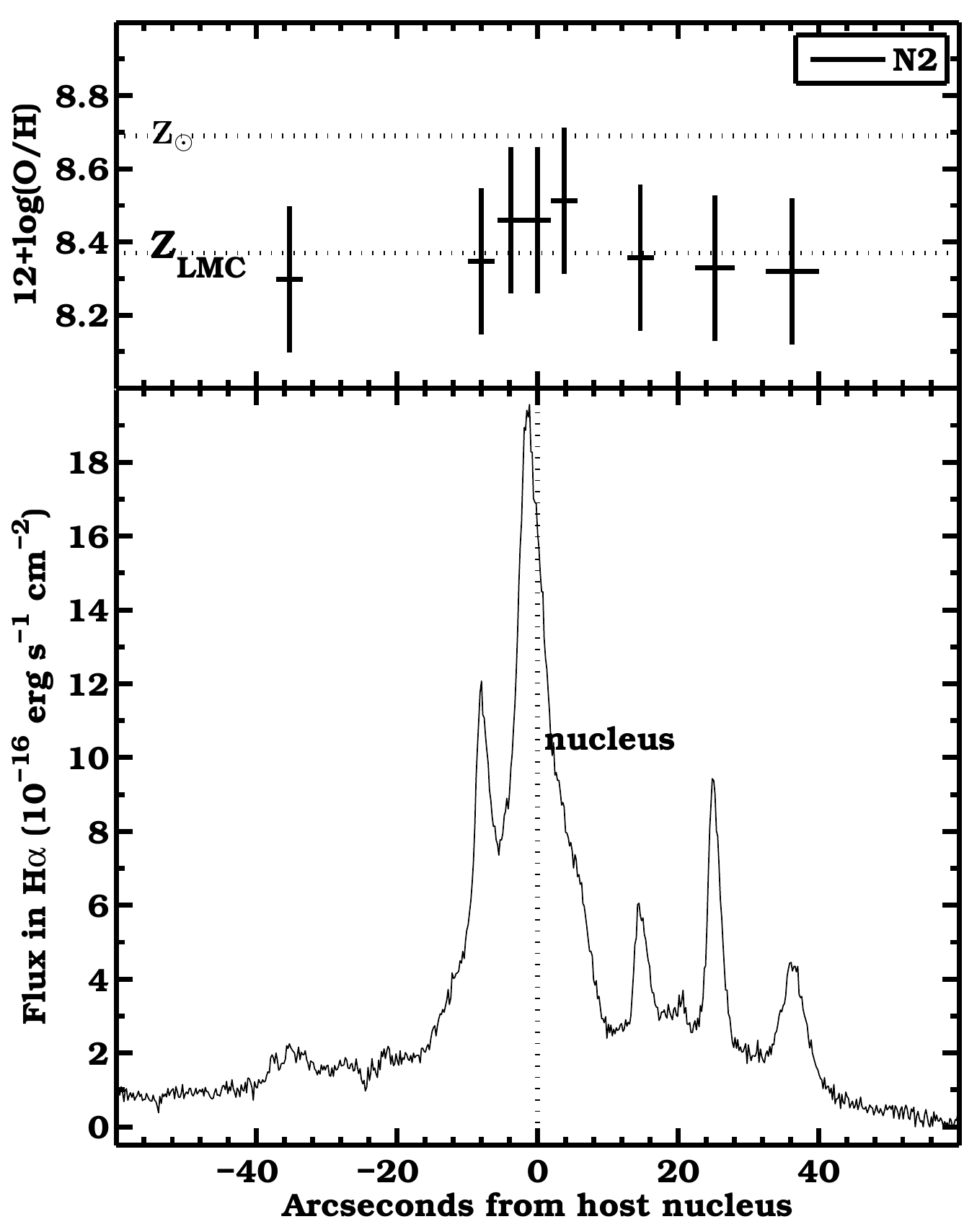} \\
\end{array}$
\includegraphics[width=11.5cm,angle=0]{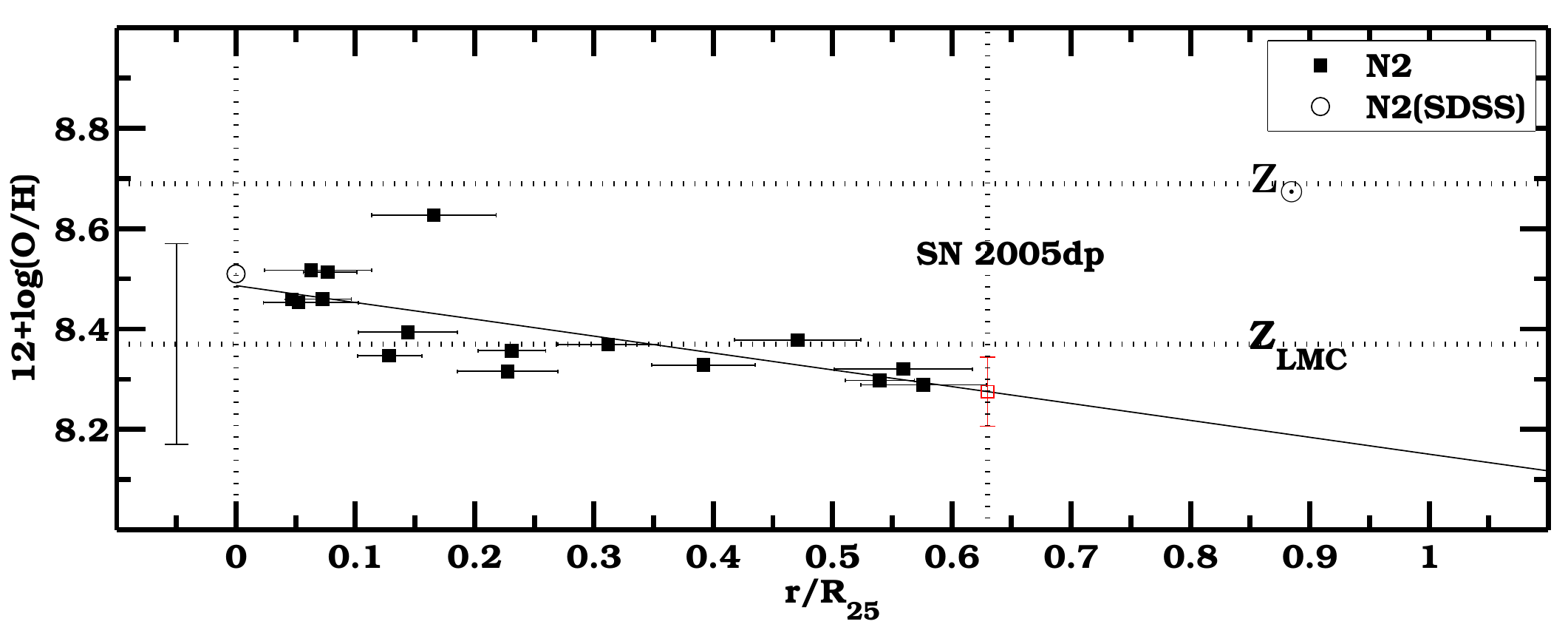}
 \caption{\textit{(Top panel)} Continuum-subtracted H$\alpha$ image of NGC~5630, see the top-right panel caption of Fig.~\ref{sn00cb} for details. Here we show the positions of two slits. \textit{(Center panels)} Flux at the H$\alpha$ wavelength along the two slits, shown as a function of the distance from the SN location and from the nucleus. See the top-left panel caption of Fig.~\ref{sn00cb} for details. \textit{(Bottom panel)} Metallicity gradient of NGC~5630, see the bottom-panel caption of Fig.~\ref{sn05ci} for details.\label{sn05dp}}
\end{figure}}

\onlfig{12}{
\begin{figure}
 \centering
\includegraphics[height=10cm,angle=90]{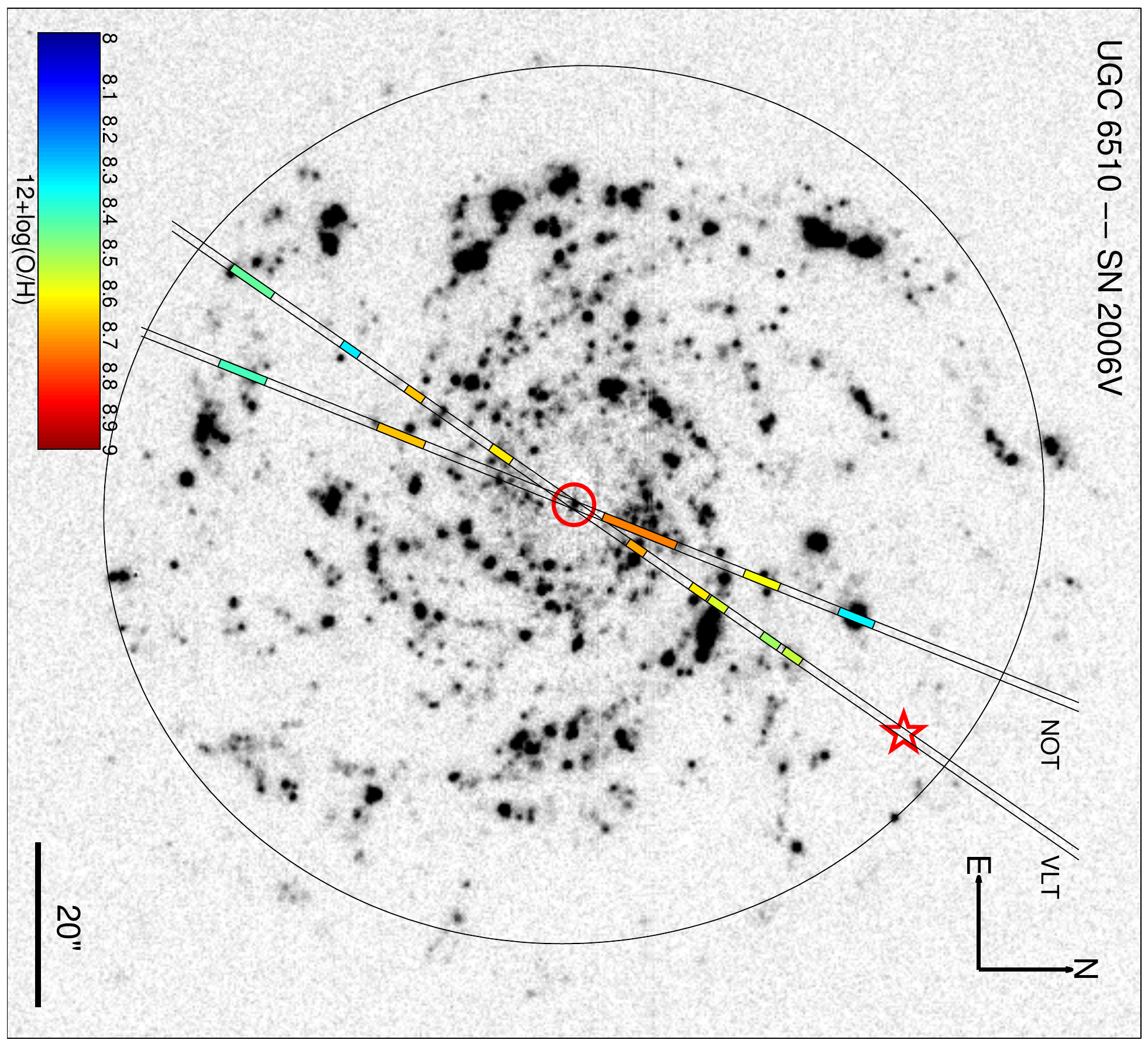} \\
 $\begin{array}{cc}
\includegraphics[width=5.cm,angle=0]{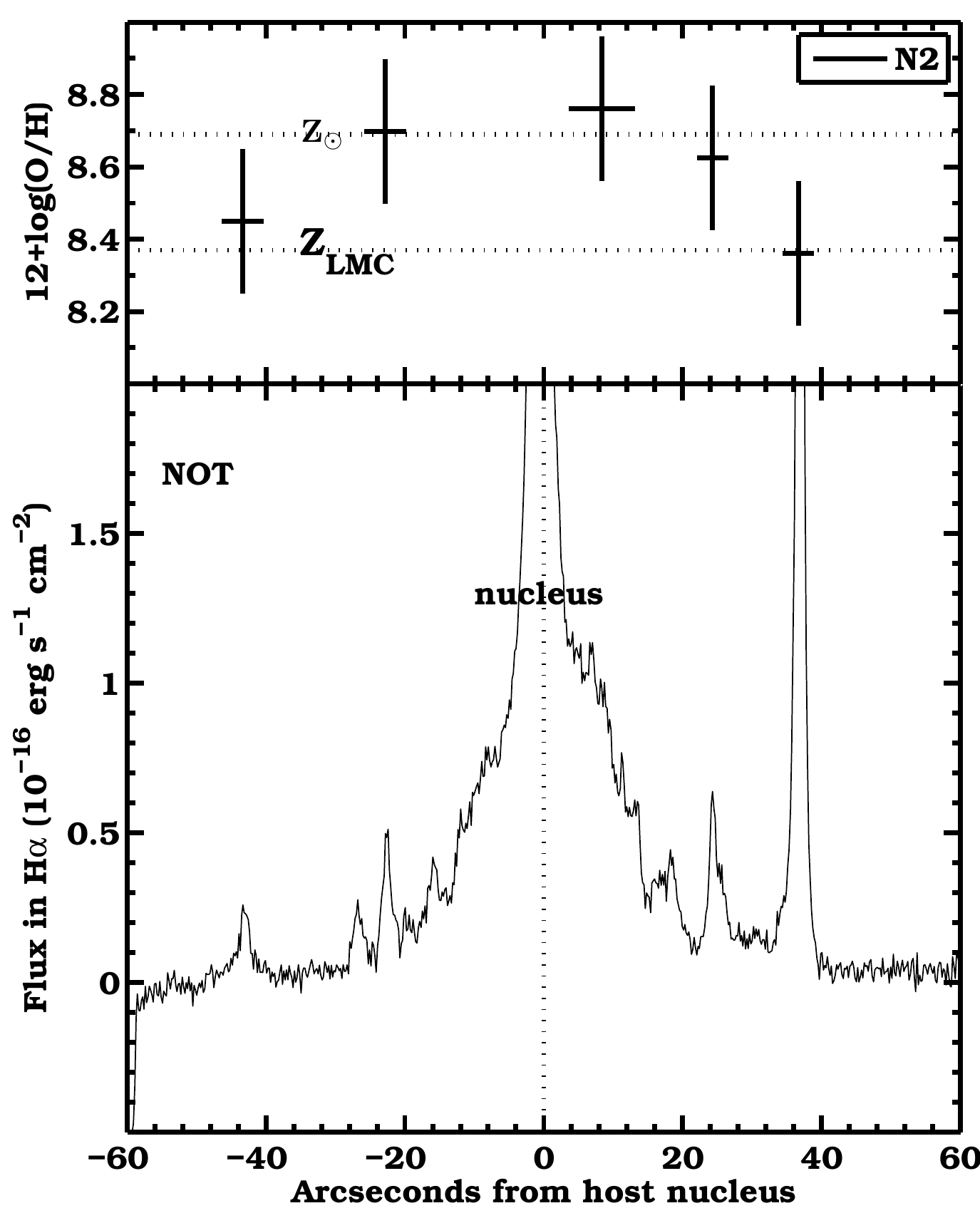}&
\includegraphics[width=5.cm,angle=0]{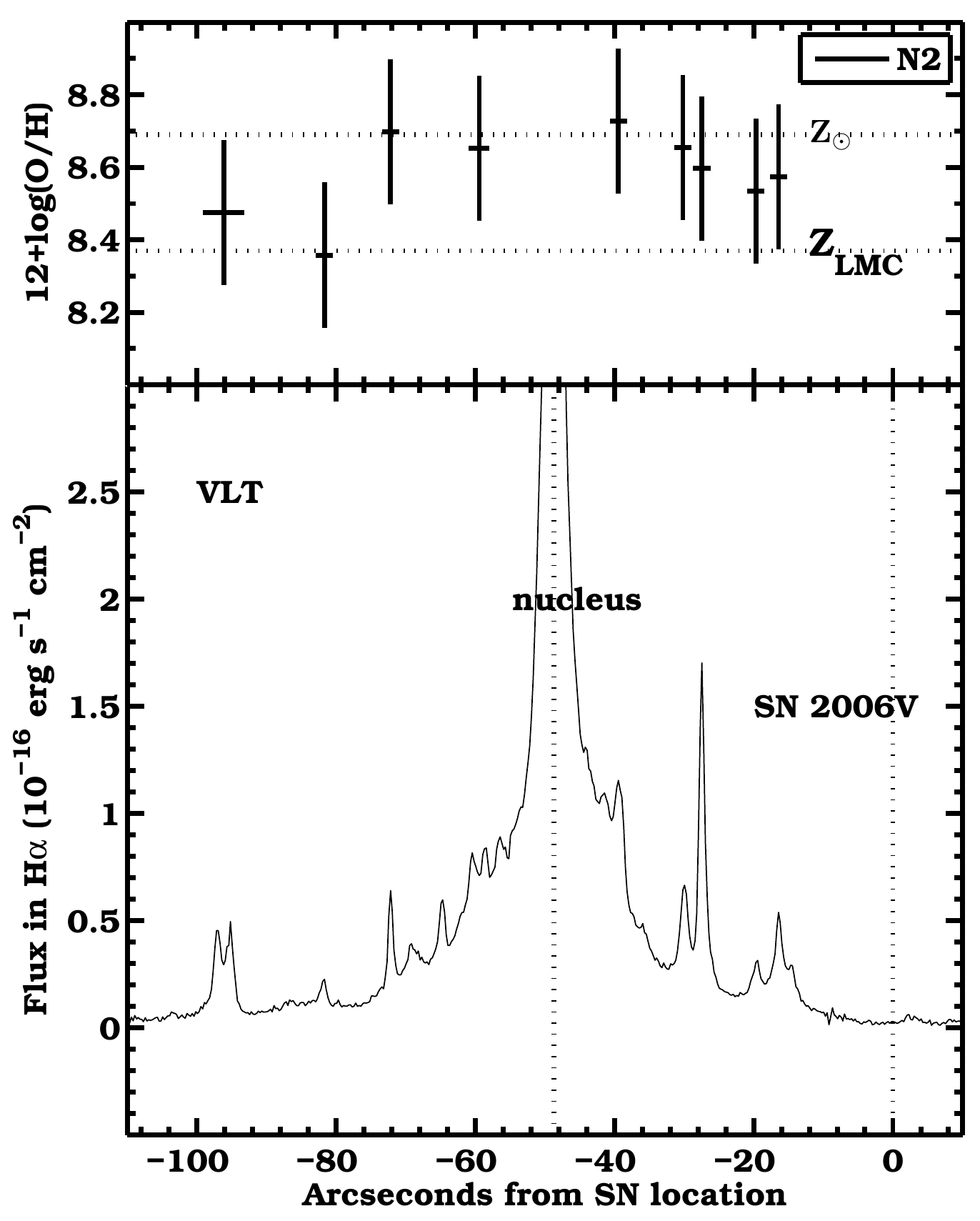}
\end{array}$
\includegraphics[width=10cm,angle=0]{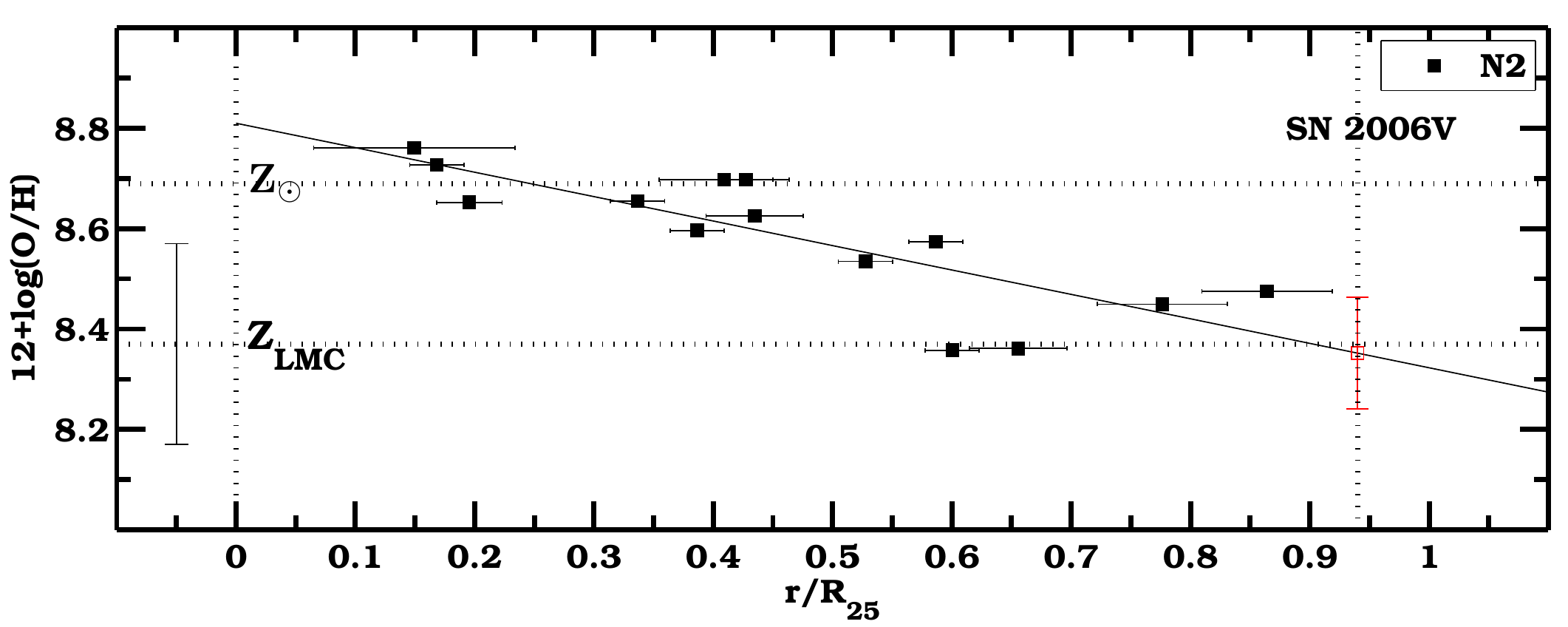}
 \caption{\textit{(Top panel)} Continuum-subtracted H$\alpha$ image of UGC~6510, see the top-right panel caption of Fig.~\ref{sn00cb} for details. Here we show the positions of two slits, one for the NOT and one for the VLT spectrum. \textit{(Center panels)} Flux at the H$\alpha$ wavelength along the two slits, shown as a function of the distance from the SN location and from the nucleus. See the top-left panel caption of Fig.~\ref{sn00cb} for details. \textit{(Bottom panel)} Metallicity gradient of UGC~6510, see the bottom-panel caption of Fig.~\ref{sn98A} for details.\label{sn06V}}
 \end{figure}}

\onlfig{13}{
\begin{figure}
 \centering$
\begin{array}{cc}
\includegraphics[width=5.7cm,angle=0]{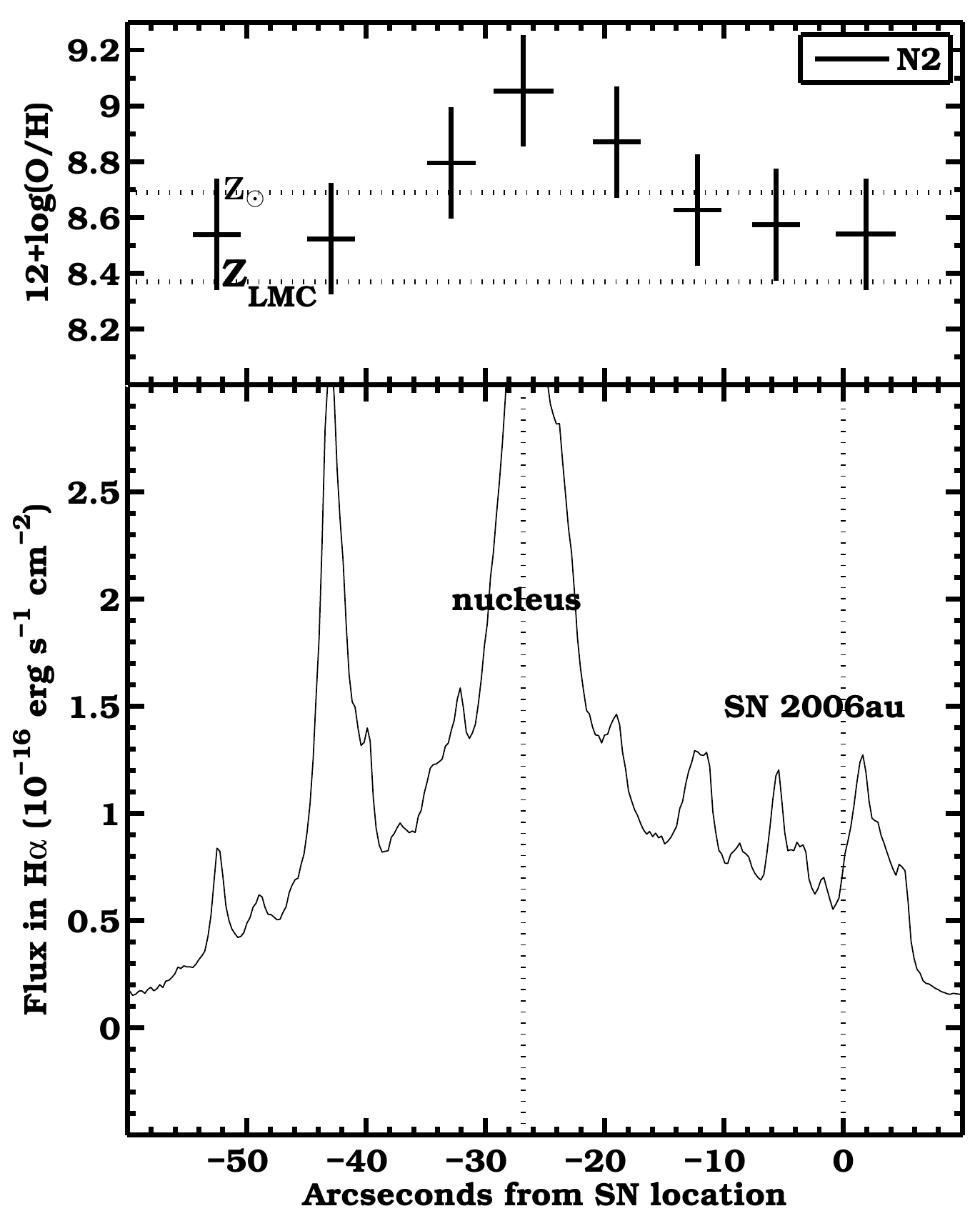}&
\includegraphics[height=12cm,angle=90]{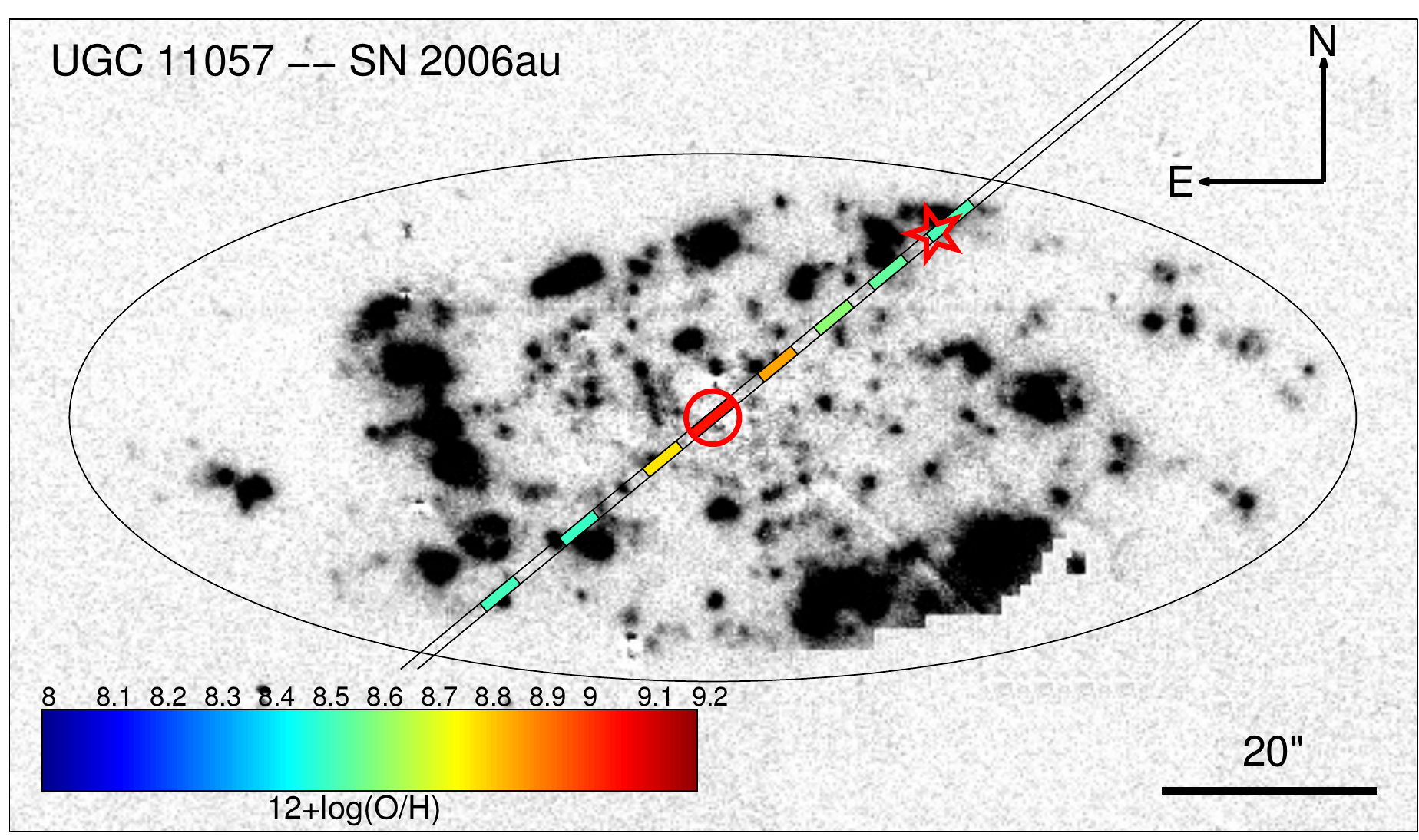}\\
\end{array}$
\includegraphics[width=18.0cm,angle=0]{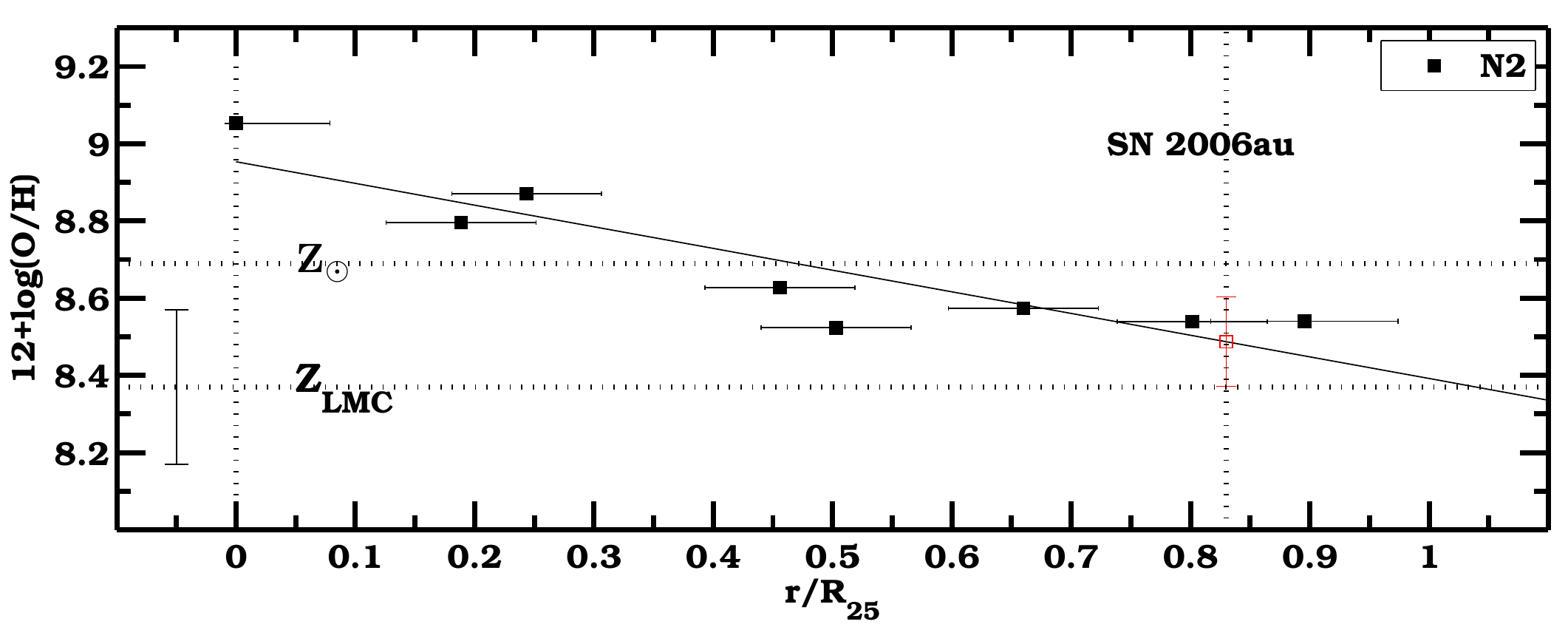} 
 \caption{\textit{(Top-right panel)} Continuum-subtracted H$\alpha$ image of UGC~11057, see the top-right panel caption of Fig.~\ref{sn00cb} for details. \textit{(Top-left panel)} Flux at the H$\alpha$ wavelength along the slit, see the top-left panel caption of Fig.~\ref{sn00cb} for details. \textit{(Bottom panel)} Metallicity gradient of UGC~11057, see the bottom-panel caption of Fig.~\ref{sn00cb} for details.\label{sn06au}}
 \end{figure}}

\begin{figure}
 \centering$
\begin{array}{cc}
\includegraphics[width=7cm,angle=0]{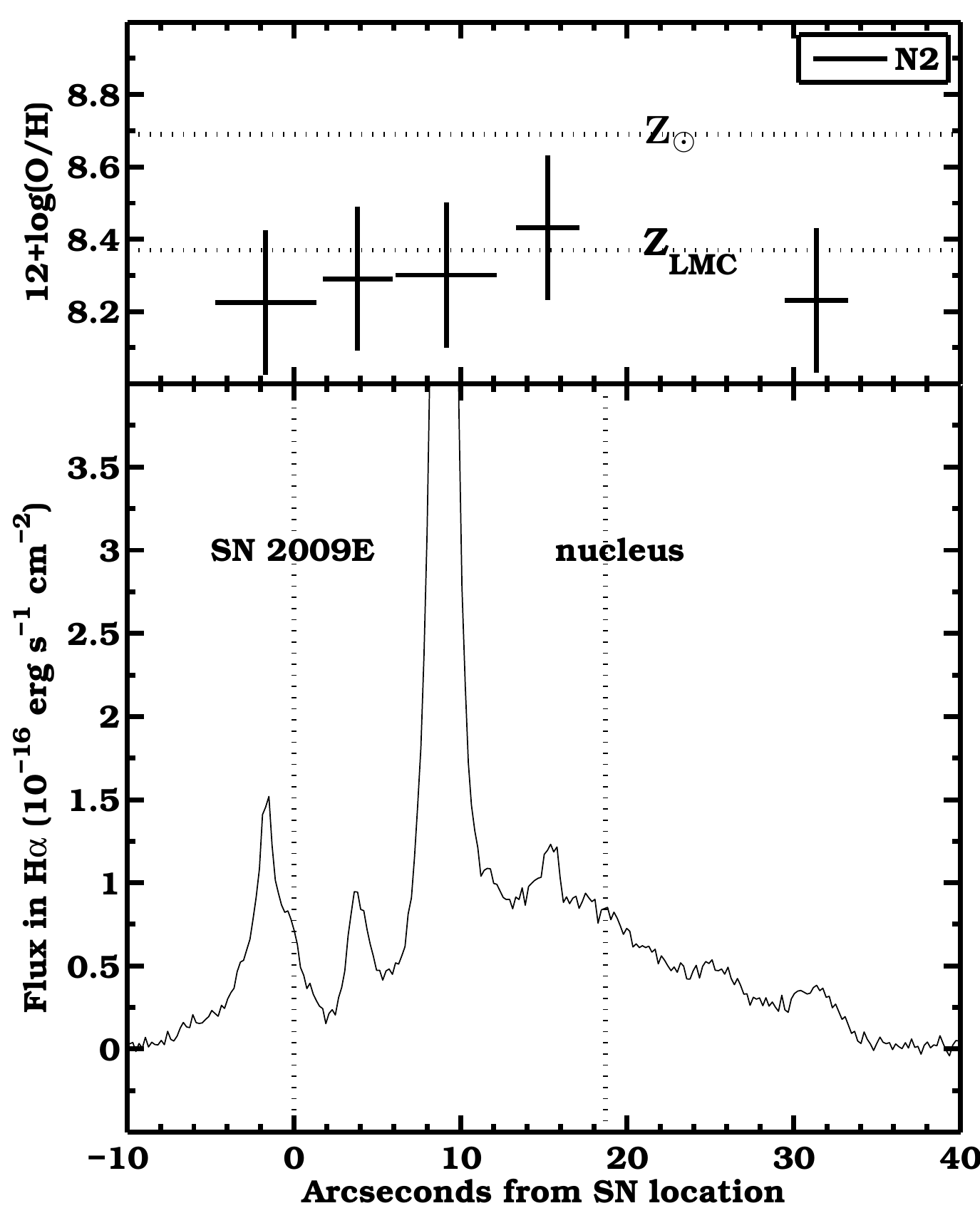}&
\includegraphics[width=8.6cm,angle=90]{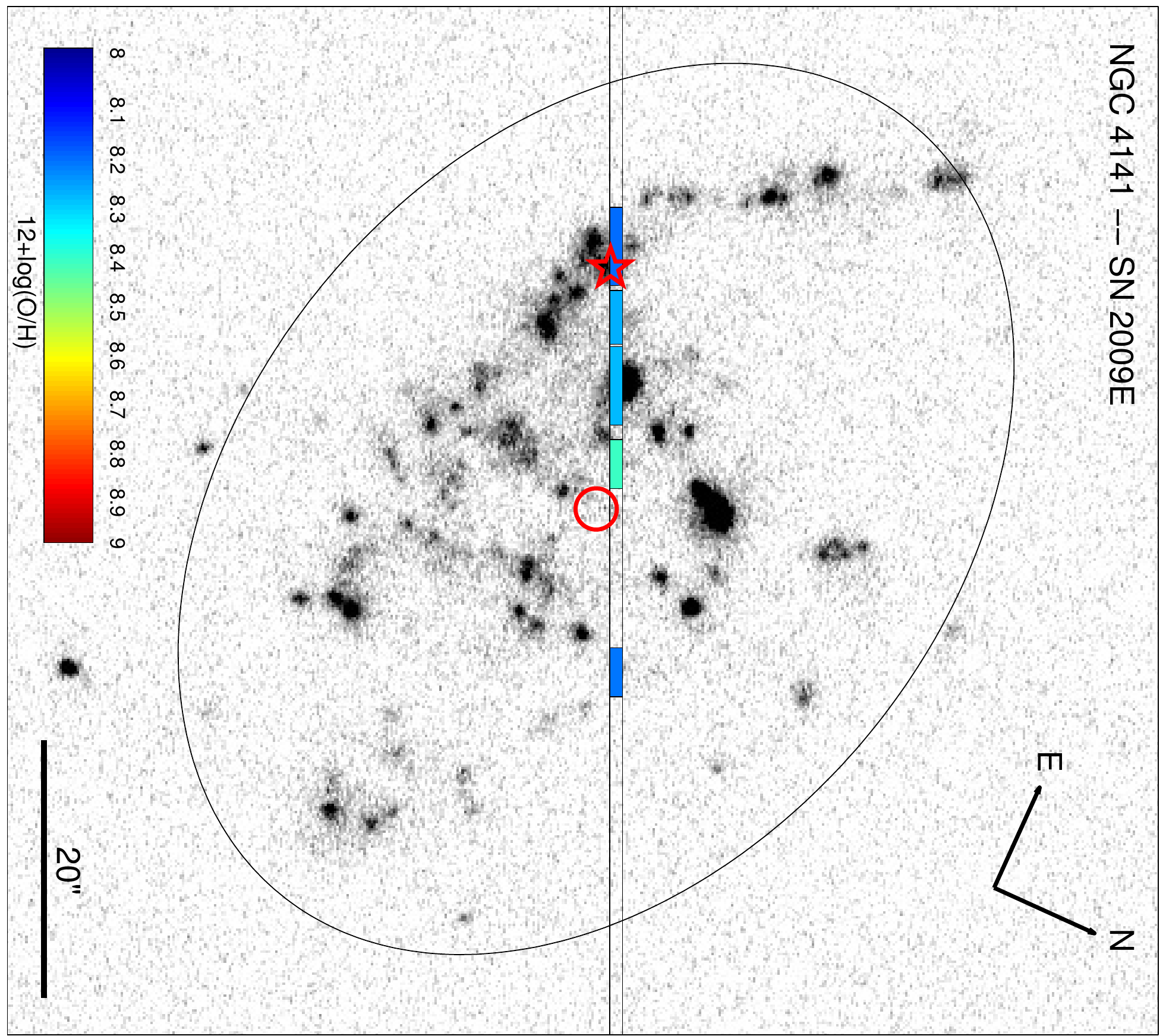} \\
\end{array}$
\includegraphics[width=15.1cm,angle=0]{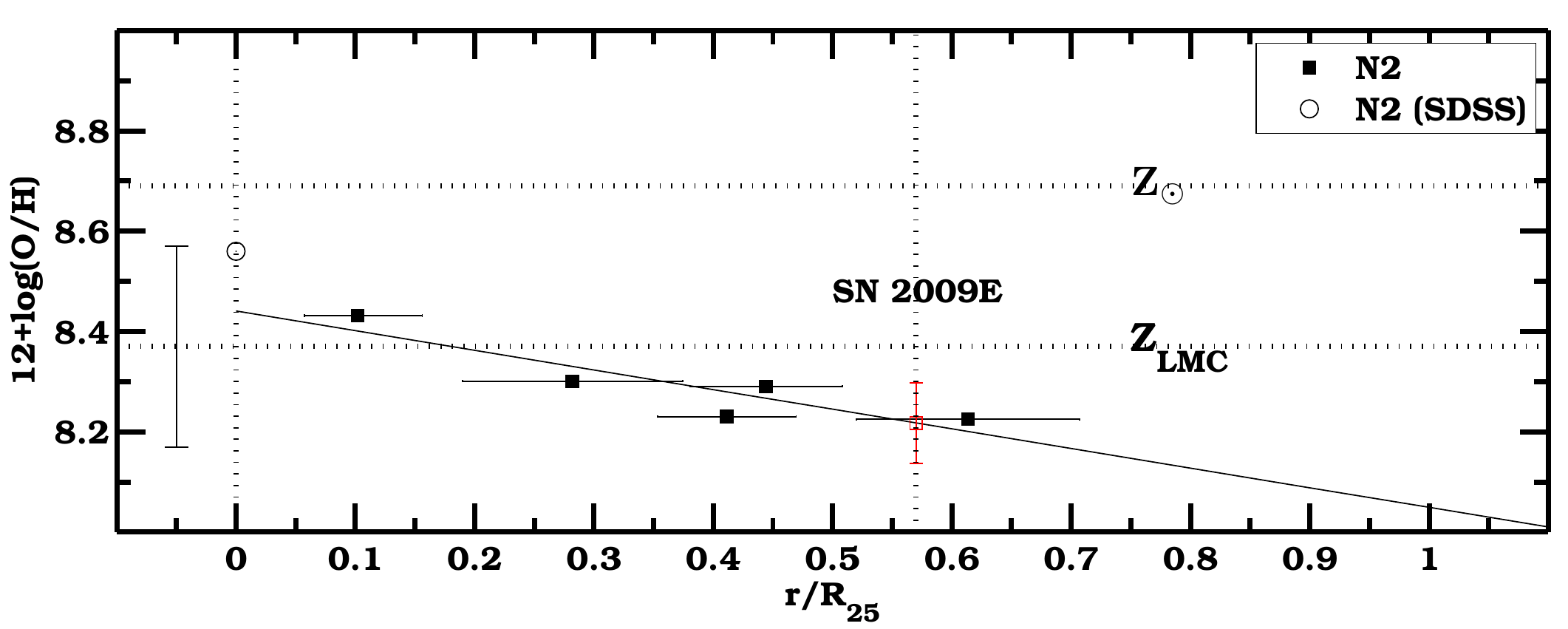}
  \caption{\textit{(Top-right panel)} Continuum-subtracted H$\alpha$ image of NGC~4141, see the top-right panel caption of Fig.~\ref{sn00cb} for details. \textit{(Top-left panel)} Flux at the H$\alpha$ wavelength along the slit, see the top-left panel caption of Fig.~\ref{sn00cb} for details. \textit{(Bottom panel)} Metallicity gradient of NGC~4141, see the bottom-panel caption of Fig.~\ref{sn05ci} for details.\label{sn09E}}
 \end{figure}

\clearpage
\begin{figure}
 \centering
\includegraphics[width=7cm,angle=0]{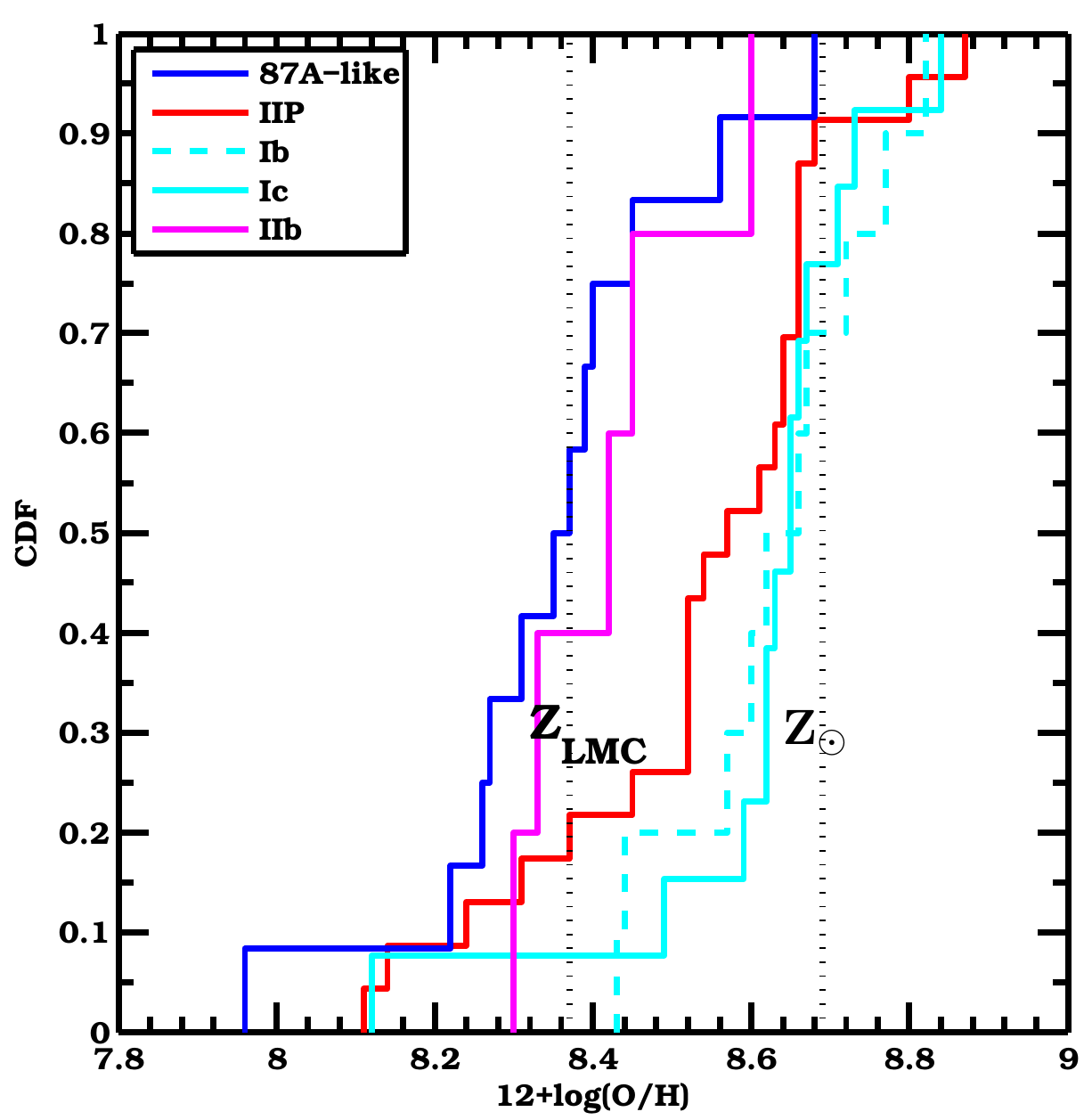}
  \caption{Cumulative distribution of the directly measured (N2 method) metallicity of our BSG~SNe
  and of samples of SNe~IIP, Ib, Ic from \citet{anderson10} and of SNe~IIb from \citet{sanders12}. BSG~SNe tend to explode at LMC metallicity (12$+$log(O/H)~$=$~8.3-8.4~dex). However, a few objects are found at almost solar metallicity (12$+$log(O/H)~$>$~8.5~dex). BSG~SNe clearly have lower metallicity than SNe~IIP.\label{cdfmetal}}
 \end{figure}

\begin{figure}
 \centering
\includegraphics[width=7cm,angle=0]{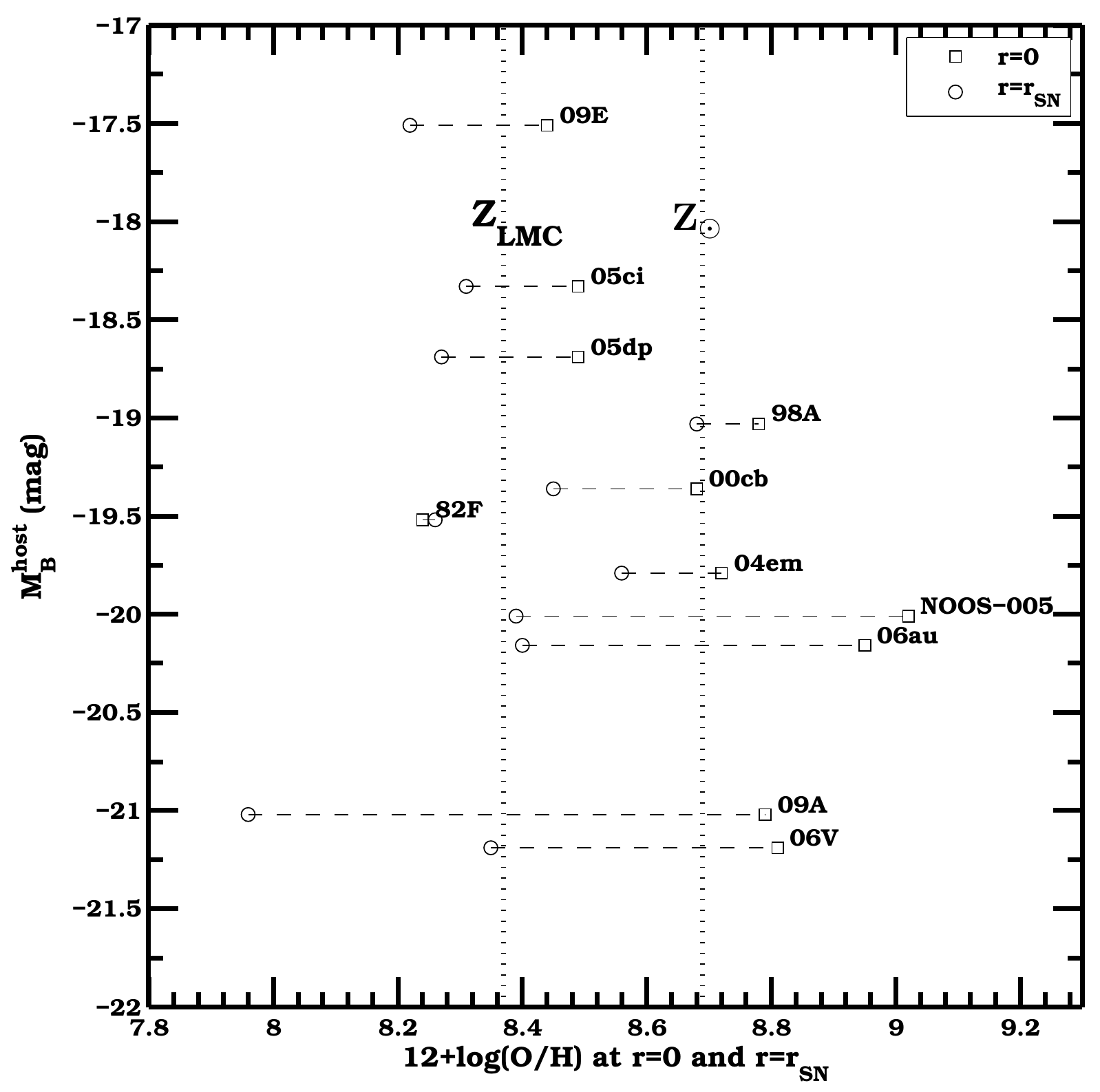}
  \caption{Directly measured metallicity at the host center and at the SN distance vs host galaxy M$_{B}$. A low central metallicity has been found for low-luminous galaxies, a higher metallicity for the luminous ones. This is consistent with previous studies concerning the luminosity-metallicity relation. The metallicity gradient explains why also the SNe hosted by bright galaxies were found at sub-solar metallicity. \label{mbvscentrmet}}
 \end{figure}

\begin{figure}
 \centering
\includegraphics[width=7cm,angle=0]{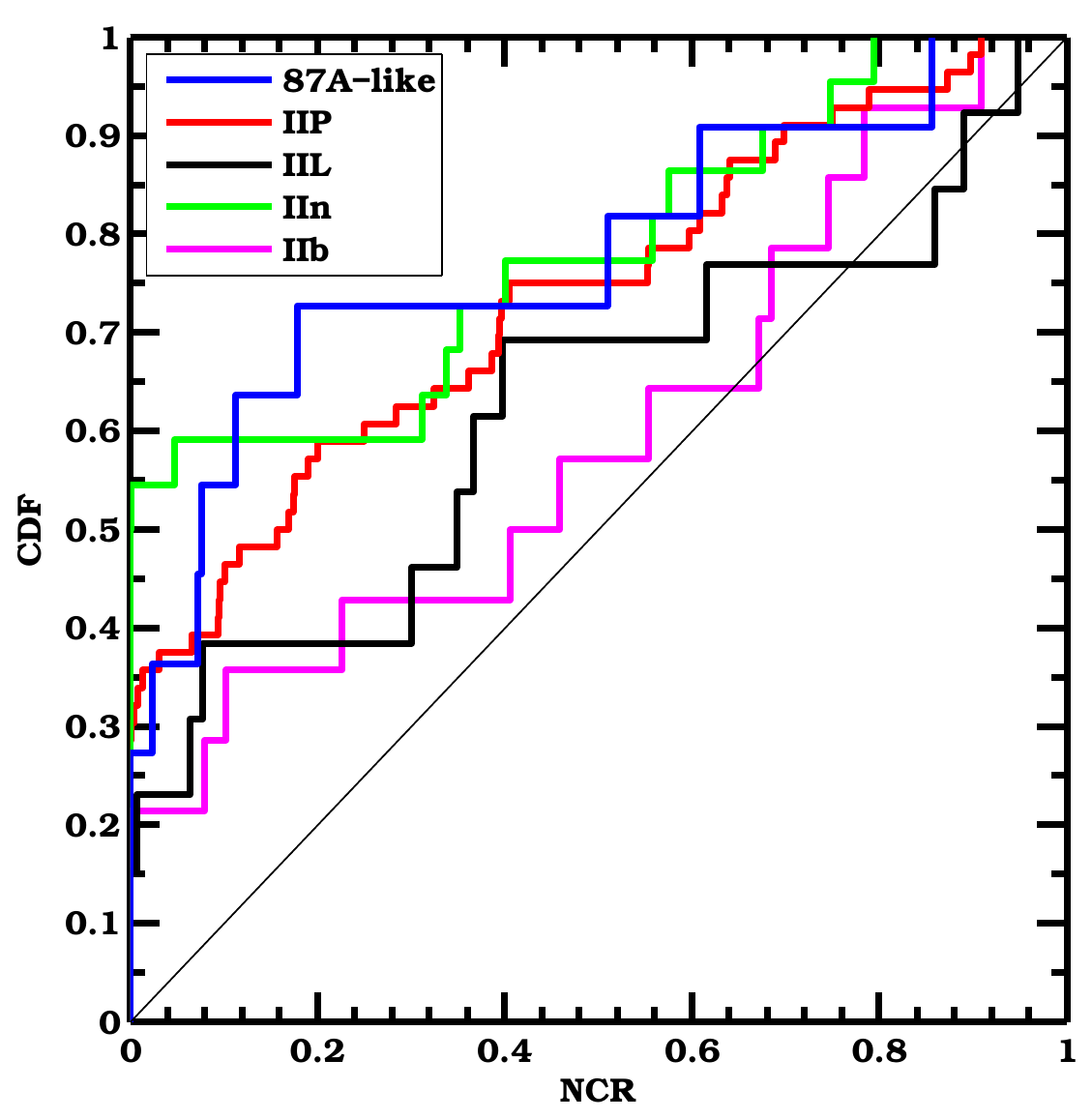}
  \caption{Cumulative distribution of the NCR values for our BSG~SNe and for samples of different CC~SN types from \citet{anderson12}. BSG~SNe seem to have a degree of association to star-forming regions similar to that of SNe~IIP and IIn.\label{NCRfig}}
 \end{figure}

\end{document}